\documentclass{report}
\usepackage{enumitem}
\usepackage[bottom]{footmisc}
\usepackage{graphicx}
\usepackage{subcaption,float}
\usepackage{hyperref}
\usepackage{xspace}
\usepackage[T1]{fontenc}
\usepackage[utf8]{inputenc}
\usepackage{xcolor}
\usepackage{booktabs}
\usepackage[many]{tcolorbox}
\usepackage{enumitem}
\usepackage{circledsteps}
\usepackage{authblk}
\usepackage{titling}
\usepackage[normalem]{ulem}
\usepackage[
  backend=biber,
  sorting=nyt,
  style=numeric-comp,  %
  maxbibnames=99,
  maxnames=99
]{biblatex}
\usepackage{cleveref}
\usepackage[ruled,lined]{algorithm2e}
\usepackage{longtable}
\usepackage{colortbl}
\usepackage{ragged2e}
\usepackage{makecell}

\setlist[itemize]{itemsep=0.5mm}

\definecolor{ForestGreen}{RGB}{34,139,34}

\definecolor{Aquamarine}{HTML}{00B5BE}

\definecolor{Fuchsia}{RGB}{255,0,255}

\definecolor{maryamcolor}{RGB}{139,0,139}

\definecolor{sencolor}{RGB}{85,107,47}

\definecolor{samcolor}{RGB}{255,140,0}

\definecolor{OrangeRed}{RGB}{255,69,0}

\definecolor{BernRed}{RGB}{228,0,70}

\newcommand{\sMiLe}{S\textbf{M}i\textbf{L}e Bank\xspace}
\newcommand{\MedicaL}{MedicaL Inc.~}

\newcommand{\progP}{\textsf{Program}\xspace}
\newcommand{\inputP}{\textsf{Input}\xspace}
\newcommand{\outputP}{\textsf{Output}\xspace}
\newcommand{\intentP}{\textsf{User Intent}\xspace}

\newcommand{\Props}{\textsf{Props}\xspace}

\newcommand{\adv}{\ensuremath{\mathcal{A}}}

\newtheorem{example}{Example}

\definecolor{keyInsightColor}{HTML}{DDCC77}

\definecolor{keyTakeawayColor}{HTML}{DDCC77}   %
\definecolor{keyTakeawayBackground}{HTML}{FFFFF0} %

\definecolor{questionColor}{HTML}{00BFFF}   %
\definecolor{questionBackground}{HTML}{F0FFFF} %

\definecolor{researchQuestionColor}{HTML}{E68A8A} %
\definecolor{researchQuestionBackground}{HTML}{FFF0FF} %

\definecolor{highlightColor}{HTML}{2E8B57} %
\definecolor{highlightBackground}{HTML}{F0FFF0} %

\definecolor{mythColor}{HTML}{852E8B} %
\definecolor{mythBackground}{HTML}{DFD3E6} %

\newlist{keytakeaways}{itemize}{1}
\setlist[keytakeaways,1]{%
  label=\(\triangleright\), %
  leftmargin=1.4em,
  itemsep=2pt plus 1pt minus 1pt,
  topsep=4pt plus 1pt minus 1pt
}

\newtcolorbox[auto counter, number within=section]{keyInsight}[1][]{%
  enhanced, breakable,
  colback=keyTakeawayBackground,
  colframe=keyTakeawayColor,
  coltitle=black,
  boxrule=0pt, frame hidden,
  borderline west={3pt}{0pt}{keyTakeawayColor},
  sharp corners,
  left=10pt,right=10pt,top=8pt,bottom=10pt,
  before skip=18pt plus 2pt minus 2pt,
  after  skip=10pt plus 2pt minus 2pt,
  fonttitle=\bfseries,
  title={Key Takeaway~\thetcbcounter\ifstrempty{#1}{}{:\ #1}},
  attach boxed title to top left={yshift=-2mm, xshift=8pt},
  boxed title style={colback=white,colframe=keyTakeawayColor,boxrule=0.5pt,sharp corners,
                     left=8pt,right=8pt,top=2pt,bottom=2pt},
  before upper=\vspace{4pt},
}

\newtcolorbox[auto counter, number within=section]{researchQuestion}[1][]{%
  enhanced, breakable,
  colback=researchQuestionBackground,
  colframe=researchQuestionColor,
  coltitle=black,
  boxrule=0pt, frame hidden,
  borderline west={3pt}{0pt}{researchQuestionColor},
  sharp corners,
  left=10pt,right=10pt,top=8pt,bottom=10pt,
  before skip=18pt plus 2pt minus 2pt,
  after  skip=10pt plus 2pt minus 2pt,
  fonttitle=\bfseries,
  title={Research Question~\thetcbcounter\ifstrempty{#1}{}{:\ #1}},
  attach boxed title to top left={yshift=-2mm, xshift=8pt},
  boxed title style={colback=white,colframe=researchQuestionColor,boxrule=0.5pt,sharp corners,
                     left=8pt,right=8pt,top=2pt,bottom=2pt},
  before upper=\vspace{4pt},
}

\newtcolorbox[auto counter, number within=section]{questionBox}[1][]{%
  enhanced, breakable,
  colback=questionBackground,
  colframe=questionColor,
  coltitle=black,
  boxrule=0pt, frame hidden,
  borderline west={3pt}{0pt}{questionColor},
  sharp corners,
  left=10pt,right=10pt,top=8pt,bottom=10pt,
  before skip=18pt plus 2pt minus 2pt,
  after  skip=10pt plus 2pt minus 2pt,
  fonttitle=\bfseries,
  title={Question~\thetcbcounter\ifstrempty{#1}{}{:\ #1}},
  attach boxed title to top left={yshift=-2mm, xshift=8pt},
  boxed title style={colback=white,colframe=questionColor,boxrule=0.5pt,sharp corners,
                     left=8pt,right=8pt,top=2pt,bottom=2pt},
  before upper=\vspace{4pt},
}

\newtcolorbox[auto counter, number within=section]{highlight}[1][]{%
  enhanced, breakable,
  colback=highlightBackground,
  colframe=highlightColor,
  coltitle=black,
  boxrule=0pt, frame hidden,
  borderline west={3pt}{0pt}{highlightColor},
  sharp corners,
  left=10pt,right=10pt,top=8pt,bottom=10pt,
  before skip=18pt plus 2pt minus 2pt,
  after  skip=10pt plus 2pt minus 2pt,
  fonttitle=\bfseries,
  title={{#1}{}{}},
  attach boxed title to top left={yshift=-2mm, xshift=8pt},
  boxed title style={colback=white,colframe=highlightColor,boxrule=0.5pt,sharp corners,
                     left=8pt,right=8pt,top=2pt,bottom=2pt},
  before upper=\vspace{4pt},
}

\newtcolorbox[auto counter, number within=chapter]{myth}[1][]{%
  enhanced, breakable,
  colback=mythBackground,
  colframe=mythColor,
  coltitle=black,
  boxrule=0pt, frame hidden,
  borderline west={3pt}{0pt}{mythColor},
  sharp corners,
  left=10pt,right=10pt,top=8pt,bottom=10pt,
  before skip=18pt plus 2pt minus 2pt,
  after  skip=10pt plus 2pt minus 2pt,
  fonttitle=\bfseries,
  title={Misconception~\thetcbcounter\ifstrempty{#1}{}{:\ #1}},
  attach boxed title to top left={yshift=-2mm, xshift=8pt},
  boxed title style={colback=white,colframe=mythColor,boxrule=0.5pt,sharp corners,
                     left=8pt,right=8pt,top=2pt,bottom=2pt},
  before upper=\vspace{4pt},
}

\newtcolorbox{execbox}[1]{
    colback=blue!5!white,    %
    colframe=blue!75!black,  %
    fonttitle=\bfseries,      %
    colbacktitle=blue!75!black, %
    title=#1,                 %
    enhanced,                 %
    attach boxed title to top left={yshift=-2mm, xshift=2mm}, %
    boxed title style={size=small,colback=blue!75!black},
    sharp corners,            %
    boxrule=0.5mm,            %
}

\title{\textbf{Crypto x AI, AI x Crypto: A Survey}}
\date{21 August 2026 \\ v1.1}

\begin{document}

\begin{titlepage}
  \begin{center}
    \begin{figure}
        \centering
    \includegraphics[width=.75\linewidth]{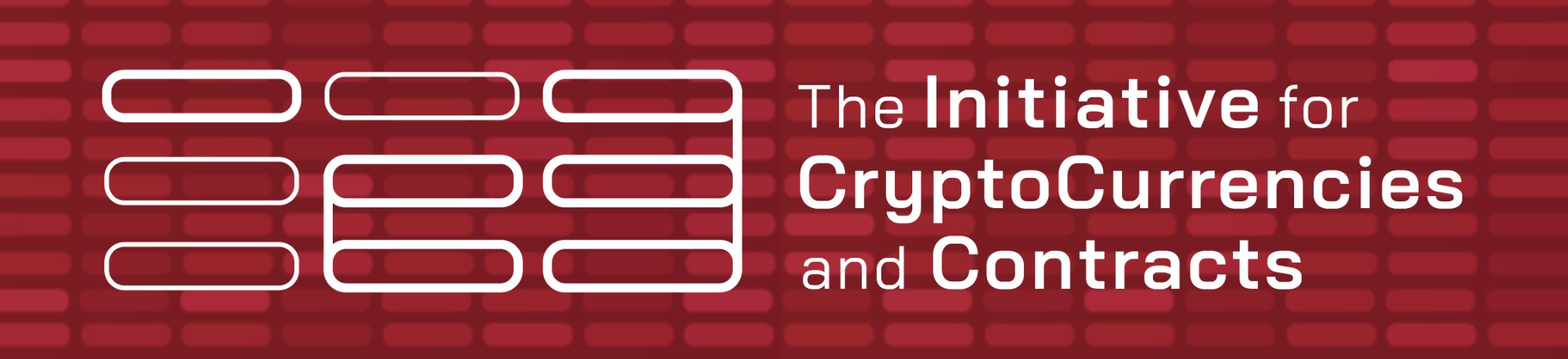}
        \label{fig:placeholder}
    \end{figure}
    \vspace*{1.5in}
    {\huge \thetitle} \\[1em]
    {\large \thedate} \\[2em]
    \vspace*{0.25in}
    \textbf{Editors:} Giulia Fanti$^{1,3}$ and Ari Juels$^{1,4}$ \\[1em]
    \vspace*{0.25in}
    \textbf{Authors:} \\[0.5em]
    Sarah Allen$^{1,5}$, 
    Pranay Anchuri$^{6}$, 
    James Austgen$^{1,4}$,
    Maryam Bahrani$^7$, 
    Samuel Breckenridge$^{1,4}$,
    Aaron Buchwald$^2$, 
    Christian Cachin$^{1,9}$, 
    Andrés Fábrega$^{1,4}$,
    Jared Fernandez$^3$,
    James Hsin-yu Chiang$^{1,12}$,  
    Marwa Mouallem$^{1,8}$, 
    Roi Bar-Zur$^{14}$, 
    Neil DeSilva$^1$,
    Ittay Eyal$^{1,8}$, 
    Giulia Fanti$^{1,3}$, 
    Ari Juels$^{1,4}$, 
    Andrew Miller$^{1,13}$,   
    Christian Sillaber$^{9}$,
    Dani Vilardell$^{1,4}$, Pramod Viswanath$^{11}$,
    Wenhao Wang$^{1,10}$, 
    Matt Weinberg$^{1,11}$, 
    Sen Yang$^{1,10}$, 
    Jianzhu Yao$^{11}$, 
    and Fan Zhang$^{1,10}$\\[10em]
    $^1${Initiative for CryptoCurrencies and Contracts (IC3)},
    $^2${Ava Labs},
    $^3${Carnegie Mellon University},
    $^4${Cornell Tech},
    $^5${Flashbots},
    $^6${Offchain Labs},
    $^7${Ritual Labs}, 
    $^8${Technion}, 
    $^9${University of Bern},
    $^{10}${Yale University},
    $^{11}${Princeton University},
    $^{12}${ETH Zurich}, 
    $^{13}${Teleport; Flashbots(X)},
    $^{14}${Tel Aviv University}
      
  \end{center}
\end{titlepage}

\renewcommand{\thechapter}{\Alph{chapter}} 
\renewcommand{\thesection}{\Alph{chapter}-\arabic{section}} 
\renewcommand{\thesubsection}{\thechapter-\arabic{section}.\arabic{subsection}}

\nopagecolor
\begin{execbox}{Executive Summary}
    \vspace{0.2in}
    The intersection of crypto x AI is spawning papers, products, online posts, and companies.  
    All the surrounding buzz, though, obscures what exactly has been done, what the opportunities and challenges are, and what open questions deserve attention. 
    This survey paper asks what AI can do for blockchain-based technologies (broadly construed as ``crypto'') (crypto x AI), and vice versa (AI x crypto). We systematize existing work, summarize key takeaways, highlight open research questions, and offer a perspective on pervasive industry misconceptions, concluding that AI and crypto are still in the very early stages of meaningful integration. Highlights include: 
    
    \vspace{2mm}
    \textbf{Where we are today:} 
    \vspace{-1mm}
    \begin{enumerate}
        \item \textbf{Crypto x AI:} \emph{AI can help analyze and detect important properties of existing crypto transactions, events, and protocols.} 
        A wide swath of work has explored AI-based approaches for detecting fraudulent or buggy smart contracts and protocols, for instance. These techniques have classically used simple machine learning approaches, and they are most effective in controlled settings with ample training data.
        \item \textbf{AI x Crypto:} \emph{Crypto tools offer new ways to \emph{secure} and \textit{govern} AI pipelines.}
        Several tools that are widely used in the crypto community---including zero-knowledge proofs and trusted computing---can be repurposed to make AI outcomes less susceptible to tampering. Other appealing ideas from the crypto community, such as decentralized governance and infrastructure management, have yet 
        to see real adoption in the mainstream AI community.

    \end{enumerate}

    \vspace{2mm}
    \textbf{Proof points needed by the crypto community:}
    \vspace{-1mm}
    \begin{enumerate}
        \item \emph{Decentralized AI solutions need more rigorous and direct cost comparisons to their centralized counterparts. }
        The crypto x AI industry has largely focused on demonstrating the feasibility of training large models in a decentralized fashion. While decentralization has its own merits, opportunities to compete with centralized AI platforms on cost for specific use cases and regimes need to be better quantified.
        \item \emph{Crypto rails need more rigorous articulation and demonstration of their utility for agentic payments over centralized alternatives.} While crypto lacks significant traction in the payments sector, agentic payments hold promise for their low fees and avoidance of the human-ownership model of traditional financial infrastructure. The crypto community should capitalize on this opportunity---and on growing crypto activity by traditional finance companies in payments and new agentic pilot projects---by quantitatively demonstrating the benefits of crypto for agentic payments, rather than  only demonstrating feasibility. 
    \end{enumerate}

    \vspace{2mm}
    \textbf{Research challenges:}
 \vspace{-1mm}
    \begin{enumerate}
        \item \emph{AI security requires system-level defenses.} The AI community generally tackles safety and security problems \textit{at the level of AI models}, designing guardrails and defenses around input / output semantics. As agents gain autonomy and access to infrastructure, this approach will prove inadequate. Crypto tools, including verified execution and authenticated pipelines, can help provide systems-layer assurances that model-level defenses cannot. 
\item \textit{Combining crypto with AI creates new threat actors and vectors.} AI applications such as investment-portfolio management create an unavoidable privacy vs.~fairness tension, while merging AI agents with decentralization and cryptocurrency can create dangers such as \textit{unstoppable autonomous agents} or rogue smart contracts. Understanding which threats are realistic and what mitigations will be effective are both research imperatives.
    \end{enumerate}

    \vspace{2mm}

\end{execbox}

\newpage

\tableofcontents

\chapter{Introduction}
\label{chap:intro}

\label{sec:intro}

In the space of emerging technologies, crypto(currencies)
and  artificial intelligence (AI) have  received perhaps unprecedented levels of attention, excitement, hype, and skepticism \cite{alles2023hope,ai-investment,ai-crypto-investment,ai-crypto-investment2,paradise}. 
Today, there exist myriad proposals to revolutionize crypto with AI and vice versa \cite{coinbase-agentic,cryptoslate,forbes-ai-crypto,kshetri2025building}. 
For observers, it can be challenging to tease apart the real use cases and understand when and how AI and crypto fit together. 

This survey paper presents a unified framework for categorizing the connections between AI and crypto. 
We will show how existing research maps to our proposed framework, and what major research questions remain unanswered. Additionally, we aim to highlight popular trends that \emph{do not} fit into our framework, and/or we view as unrealistic use cases for the time being. Throughout the survey, we use Crypto x AI to mean AI applied to crypto; AI x Crypto means crypto applied to AI. 

\paragraph{What are ``crypto'' and ``AI''?} 

We use the term ``crypto'' in roughly three senses. First, historically ``crypto'' has been an abbreviation of  ``cryptography,'' and while it originally referred to hidden messages, it now refers to a broad range of techniques for securely storing, transmitting, and computing on (possibly) confidential data. The cryptography toolbox includes digital signatures, and threshold or multi-party computation. Blockchain developers have been among the earliest widespread adopters of certain advanced cryptographic tools such as these and have especially propelled the evolution of  \textit{zero-knowledge} (ZK) proofs, which enable users to prove knowledge of secrets without disclosing them.\footnote{ZK proofs are used for ``privacy coins,'' such as Zcash, where they prove that secret transactions were correctly executed. Often, though, ZK in crypto circles is technically misapplied to  ``succinct'' proofs---compact proofs (often in the form of what are called STARKs or SNARKs) used to prove valid processing of transactions. The confusion arises because STARKs and SNARKs can be ZK.} All of these cryptographic tools have application to securing and mediating the use of AI as we’ll examine throughout this survey.

We will also include under the banner of ``crypto''  the closely-aligned technology of \textit{trusted computing}. This term refers to special computing environments---often backed by special-purpose hardware---that aim to secure software applications by preventing tampering and leakage of secrets (although with important provisos). They are increasingly important in blockchain applications, which is in turn helping to fuel their growth and their use in non-blockchain applications, as we explain in this survey.\footnote{E.g., today nearly half of blocks in Ethereum are constructed using trusted computing within block-building infrastructure~\cite{buildernet2024}.}

Second, following the popularity of Bitcoin, Ethereum, and other blockchain-based systems over the last decade, crypto is short for ``cryptocurrencies.'' This is an entire economic layer built out of cryptographic primitives, including tokens, stablecoins, decentralized finance (DeFi), as well as the ecosystem of exchanges and other service providers built around them. Several of the characteristic features here are directly relevant for AI, such as irreversibility, the ability to create accounts and transfer funds without registration or identification, and programmable settlement rules.

Third, ``crypto'' can be understood as a cultural movement with a set of values revolving around permissionless innovation, resilience through decentralization, and avoiding reliance on trusted third parties and intermediaries. In this sense, crypto is a continuation of the ``cypherpunk'' tradition that produced end-to-end encrypted messaging and BitTorrent, for example, and now informs the engineering tradeoffs made in cryptocurrencies and modern cryptography-based systems. The ambiguous nature of these values is a recurring source of tension: many projects in the cryptocurrency ecosystem are ``decentralized in name only''~\cite{peirce2021}. At the same time, it has been at least partially successful: the U.S. regulatory architecture has gradually adapted around it, with FinCEN's 2013 guidance distinguishing ``users'' of virtual currency (miners spending what they mined; ordinary holders) from ``exchangers'' and ``administrators,'' classifying only the latter as money transmitters~\cite{fincen2013}.

The term ``AI'' is similarly nebulous, but denotes systems that can perform tasks typically requiring human intelligence, such as  reasoning, problem-solving, understanding language, recognizing patterns, and making decisions. AI encompasses many approaches and techniques, but in this survey, we focus primarily on  \textit{machine learning} (ML). ML systems learn from data or an environment to achieve a target goal, rather than being explicitly programmed for a specific set of tasks. Accordingly, we use the terms AI and ML in this survey interchangeably.

\subsection*{Why do we need another survey? }
Many prior works explore the intersection of AI and crypto \cite{van2024humans,amirzadeh2022applying,yang2022fusing,zuo2023survey,salah2019blockchain,pathak2023qualitative,hussain2021artificial,srivastava2024blockxai,pandl2020convergence,bellagarda2022updated}. 
Most of these resources are focused on specific domain areas, such as finance and trading \cite{amirzadeh2022applying}, smart environments and the metaverse \cite{yang2022fusing,fadi2022survey,shen2023blockchains,zuo2024exploring}, and even enhancing the security of wireless communications systems such as 6G~\cite{zuo2023survey,pathak2023qualitative}, to name a few. 
Our goal is to take a broader view, focusing on the ecosystem at large rather than a specific vertical; 
at the same time, we acknowledge that many of the most widely-deployed applications of blockchains are centered around finance and cryptocurrency \cite{hu2019market,labazova2019hype}.

Among general-purpose surveys, the majority either focus on the impact of blockchains on AI~\cite{salah2019blockchain,hussain2021artificial,srivastava2024blockxai} or the impact of AI on blockchains~\cite{zebari2025comprehensive,karim2025ai,abdelhamid2024review,soori2023ai}, with relatively few covering both \cite{bellagarda2022updated,pandl2020convergence}. 
Further, the surveys that cover both directions do not explicitly separate future directions in terms of infrastructure vs. applications, and they focus on blockchains in a precise sense of the word, considering only distributed ledger technology (DLT)-based deployments. 

This survey takes a broader view, giving rise to a few key differences or traits:
\begin{enumerate}
    \item We discuss the potential impact of AI on crypto and vice versa. To this end, we frame both technologies as \emph{middleware} between humans and automated decision-making pipelines. Our framework explicitly separates how AI and crypto can be used at different layers of the crypto stack, rather than jointly listing different use cases at the same level of abstraction. 
    \item We use ``crypto" as an umbrella term, including not just blockchain stacks and applications, but decentralized systems built on trusted computing tools and technologies. 
    \item Most prior surveys came out before the explosive growth of generative and agentic AI, which we include in our survey. 
    \item In addition to a survey of existing literature, we share our collective views and interpretation of current trends, both in the research and industry domains, that  connect crypto and AI. In this spirit, we aim to identify promising research directions in a wide field of contenders.   
\end{enumerate}

\section{A Framework for  AI-Crypto Interactions}
\begin{figure}[!ht]
    \centering
    \includegraphics[width=0.85\linewidth]{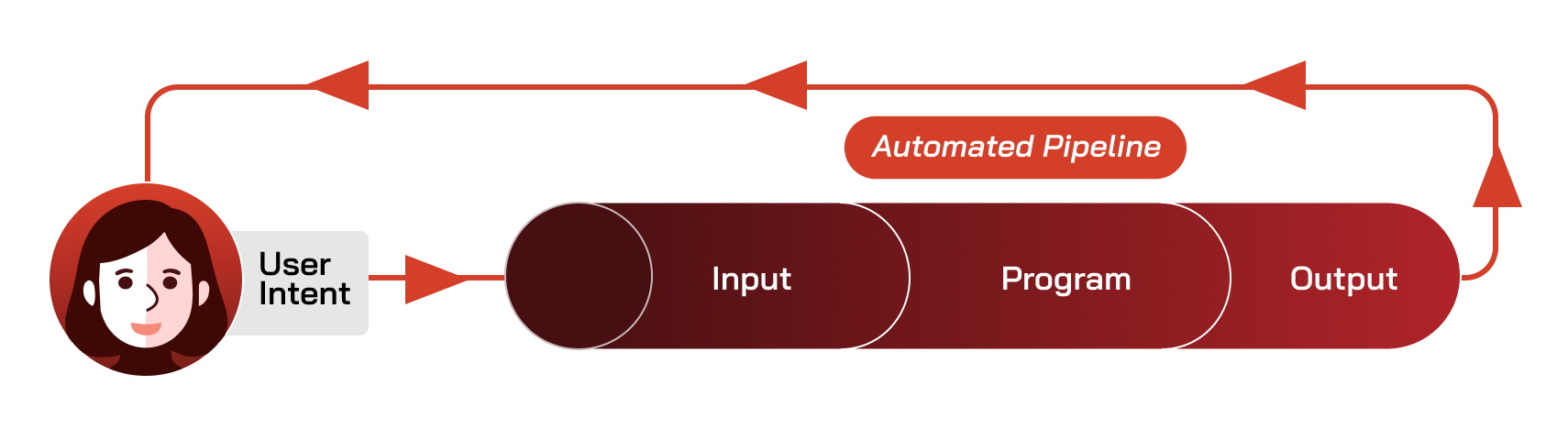}
    \caption{A standard automated decision-making pipeline. A human \intentP (e.g. ``I want to stop my car at stop signs") is translated into a \progP that processes structured \inputP (e.g., video streams from a car) and produces \outputP decisions (e.g., stop the car). This processing is handled autonomously by \progP.}
    \label{fig:pipeline}
\end{figure}

Many important decision-making pipelines translate human intentions into an automated processing pipeline, as shown in Figure \ref{fig:pipeline}.  

\bigskip
\noindent 
Let us consider the following example:  a user wants to ensure that their self-driving car stops at stop signs. We call this goal 
a human intent (denoted \intentP in \Cref{fig:pipeline}). 
To automatically fulfill a human intent, we must first specify a \progP that decides when to stop the car. 
This program should take \inputP from the environment; in this case, the input might be a sensor stream of the environment surrounding the car (e.g., video, LIDAR).
The program uses its input(s) to attempt to fulfill the user's intent: if the program ``sees" a stop sign, it produces the \outputP signal for the car to stop. 
Note that in the classical Sense-Think-Act framework of robotics and artificial intelligence \cite{russell2020aima}, the \outputP can be viewed as the output of the ``Think" component of the loop---e.g., the signal to an actuator. We can summarize this decision-making pipeline as follows: 
\begin{itemize}
    \item \intentP = ``I want to stop my car at stop signs''
    \item \inputP = Sensor readings of current surroundings of the car 
    \item \progP = Checks if there’s a stop sign in the input 
    \item \outputP = ``Send a STOP code if there is a stop sign''
\end{itemize}

\paragraph{The Role of Trust.}
In a computer-assisted decision-making pipeline, we may not trust any of the links in this chain. That is, we may not trust that our program accurately reflects human intent, we may not trust that our program is running on the inputs we think it is, and we may not trust that the output is computed from the program and inputs we think it is. Hence, a central question is:

\bigskip
\begin{questionBox}[Decision-making pipeline utility]
\begin{center}
    \noindent \textbf{How do we ensure that decision-making pipelines are both useful and trustworthy? }
\end{center}
\end{questionBox}

\bigskip

\noindent Two important technologies that can help with this question are \textbf{Trusted Computers}, often aided by \textit{decentralization}, and \textbf{AI Models}. These will be the focus of our survey.

\subsection{Trusted Computers}
\label{subsec:trusted_computers}

In recent years, trusted computers have become increasingly prevalent. A trusted computer is a system that executes programs with the aim of guaranteeing that a (correctly constructed) program does what it is told, and/or we can verify that the program did what it was told after the fact.  

Examples of trusted computers include: 

\begin{itemize}
    \item \textbf{Trusted execution environments (TEEs)}:
    TEEs are dedicated computational modules that provide isolation and other security guarantees; notably, they are included as part of the system on chip (SoC), which gives them flexibility over other types of trusted hardware \cite{sabt2015tee}.

    \item \textbf{Verifiable Computing (a.k.a.~SNARKs, or ZK)}:
    Verifiable computing refers to cryptographic techniques for proving that a given computation was carried out correctly. Verifying the proof is typically much less costly than carrying out the computation from scratch.
    These are also known as ``snarks," or ``zkVMs" depending on other secondary properties. 
    
    \item \textbf{Blockchains}: 
    A blockchain is a decentralized system that is endowed with the ability to execute and commit to specific types of computations, such as processing transactions and/or executing blockchain programs, known as smart contracts \cite{nakamoto2008bitcoin,wood2014ethereum}. Like TEEs and Verifiable Computing, blockchains are designed to ensure correct execution of programs. Unlike TEEs and Verifiable Computing, blockchains are run over a group of nodes who collectively come to consensus over the state of the system. Blockchains also do \textit{not} natively enforce confidentiality.  
\end{itemize}

Different implementations of trusted computers have different implications for security and performance, which we discuss in \Cref{sec:basics}. However, the core goal is similar: to verify the state of a computation with strong assurance. 

\subsubsection{Properties}
\label{subsubsec:properties}
At their core, trusted computers can offer three main security properties: \textbf{confidentiality}, \textbf{integrity}, and \textbf{availability}. (The classic CIA triad of computer security.) Different trusted computers achieve different subsets of these properties.

\paragraph{Integrity.}
In the computer security literature, \textbf{integrity} means ensuring that computations or communications have not been tampered with. For instance, maybe we are worried that \progP or \inputP got corrupted. Trusted computers can provide two important notions of integrity: 

\begin{itemize}
\item \textbf{Computational Integrity:} 
It is not uncommon for organizations to claim they are running one program, when in reality they are (suspected of) running  a different one \cite{batterygate,tesla-autopilot}. 
Computational integrity ensures that the trusted computer actually ran the \progP that it claims; it solves the following problem: \textbf{``I don’t trust that \textbf{\outputP} = \textbf{\progP\hspace{-1mm}(\inputP)}.''}
\bigskip

A trusted computer can give assurance in the following form: 

\begin{figure}[h!]
    \centering
    \includegraphics[width=\linewidth]{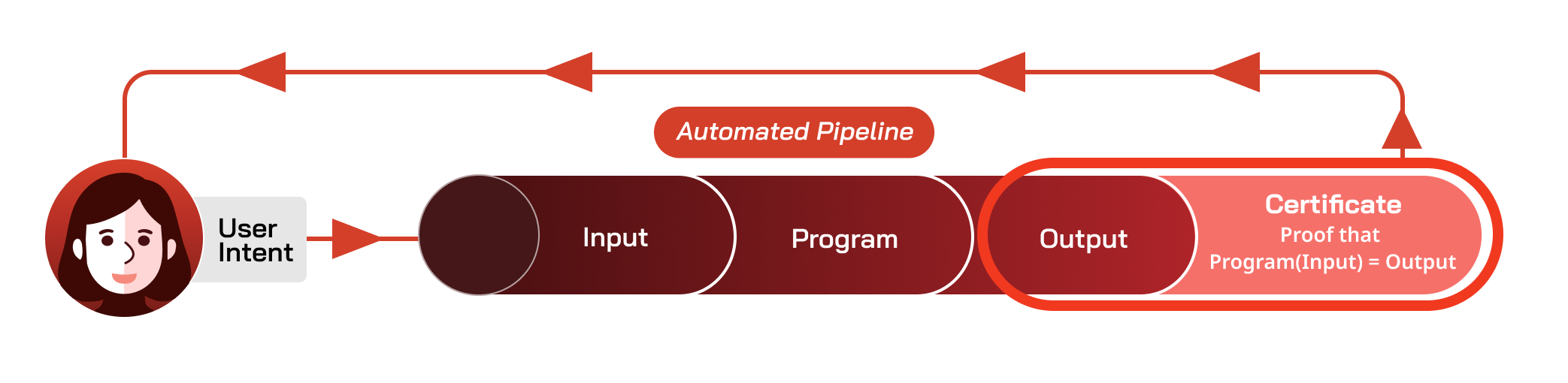}
    \caption{A decision-making pipeline implemented with a trusted computer can give an assurance that the output was actually obtained by running \progP on \inputP.}
    \label{fig:cert}
\end{figure}

Namely, it can output a certificate attesting to the fact that \outputP was produced by running \progP on \inputP. 
This certificate comes with \emph{cryptographic} guarantees, which ensure under standard cryptographic assumptions that the certificate can be forged only with negligibly small probability. 

\item \textbf{Data Integrity:} It is common in decision-making pipelines for \progP not just to ingest input from a user, but also to \textit{fetch data from the web}. A trusted computer can ensure that \progP fetches such web data correctly. This provides \textbf{data integrity}. In blockchain terminology, the part of a trusted computer that performs this operation is often referred to as an \textit{oracle}.

An oracle can ensure only that \progP retrieves data from a particular source. If \progP is designed to fetch a particular weather report (e.g., weather in NYC) from a particular website (e.g., \texttt{www.trustyweather.com}), the oracle ensures that it in fact comes from that site. The oracle cannot ensure that the data itself is correct (e.g., that \texttt{www.trustyweather.com} reports correctly that there's rain in NYC), but the trustworthiness of the source can often serve as a strong proxy for the trustworthiness of its data. 
\end{itemize}

\paragraph{Confidentiality.} 
Certain types of trusted computers, including TEEs (mentioned above) and some privacy-aware blockchains, can additionally endow parts of a decision-making pipeline with \textbf{confidentiality}. This notion is sometimes called \textit{confidential computing}~\cite{russinovich2024confidential}, a term we will use throughout this survey. 
TEEs, for example, can conceal \intentP, \inputP, \progP, and \outputP---indeed, any or all parts of a pipeline. We can think of the trusted computer as a ``black box'' that is programmed to conceal the pipeline by default and just reveal selected data  to certain users. Schematically, ``black-box'' privacy with a TEE---which, as a trusted computer, also enforces integrity---adds confidentiality as in~\Cref{fig:confcomp}:

\begin{figure}[h!]
    \centering
    \includegraphics[width=1\linewidth]{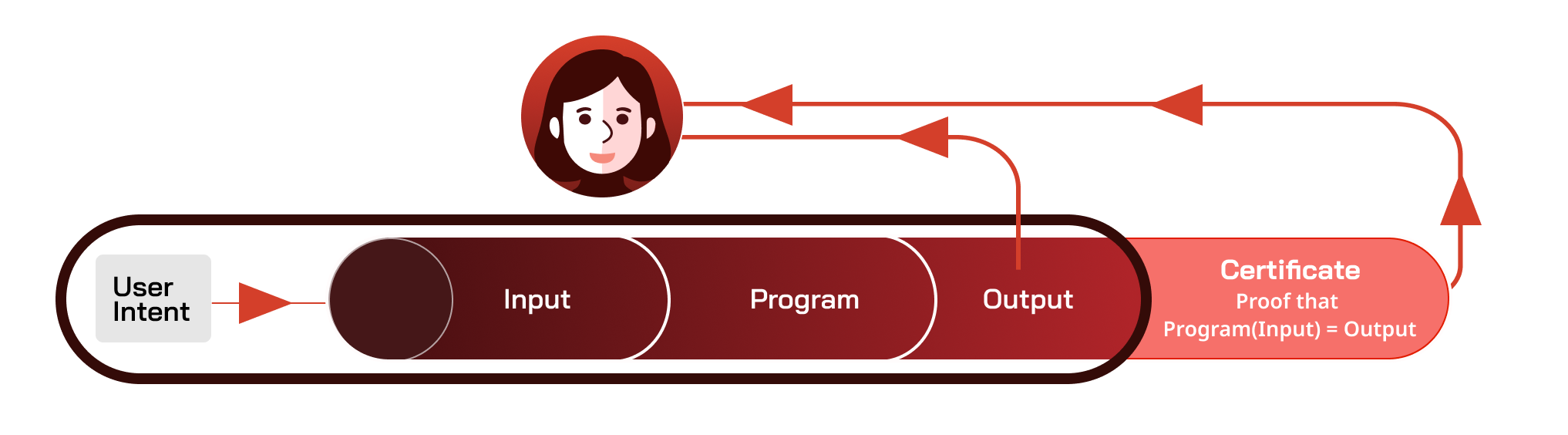}
    \caption{Trusted computer enforcing privacy in a decision-making pipeline. The user is only able to access pre-defined parts of the pipeline, such as \outputP.}
    \label{fig:confcomp}
\end{figure}

\paragraph{\textbf{Availability.}} A key feature of blockchains is their robust \textit{availability}. Blockchains are designed to have high uptimes, in some cases approaching 100\% over a period of years. (Bitcoin, for example, has had no major outages for well over a decade to date, while other chains have had repeated ones~\cite{helius2024solanaoutages}.) Most distinctively, however, blockchains also in principle provide \textit{censorship resistance}, meaning that they are available to users in highly adversarial environments that may include adversarial control of blockchain entities, e.g., a subset of validators.  

\bigskip
Hence, trusted computers help to solve the following problem:

\begin{highlight}[Key Trusted Computing Use Case (High-Level)]
    A program \emph{user} does not trust whether a program output was obtained correctly. Additionally, the program \emph{owner} may  want to hide the internal state of the program from the user. Trusted computing provides assurances of correctness and/or confidentiality to the user, as well as downstream applications of the program output. It may also ensure that a program is consistently available to the user.
\end{highlight}

\subsection{Decentralization}
\label{subsec:decentralization}

Blockchains are an important category of trusted computers whose key distinguishing feature is their \textit{decentralization}. A public blockchain is not truly a blockchain without this property. Decentralization, however, is not a necessary feature of crypto as we define it in this survey: Two types of trusted computer described in~\Cref{subsec:trusted_computers}---TEEs and ZK---are agnostic to whether a system is centralized or decentralized. It is important therefore to explain what decentralization is and what special properties it can impart to trusted computers and trusted-computing ecosystems. 

There are many different definitions of decentralization used in the crypto community today, as well as many different decentralization metrics~\cite{ovezik2025sok,fabrega2025vbe}. Informally, however, decentralization means that no single entity (or small cluster of entities) can exert meaningful control over a system. 

In a blockchain, the ``control'' in question often corresponds to \textit{censorship}, meaning the ability to prevent the on-chain inclusion of a target valid transaction or class of valid transactions. Conversely, decentralization in this context is called \textit{censorship resistance}, and corresponds to a universal notion of the \textit{availability} discussed in~\Cref{subsubsec:properties}. Censorship resistance means that a user can always cause a valid transaction to be included on a blockchain in a timely way. While a seemingly narrow property, availability is in fact quite powerful: It means that a system remains open to all users and that user assets cannot be confiscated by blocking transactions (as happens in centralized systems, e.g., traditional banking). 

Another important notion of blockchain decentralization regards \textit{governance}, the question of how decisions are made regarding the management of a blockchain or applications running on it. Such governance has been a topic of particularly active interest in the context of decentralized autonomous organizations (DAOs), which typically take the form of communities organized around smart contracts (see, e.g.,~\cite{sharma2024unpacking}). Membership in a DAO is a function of ownership of DAO tokens and decisions of consequence---roles within the DAO, technical upgrades or modifications, etc.---are determined by token-weighted votes. Governance mechanisms continue to evolve. As an example, there has recently been call to use AI agents to vote on behalf of DAO members as a means to avoid the decision-fatigue that many perceive as undermining governance effectiveness~\cite{rodrigues2026buterin}. 

Finally, decentralization works in the service of both \textit{integrity} and \textit{availability} as expressed in~\Cref{subsubsec:properties}. Lack of centralized control also means lack of a single point of failure. Put another way, to gain control of a highly decentralized system---in order to compromise either its integrity or availability---an adversary must compromise a multiplicity of entities, which is typically harder to orchestrate than a focused attack on one entity. 

As we explain in this survey, the properties brought about by decentralization have a number of potential uses in AI settings.

\begin{highlight}[Key Decentralization Use Case (High-Level)]

A system (be it a platform or technology) is subject to concentrated control: a single or small set of entities can dictate who gets to use it and how it evolves. Decentralization technologies can instead help ensure broad access to the system's resources and decision-making processes. 
\end{highlight}

\subsection{AI Models}

In parallel with the rise of trusted computing, AI and machine learning (ML) have created an upheaval in the world of technology and society at large. AI models can achieve many end goals, but for our purposes, we view them as translating a user’s intent into programs (i.e., a pipeline of \inputP $\to$ \progP $\to$ \outputP) that realize the user’s intent. Previously, the task of designing \progP would have been accomplished with a laborious manual process of software design and engineering, based on domain knowledge and many iterations. AI allows us to instead learn by example: we can use data that reflects our intent to define a program to execute it. 
Note that in this survey, we focus only on ML models trained from data and/or environments, such as discriminative, generative, or reinforcement learning models;  broader interpretations of AI (e.g., including those based exclusively on classical rule-based systems) are considered out of scope.

For instance, a user may know that they want to stop at stop signs, but they may not know exactly how to define a \progP that can take an image from a dashboard camera and identify a stop sign. AI can learn that program from representative (\inputP, \outputP) pairs. In doing so, it provides a different interface for users to translate human intents into a computing pipeline. 
Some examples of such translation include: 
\begin{itemize}
    \item \textbf{Discriminative models} can be viewed as programs that translate an input (e.g., an image) into a conditional output (e.g., a label). A model architecture can be viewed as the class of possible programs (functions) that can be learned; we use data and ML techniques to learn model weights, thereby specifying \emph{which} element of the function class is best. Hence, we are using AI to translate an intent (``label an image'') into a program (a trained model that maps images to labels).
    \item \textbf{Generative models} instead capture a different intent: to produce samples from a given \emph{unlabeled} data distribution. As before, we can train models from data to satisfy this intent, thereby learning \progP. 
    \item \textbf{Reinforcement learning} does not (classically) directly learn from data, but rather from an environment and a reward function. However, it has the same property that a human intent (e.g., ``learn to win any chess game") is translated into a program (i.e., a chess-playing strategy) using ML. 
\end{itemize}

\begin{figure}
    \centering
    \includegraphics[width=1\linewidth]{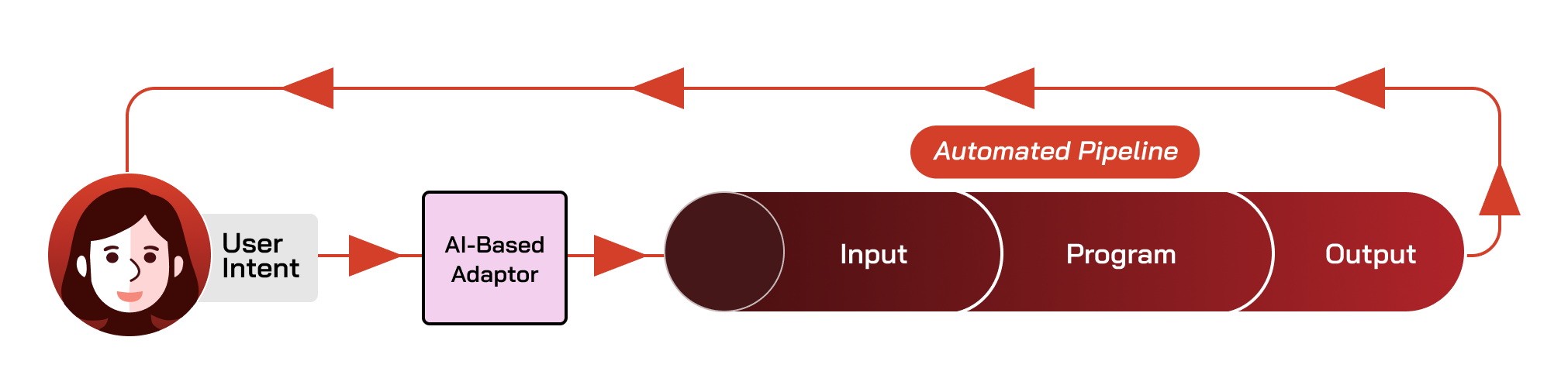}
    \caption{AI can help translate human intents  (typically coupled with data and/or information about an environment) into a well-specified \progP.}
    \label{fig:ai-use-case}
\end{figure}

Hence, in this survey, we will think of AI as solving the following problem, illustrated in \Cref{fig:ai-use-case}:

\begin{highlight}[Key AI Use Case (High-Level)]
    The program \emph{owner} does not know how to define \progP to accurately reflect the human intent. AI can help to translate human intents, expressed via examples or natural language for instance, into programs.
\end{highlight}

\subsection{Unifying Framework: Crypto and AI as Middleware}
\begin{figure}[t]
    \centering
    \includegraphics[width=\linewidth]{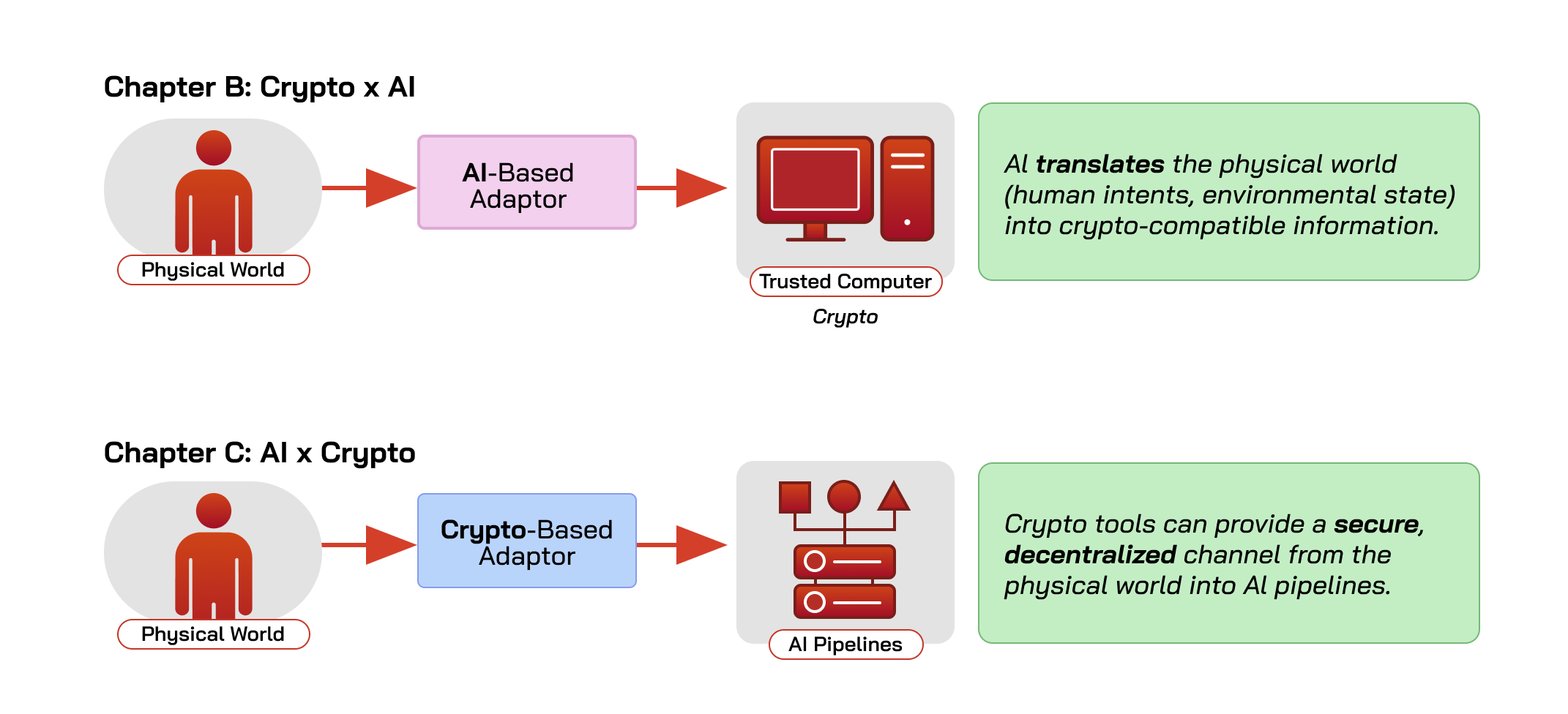}
    \caption{In this survey, we argue that crypto and AI can act as middleware for each other, connecting the physical world to virtual systems (e.g., a trusted computer or an AI pipeline). However, the form of these interfaces looks very different in the two directions, as we explore in~\Cref{chap:ai4block,chap:block4ai}.}
    \label{fig:overview}
\end{figure}

At face value, AI and crypto are solving very different problems. 
However, they have an important property in common: both technologies can be viewed as \emph{middleware} in decision-making pipelines. 
In the classical computer science literature, middleware refers to software that sits between complex systems, enabling communication and interaction between those systems \cite{middleware}. 
An example of classical middleware would be a middlebox in a network that inspects, transforms, and filters packets according to rules specified by the network administrator \cite{gember2012toward}. 
In our case, we can abstractly view both AI and crypto (both the infrastructure and applications) as middleware between humans and computer systems. 
We illustrate this relation in Figure \ref{fig:overview}.

\paragraph{AI as middleware between the physical world and blockchains.}
We saw previously that AI can help to translate a human user's intent into a program that executes the intent (e.g., \Cref{fig:ai-use-case}); in other words, it acts as translation middleware. 
This  capability is particularly useful in the context of blockchains, which are notoriously difficult to use.  
AI could significantly lower barriers to the design, analysis, and even simple use of existing blockchains. The key is to deploy it appropriately at the interfaces between humans and blockchains. Examples of promising application categories include:
\begin{itemize}
    \item \textbf{Constructive Tasks (Infrastructure):} Today, humans design algorithms for the blockchain technology and governance stack using laborious, manual processes. AI could help accelerate this process by proposing or discovering foundational algorithms across the blockchain stack based on desired properties expressed by a human. If successful, human designers could simply \emph{evaluate} a proposed design, rather than designing it from scratch.
    \item \textbf{Constructive Tasks (Application):}  Today, using blockchains---particularly in cross-chain settings---is generally a difficult and painful process. Users regularly implement smart contracts that do not reflect their initial intent---sometimes with disastrous consequences \cite{dao}. AI could assist in this problem by translating human preferences into proposed smart contracts, as well as searching for security vulnerabilities or logic flaws. 
    \item \textbf{Analytic Tasks:} Today, a great deal of efforts goes towards analyzing blockchain transactions to understand the current state of the system, both in terms of macro properties and individual trades (e.g., for fraud detection). AI could help facilitate  analysis of blockchain dynamics, given  high-level conditions and large quantities of data.
\end{itemize}
We discuss this categorization of AI applications in \Cref{chap:ai4block}.

\paragraph{Crypto as middleware between computing platforms and the physical world.}

We saw previously that trusted computing can give humans assurances of correctness (possibly under confidentiality constraints), while decentralization can help ensure availability. 
Crypto can thus be viewed as middleware between an automated decision-making pipeline and a human (or another program) who must \emph{trust} the output and availability of the pipeline.
Indeed, this kind of trust mediation is conceptually similar to what security-oriented middleware does in computer systems: it (attempts to) ensure that information reaching a target is trustworthy and arrives reliably. 
In the AI space, examples of promising applications include:
\begin{itemize}
    \item \textbf{Trustworthy data:} Today, data for ML models is often used in training without ensuring that the data sources are trustworthy. However, this can lead to system-critical failures in high-stakes applications like medical diagnosis, infrastructure management, or security. Hence, trusted computing offers middleware between AI data providers and data  users, allowing the recipient to trust the provenance of data. 
    \item \textbf{Trustworthy computation:} Similarly, training processes are typically conducted behind closed doors, and it can be difficult to discern whether a training process was conducted correctly. Trusted computing offers a powerful method for guaranteeing to downstream users that models were trained according to pre-defined specifications. To help ensure responsible creation of the specifications themselves, decentralization, in the form of AI governance, can serve as a vehicle for community input and oversight.     
    \item \textbf{Private data:} Finally, a major use case for trusted computing is the ability to conduct trustworthy operations without leaking information about data inputs. This is a massive problem in the AI space today, as many companies want models that are tailored to proprietary, internal data. Trusted computing is one of the most practical solutions for obtaining trustworthy results without exposing data in plaintext to the party executing model training. In this example, the necessary notion of ``trust" encompasses both the correctness of the program output, as well as privacy guarantees over the inputs. 
\end{itemize}
We discuss this categorization of crypto applications in \Cref{chap:block4ai}.

\subsection{Survey Roadmap}

This viewpoint of crypto and AI as middleware will inform the kinds of applications and research questions we consider. We divide the survey into two main components:

\begin{enumerate}
    \item \textbf{Crypto x AI}: How can AI systems amplify the capabilities of crypto systems?
    \item \textbf{AI x Crypto}: How can crypto systems help secure AI systems and give them new capabilities?
\end{enumerate}

This survey is split into chapters. 
We begin with a more detailed preliminary presentation of the relevant technologies in the remainder of \Cref{chap:intro}.
As illustrated in \Cref{fig:overview}, \Cref{chap:ai4block} discusses the benefits that AI can bring to crypto, including by enhancing our ability to analyze and use blockchain systems. \Cref{chap:block4ai} discusses the ways in which crypto can improve AI systems, particularly related to bolstering their decentralization, security, and privacy. 
We conclude in~\Cref{chap:conclusion} by presenting our viewpoints on common misconceptions (or incomplete characterizations) that persist in the AI x crypto community.

\section{Basics of Trusted Computing}
\label{sec:basics}

This section provides an overview of the goals of trusted computing and the various ways to achieve its various properties.

\subsection{Trusted computing example: Inference with a single user}
\label{subsec:single_user_inf}

As an example to illustrate how trusted computing can be used in conjunction with AI systems, we consider a simple scenario in which a \textit{single user} provides inputs to an ML model for \textit{inference}.

\begin{figure}[h!]
    \centering
    \includegraphics[width=0.8\linewidth]{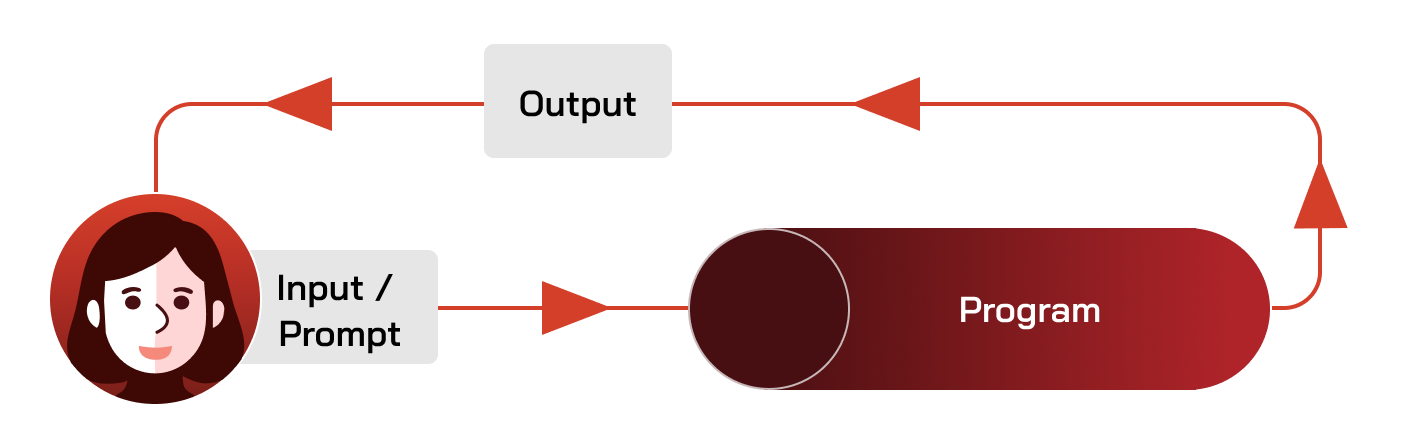}
    \caption{User furnishing a prompt to a model.}
    \label{fig:userprompt}
\end{figure}

\noindent Suppose, for example, that I input this prompt to an online LLM platform such as ChatGPT or Claude: 

\bigskip
\noindent \texttt{My shadow is the wrong color. What is this medical condition called?}
\bigskip

There is no privacy concern when it comes to me, the user. I provided the prompt, so it would not make sense to try to keep it private from me. The same goes for whether I should trust the prompt. Since it came from me, I am sure that it matches my intended use of the LLM. 

However, even in this common scenario, there \textit{are} two natural security concerns. Broadly, they are whether I can \textbf{trust} the output of the LLM platform and whether \textbf{privacy} is enforced for both the prompt and the output, as well as for the AI system itself. 

\subsubsection*{Output integrity}

Suppose that I receive this response to my prompt above:

\bigskip

\smallskip
\noindent\texttt{FINAL NOTICE: Medical Alert Regarding ``Umbrachromotosis''}

\smallskip\noindent    \texttt{Umbrachromotosis: Your shadow is the wrong color.}

\smallskip
\noindent\texttt{Case ID: DBB-44721 * Patient Status: URGENT}

\smallskip
\noindent\texttt{Our automated diagnostic flagged indicators of Umbrachromotosis. Untreated, this \\ condition may escalate and can be fatal within 7-14 days.
Good news: a cure is \\ available, but for one day only.}

\smallskip
\noindent\texttt{To unlock your personalized cure and secure a same-day consultation, send a deposit of 0.24 ETH immediately to \texttt{0x728877d47ac48dbd17a1e95f7b1dec20be6fb8d6}.}

\smallskip
\noindent\texttt{Time remaining: 02:13:57.}
\bigskip

A properly trained LLM shouldn't respond like this and solicit cryptocurrency payments! Seeing this response, you would naturally wonder if 
the LLM was corrupted somehow and hallucinated or whether it has been hacked by someone trying to steal cryptocurrency from users. In other words, for

\begin{itemize}
    \item \inputP = My prompt
    \item \progP = A properly trained and deployed AI system
    \item \outputP = The response I receive
\end{itemize}

\noindent the concern would be that:

\begin{center}
\textbf{\outputP} $\neq$ \textbf{\progP\hspace{-1mm}(\inputP)}.
\end{center}

This is exactly the problem that a trusted computer can help address. Recall from~\Cref{subsec:trusted_computers} that a trusted computer generates a certificate showing that \outputP = \progP(\inputP). An AI system can be designed to send such a certificate along with its response to a prompt. (See~\Cref{fig:cert}.) In this way, a user obtains assurance that a response is truly generated by \progP---a \textit{particular} AI system---ingesting \inputP, the user's prompt.

\begin{keyInsight}[Output integrity]

To trust the \outputP resulting from an \inputP to a program \progP such as an ML model, a user must have assurance that no tampering occurred, meaning:

\begin{center}
    \outputP = \progP\hspace{-1mm}(\inputP). 
\end{center}

Trusted computers can output certificates that attest to this property. 

\end{keyInsight}

\subsubsection*{User privacy}

In our running single-user example, even if the LLM platform seems to be working correctly, I might still be concerned about the \textit{privacy} of my prompt---and the resulting output---from the organization that runs the LLM platform. I might not want others to know about my medical condition. 

Today, users rely on the organizations running AI systems to enforce sound privacy policies. But this approach doesn't always work. Recently, some users were dismayed to find that they had inadvertently published their ChatGPT sessions and  made them available for indexing by search engines. In one embarrassing case, where a user asked to have a resume rewritten, a journalist identified the user, tracked the results on LinkedIn, and reported that the user didn't get the job~\cite{techcrunch-chatgpt-queries-indexed-2025}.

Even if prompts and/or outputs aren't publicly revealed, they are still vulnerable to abuse by the operator of the AI system. For example, the names and medical conditions of users could in principle be sold to advertisers or insurers on the sly.

\begin{keyInsight}[Trusted computing and user privacy]
Trusted computing (specifically, trusted \textit{confidential} computing) offers a key technical, practical approach to protecting the privacy of inputs and outputs when users query ML models.
\end{keyInsight}

\paragraph{Note on integrity:} Integrity is critically important for enforcing privacy. Suppose that I send my medical query to what I think is a trustworthy ML model execution environment, but it instead goes to an ML environment that is programmed to forward prompts to an insurance company. It doesn't matter in this case if the \textit{interaction}---i.e., the channel between the ML model and me---remains private, because the \textit{model environment} itself breaks privacy. (It's as though I've whispered a secret to a gossipy confidante.)

\subsection{Trusted Computing via Hardware and Replication}
\label{sec:tee}

Having discussed what trusted computing is for, we turn now to how it is realized:  first via hardware and replication, then  via cryptography (\Cref{sec:zkps_and_MPC}).

\subsubsection{Blockchains}

The functionality of a blockchain is best explained as a ``bulletin board": a public append-only log of messages called transactions, with nodes running a consensus protocol to reach agreement in the ordering of the transactions~\cite{nakamoto2008bitcoin, wood2014ethereum}.  
Two security properties make a blockchain useful as infrastructure: \emph{consistency} (safety), meaning that every viewer sees the same finalized record (so that a transfer of a digital asset cannot result in a double-spend), and \emph{availability} (liveness), meaning submitted transactions eventually get included. 
The strong notion of availability against adversarial parties (also known as ``censorship resistance'') is what makes blockchains appealing as infrastructure.
These properties are ensured in part by cryptographic techniques, but also in part from economic incentives: a native token compensates validators through fees and, in proof-of-stake systems, imposes penalties for dishonest behavior via slashing. 

A blockchain becomes a \emph{programmable trusted computer} by interpreting the messages on the bulletin board according to deterministic rules. Asset transfers require a signature from the private key bound to the source account. Smart-contract languages such as Solidity make the rule set itself a program, so arbitrary state-transition logic for actions, elections and exchanges, can all run on top of the same finalized log. The availability and consistency properties from the underlying log transfer to  whatever functionality these programs compute.

Two fundamental limitations of blockchains motivate the rest of this section. 
First, the bulletin board is public by default, so any input or computed state is visible to all observers; confidentiality requires additional cryptography, such as zero-knowledge proofs or TEE-based execution. For example, Aztec, Aleo, and Penumbra build on ZK programming models~\cite{aztecnr, aleovm2025, penumbraprotocol} while Oasis Sapphire, Phala, and Secret Network build on TEEs. 
Second, replicated computation is expensive since every validator must re-execute every transaction, and so smart contract blockchains charge fees called ``gas" to limit the on-chain computation.
Heavier computation must live off-chain, via oracles (\Cref{sec:oracles}), rollups using SNARGs (\Cref{sec:zkps_and_MPC}), or TEE-based co-processors. 

\subsubsection{Trusted Execution Environments and 
Confidential Computing}
\label{subsubsec:TEEs}
A Trusted Execution Environment (TEE) is a hardware primitive for trusted computing: a processor feature that protects a computation from the rest of the system it runs on. 
TEEs have a long history in different forms~\cite{sabt2015tee, jauernig2020tee}. Most mobile phones today contain one—used, for instance, to protect the biometric data that authenticates you to your device (e.g., a face template). 
More recently, server and hyperscaler processors such as Intel Xeon and AMD EPYC have added TEE primitives, enabling cloud-based ``confidential computing'': the industry term for protecting data while it is in use, not merely at rest or in transit.

A TEE has two main properties: isolation and remote attestation~\cite{jauernig2020tee, sabt2015tee}. Here an execution environment'' is an OS process, a virtual machine, or a container. 
In the usual execution hierarchy, a process at a lower level (a kernel ring 0'' or hypervisor ``ring -1'') has full visibility and control over processes above it; \emph{isolation} means even those lower levels cannot tamper with or inspect the TEE's computation. 
\emph{Remote attestation} means a party outside this hierarchy---for example, a process on another machine and network---can verify a signed piece of evidence binding the isolated computation's output to its program code.

\paragraph{Why confidential computing matters for ML privacy:} 
As noted in~\Cref{subsec:trusted_computers}, certain types of trusted computers (TEEs) enable confidential computing---use of an AI model without revealing inputs or outputs to the system operator.
The model executes within a TEE, whose black-box abstraction conceals model state and execution from the model operator, while inputs and outputs travel over an encrypted channel between the user and the TEE. 
In this way the flow in~\Cref{fig:pipeline} can be realized in such a way that \textit{\inputP and \outputP are concealed from the operator of the program (ML model)}. (There are important limitations and provisos that we will discuss.)

GPU-based support has made TEEs practical for AI, and it is the use case mainly envisioned by the most mature TEE platform targeting AI applications:
NVIDIA's Confidential Computing (CC).
Its H100 CC mode reached general availability in June 2024 with CUDA 12.4,
and it now spans a range of their GPUs (H100/200, Blackwell B100/200, and upcoming Vera Rubin)~\cite{dhanuskodi2023creating,nvidia2025secureai,nvidia2025cc}.
The encrypted PCIe path between a confidential CPU and a confidential GPU adds modest overhead: under 7\% for inference on an 8B-parameter Llama model, and near-zero for larger models such as 70B-parameter Llama ~\cite{zhu2024gpuconfidential}. 
The primary target is inference on cloud services: an operator can deploy a model through a cloud provider while protecting both the model and user interactions with it. 
It can similarly be used by an organization for cloud-based model training and, indeed, for a range of other trusted-computing use cases.
TEE-based confidential inference is also supported by Azure in a preview offering~\cite{russinovich2024azureconfidentialdeepdive}, described in a design paper by Anthropic~\cite{anthropic2025confidential}, and available in production from providers such as NEAR AI Cloud~\cite{nearai2025cloud}, RedPill~\cite{phalaredpill},  Venice~\cite{venice2026e2ee}, Tinfoil~\cite{tinfoil}, Chutes~\cite{chutes}, and others.

\paragraph{Limitations of Confidential Computing}
It has taken a while for TEEs to find adoption in blockchains. 
One reason is that TEEs have had as much publicity for their vulnerabilities as anything else~\cite{vanschaik2024sgxfail, chuang2026teefail}. This is for several reasons.
One is the nuanced trust model. A TEE is fundamentally a software
abstraction, and so it is designed to provide isolation and remote attestation
against \textit{software}-level attackers \cite{johnson2025scalablesgx, amdsevsnp2020}:
defending the SGX process from an attacker with kernel access, defending a
confidential VM from the hypervisor.

Even though they aren't designed for tamper resistance, it also isn't trivial to attack them, so it is tempting to rely on them more than is justified.
The trend instead has been toward TEEs optimized for
confidential computing at very low overhead, even at the cost of giving up
resistance to physical attacks. The clearest illustration is the transition
from the original client-era Intel SGX to the server-era TEEs that now
dominate (scalable SGX, Intel TDX, AMD SEV-SNP). The original SGX used
integrity-tree-protected memory encryption; the server-era designs replaced
it with deterministic AES-XTS \cite{johnson2025scalablesgx, amdsevsnp2020}.
This scales to terabytes of protected memory, at the cost that identical
plaintexts produce identical ciphertexts---the property TEE.Fail exploits
via memory-bus interposition \cite{chuang2026teefail}. Intel documents this
tradeoff directly in its own description of scalable SGX
\cite{johnson2025scalablesgx}.

Because a TEE is not by design secure against a physical attacker, the context
in which it runs must be evaluated separately from the underlying remote
attestation. Even though the TEE is instantiated by the hardware providers,
the cloud providers too must be trusted to secure the hardware against
physical attacks. This is why the confidential compute products---Azure
Confidential VMs \cite{azurecvmoverview} and Google Cloud Confidential
Computing \cite{gcpconfidentialcomputing}, among others---are essentially
collaboration products between hardware vendors and cloud operators. Recent
work in this vein \cite{rezabek2025proofofcloud, proofofcloudalliance} ties
the CVM attestation to platform-level evidence that the host is a specific
registered piece of hardware in a secured data center.

Another issue is the trust model around the hardware manufacturer. We cannot
guarantee that the manufacturer has not effectively backdoored the hardware,
and this applies not just to the processors but to the remote attestation
system itself. These are opaque, and by the design of the protocol, nothing
prevents the issuance of valid attestations from a process that a spy agency
and a manufacturer have colluded to produce. This kind of Snowden-associated
system-level collusion cannot be ruled out by the technical design. It does
not necessarily mean that such an adversary could decrypt the contents of an
already-running TEE---only that they could join a fake spy TEE to a network,
potentially without detection.

Another reason, besides these inherent issues, is that the complexity of TEEs
has meant that they often suffer from vulnerabilities due to implementation
mistakes \cite{vanschaik2024sgxfail}. The best example of this is the ÆPIC
Leak \cite{borrello2022aepic}, where a microcode implementation flaw caused
the upper bits of a private register not to be cleared when accessing undefined memory-mapped I/O (MMIO) ranges for the Advanced Programmable Interrupt Controller (APIC). This could be exploited to extract
secrets from running programs and forge remote attestations---from a
software-only attacker, well within the designed threat model.

\subsection{Trusted Computing via Cryptography}
\label{sec:zkps_and_MPC}

\paragraph{ZKPs.} A \emph{zero-knowledge proof} (ZKP)~\cite{goldwasser2019knowledge} is a cryptographic protocol in which one party (the \emph{prover}, $\mathcal{P}$) convinces another (the \emph{verifier}, $\mathcal{V}$) that a statement is true without revealing anything beyond the statement's validity. In modern practice, ZKPs are typically realized as \emph{zk-SNARGs}---zero-knowledge succinct non-interactive arguments---built by layering zero-knowledge on top of an underlying SNARG~\cite{chiesa2014succinct}. We unpack both layers below.

A SNARG allows $\mathcal{P}$ to convince $\mathcal{V}$ that a public \emph{statement} $x$ satisfies some relation $\mathcal{R}$ (for example, that $x = (u, y)$ with $y = f(u)$ for a fixed function $f$), by producing a short proof $\pi$ that $\mathcal{V}$ can verify far more efficiently than re-executing the computation. The value of a SNARG is purely this \emph{succinctness}. More expressive relations let $\mathcal{P}$ use a private \emph{witness} $w$ (e.g., a secret input), proving $(x, w) \in \mathcal{R}$ without including $w$ in the proof; on its own, however, a SNARG gives no guarantee that $\pi$ hides $w$.

A \emph{zk-SNARG} adds exactly this guarantee: $\pi$ additionally reveals nothing about $w$ beyond what the relation itself implies. Throughout this survey we use ``ZKP'' to refer to zk-SNARGs, as these are by far the most widely deployed ZKPs in practice, though they are not the only kind.

ZKPs and SNARGs see use in both blockchain and AI settings.

\begin{itemize}
    \item \textbf{Applications in blockchain.} ZKPs have seen widespread adoption in blockchain, most prominently via zkRollups~\cite{buterin2021rollups}, which reduce Ethereum gas fees by up to 99\% through projects such as zkSync Era~\cite{matterlabs_zksync-era_2023}, Starknet~\cite{starknet_io_starknet_2025}, and Polygon zkEVM~\cite{polygonlabs_zkevm_2023}. A zkRollup compresses the verification of a batch of transactions into a single short proof that can be verified efficiently. The blockchain can therefore update its state to the post-batch state without re-executing the transactions. It is worth noting that in this application the transactions themselves are public, so the zero-knowledge property is unused; these systems rely only on SNARG succinctness. The ``zk'' in ``zkRollup'' is largely a misnomer inherited from loose usage in the blockchain community.

    \item \textbf{Applications in AI.} In AI, SNARGs let a compute-rich prover $\mathcal{P}$ (say, a cloud service) perform expensive computations such as ML inference on behalf of a constrained verifier $\mathcal{V}$ (say, a smart contract or a mobile device), and prove correct execution. When the model has proprietary weights, a plain SNARG is insufficient: the proof $\pi$ may leak information about the weights. The zero-knowledge property closes this gap. $\mathcal{P}$ can first publish a \emph{commitment} $c_\theta$ to the weights $\theta$ (e.g., a cryptographic hash that binds $\mathcal{P}$ to $\theta$ without revealing it), and then prove to $\mathcal{V}$ any inference was correctly executed using the weights bound by $c_\theta$, without $\mathcal{V}$ ever seeing $\theta$. \emph{Zero-Knowledge Machine Learning} (ZKML)~\cite{zhang2020zero, chen2024zkml} is the primary, though still emerging, application area for ZKPs in AI. ZKML has already been deployed in practice in applications such as RockyBot~\cite{ModulusLabs_RockyBot_2023}, an on-chain verifiable ML trading bot developed by Modulus Labs~\cite{modulus2023zkml}. Performance costs remain prohibitive: Modulus Labs' \emph{Cost of Intelligence} benchmark~\cite{modulus2023costofintelligence} reports proving times on the order of a minute for multilayer perceptrons with only $\sim$18M parameters and 22B mult-adds, running on high-end AWS instances (AMD EPYC 7R32, 128GB RAM), several orders of magnitude away from being tractable for frontier-scale LLMs.
\end{itemize}

A zk-SNARG requires $\mathcal{P}$ to hold the full witness in the clear, so it cannot support computations whose inputs are distributed across mutually distrusting parties unwilling to share them with a single prover; for instance, model inference where the proprietary weights are held by the model owner and the private input is held by a separate user. Addressing this requires a different primitive: \emph{Multi-Party Computation} (MPC).

\paragraph{MPC.} Secure Multi-Party Computation (MPC)~\cite{yao1982protocols, goldreich2019play} enables a group of $n$ parties, each holding a private input $x_i$, to jointly compute an agreed-upon function $f(x_1, \dots, x_n)$ while revealing nothing about the individual inputs beyond what the output itself implies.

Unlike a ZKP, MPC does not produce a proof of computation integrity that an external party can later verify. A line of work known as \emph{collaborative zk-SNARGs}~\cite{ozdemir2022experimenting} bridges this gap, allowing private ZKP witness to be split across multiple provers who jointly produce a single ZKP, without any party ever reconstructing the full witness. This combines the succinctness, verifiability, and zero-knowledge guarantees of a zk-SNARG with the distributed-trust model of MPC.

MPC sees use in both blockchain and AI settings.

\begin{itemize}
    \item \textbf{Applications in blockchain.} The most widely deployed use of MPC in blockchain is \emph{threshold signing}~\cite{desmedt1992threshold}, where a private key is split among $n$ parties so that producing a signature requires a threshold of them, without any single party ever reconstructing the key. This underpins commercial MPC custody and validator key-management services. A related hybrid deployment is \emph{zkTLS}~\cite{zhang2020deco}, which combines MPC and ZKPs to let a client prove statements about TLS-protected web content to a smart contract without trusting the server.

    \item \textbf{Applications in AI.} MPC enables two families of AI applications that ZKPs alone cannot support. The first is \emph{collaborative training}, where multiple data owners (for instance, multiple hospitals holding patient records or banks holding transaction histories) jointly train a model without any party revealing its raw data to the others or to a central server~\cite{wagh2019securenn}. The second is \emph{privacy-preserving inference}, where a user evaluates a proprietary model on their own private input: MPC allows the computation to proceed without the user disclosing their input to the model provider, and without the provider disclosing the model weights to the user~\cite{kumar2020cryptflow}. Performance costs remain substantial: PUMA~\cite{dong2025puma}, the state of the art framework for MPC-based transformer inference, reports roughly five minutes per token for LLaMA-7B, several orders of magnitude slower than standard inference.
\end{itemize}

The large overhead has restricted MPC and ZKPs to a narrow set of deployments. 
Where the trust model permits, TEEs offer a much cheaper alternative for both confidentiality and verifiable execution:
confidential-computing inference overhead stays under 7\%~(\Cref{subsubsec:TEEs}), several orders of magnitude below the MPC and ZKP costs noted above. 
The difference in trust models between these makes them difficult to compare. 
TEEs require trust in the hardware manufacturer and the absence of side-channel attacks, while MPC adds a non-collusion assumption among the parties (the threshold depending on the protocol), and ZKPs rely on cryptographic hardness alone.

\begin{figure}[h]
    \centering
    \begin{subfigure}{\linewidth}
        \centering
        \includegraphics[width=1\linewidth,page=1]{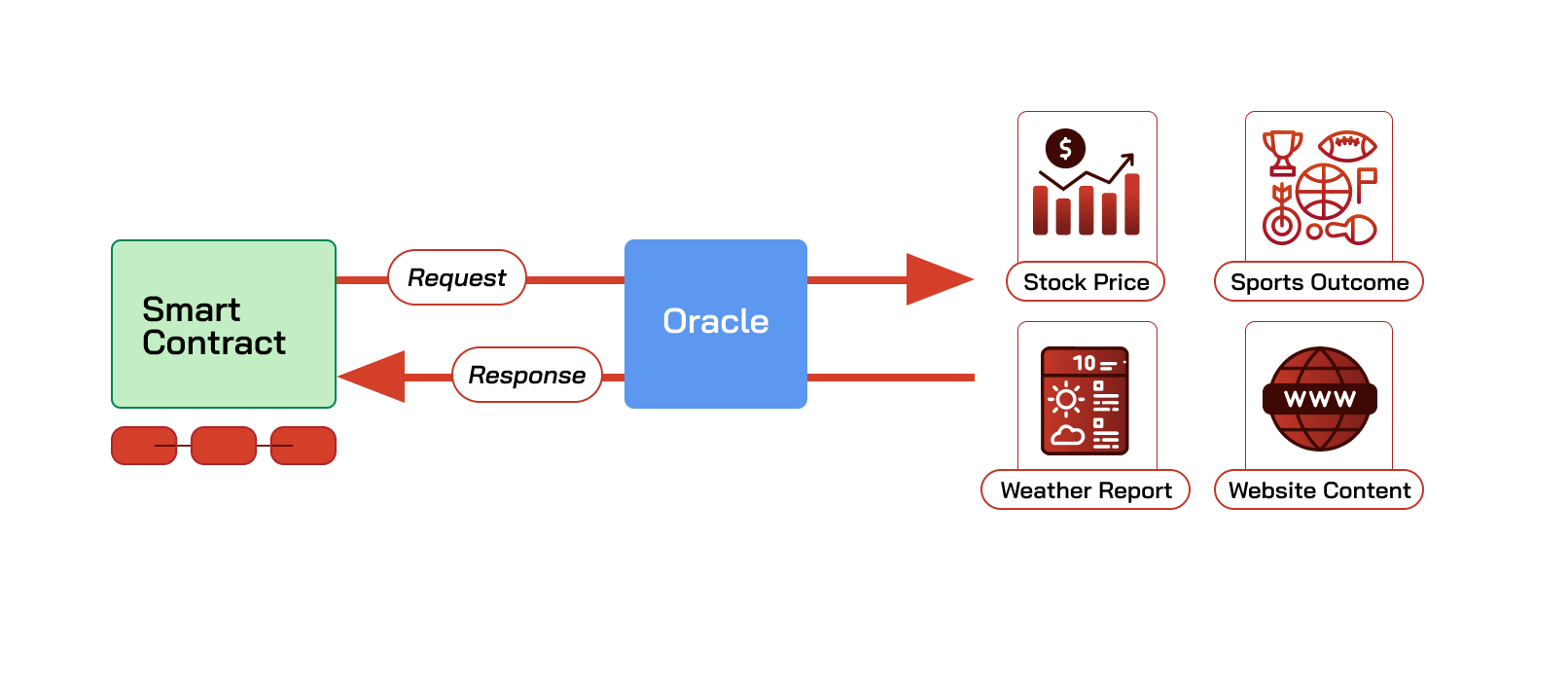}\vspace{-10mm}
\caption{Conceptual schematic of smart-contract oracles.}
\label{fig:oraclecommittee:top}
    \end{subfigure}

    \begin{subfigure}{\linewidth}
        \centering
        \includegraphics[width=\linewidth,page=2]{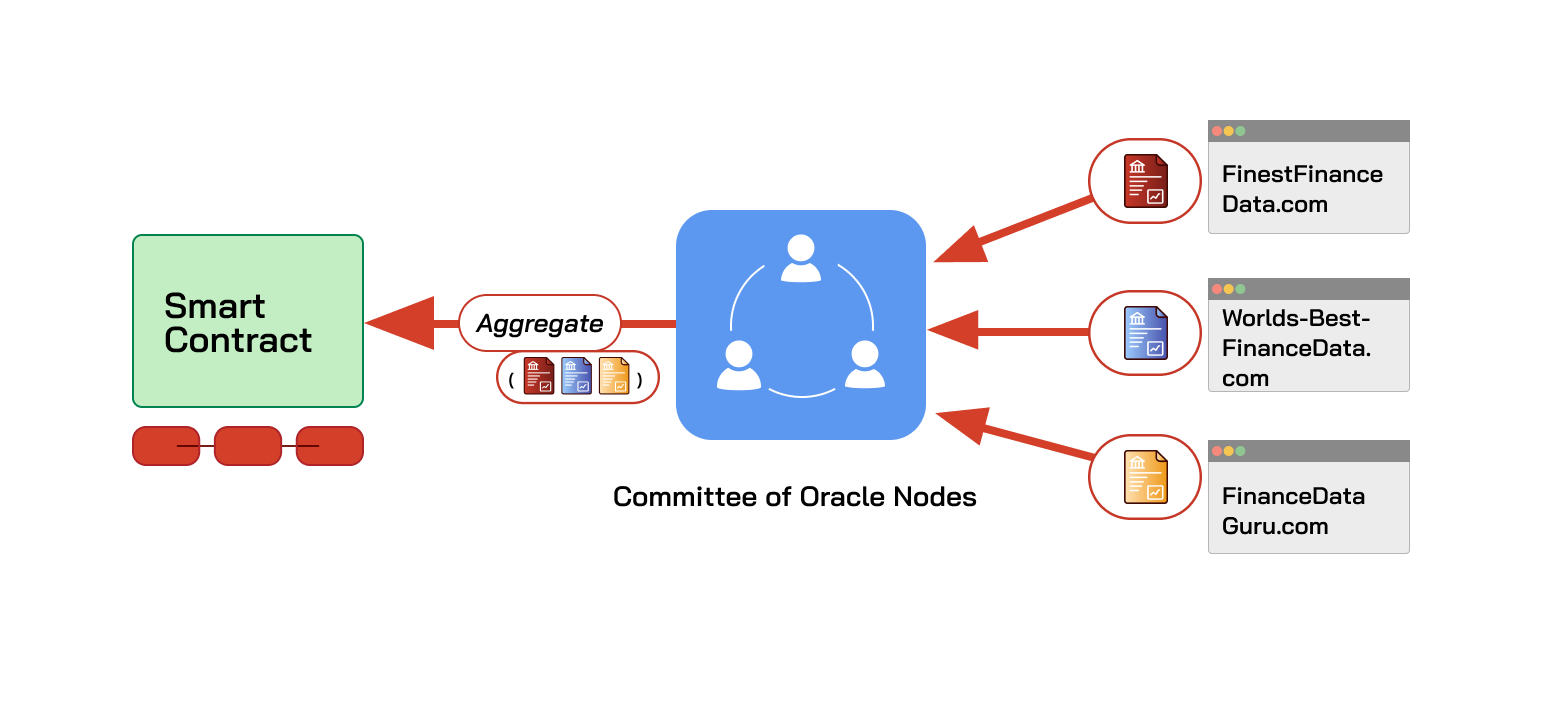}\vspace{-8mm}
        \caption{A common architecture for smart contract oracles: a committee of nodes fetches data independently, potentially from different sources, and an aggregation mechanism filters and combines data and may also filter out anomalies or outliers. This flow may be periodic or triggered upon request by the receiving smart contract.}
        \label{fig:oraclecommittee:bottom}
    \end{subfigure}

    \caption{The concept of smart contract oracles and a common committee-based architecture.}
    \label{fig:oraclecommittee}
\end{figure}

\subsection{Oracles}
\label{sec:oracles}

Oracles originate as systems that provide authenticated data to blockchains. They are also a valuable tool for AI as they can enable access to private data, which we will expand on in~\Cref{chap:block4ai}.
In this section, we first review the concept of smart contract oracles and, more relevant to AI applications, privacy-preserving oracles. We then discuss technical approaches to realizing oracles, along with their security and performance trade-offs.

\subsubsection{The Concept and Applications of Oracles}
\label{subsubsec:oracles}

\paragraph{Smart contract oracles.}
Smart contracts are autonomous programs running on top of a blockchain.
Many applications require smart contracts to access \textit{off-chain} data, such as stock quotes (for tokenizing stocks), sports results (for prediction markets), flight status (for delay insurance, such as AXA's Fizzy~\cite{axzfizzy}), and so on.
Since smart contracts can only access what is already on the blockchain, off-chain data must be pushed to the blockchain by a system called an \textit{oracle}, as shown in~\Cref{fig:oraclecommittee:top}.
As oracles are a critical part of the smart-contract stack, an entire industry is focused on building robust, efficient, and secure oracle systems, including Chainlink~\cite{chainlink}, RedStone~\cite{redstone}, Chronicle~\cite{chronicle}, Witnet~\cite{witnet}, UMA Optimistic Oracle~\cite{uma}, Tellor~\cite{tellor}, Band Protocol~\cite{band}, Pyth Network~\cite{pyth}, API3~\cite{api3}, Supra~\cite{supra}, and Gas Network~\cite{gasnetwork}.

An essential security property of oracles is \textit{authenticity}, i.e., that an oracle faithfully relays data from the specified source without tampering with it or lying about its origin.
As depicted in~\Cref{fig:oraclecommittee:bottom}, a common design is to have a committee of multiple oracle servers fetch data independently, and use an aggregation mechanism to  filter out potential malicious inputs and produce the final result. 
This approach guarantees authenticity, assuming that a fraction of the committee is malicious (e.g., less than one-third)~\cite{breidenbachChainlinkOCR}.

\paragraph{Privacy-preserving oracles.}

In addition to authenticity, \textit{privacy-preserving oracles} can relay information derived from private data that is not directly accessible to a regular oracle, such as a user's bank statement, credit report, or age information.
As we will explore in~\Cref{chap:block4ai}, privacy-preserving oracles enable ML training or fine-tuning on \textit{private web data} obtained directly from users, without requiring a special arrangement with the original data holder. At the same time, users enjoy strong privacy protection, as they can control which data to share.

\begin{figure}[h]
    \centering
    \includegraphics[width=\linewidth,page=3]{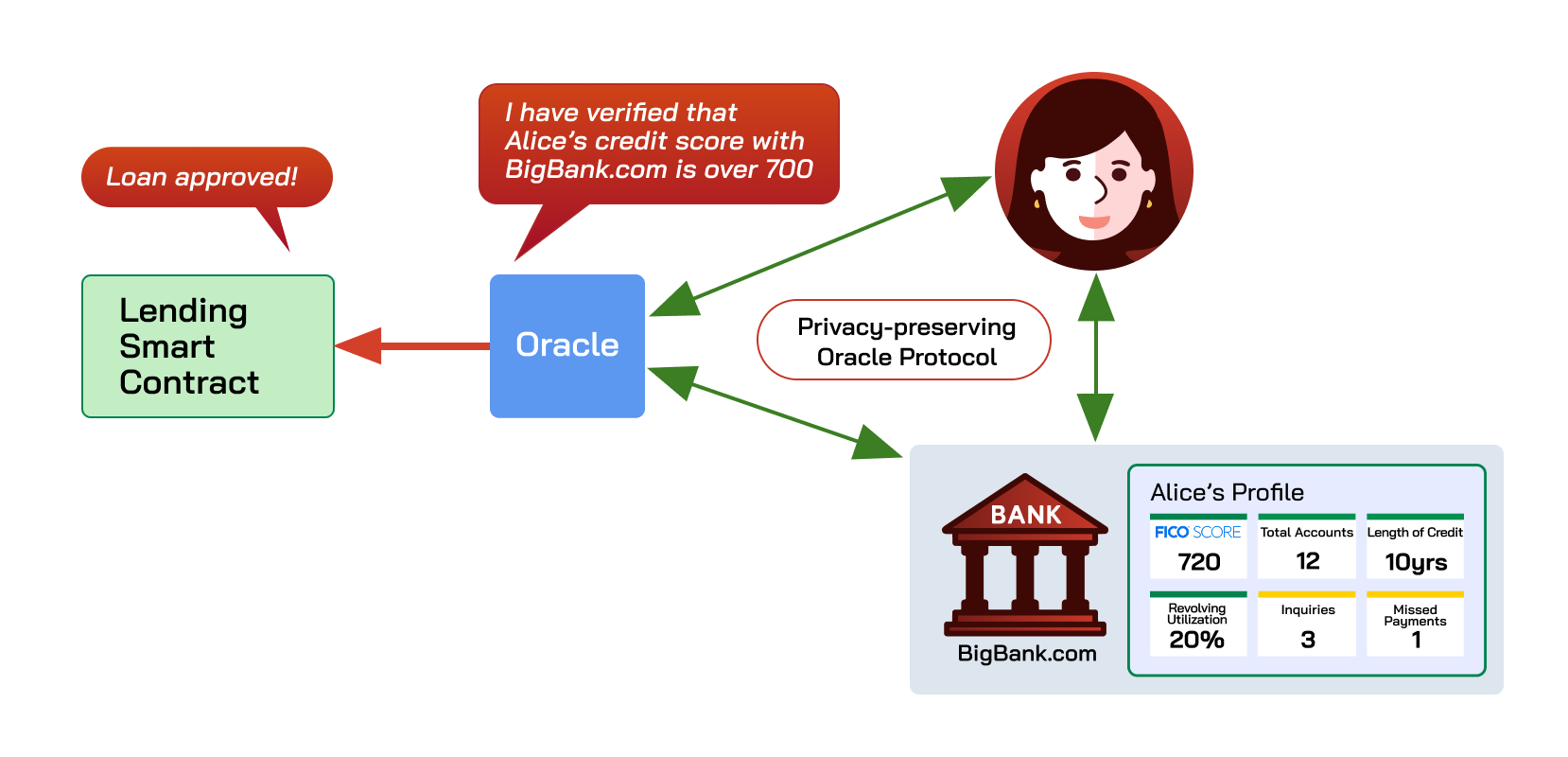}
    \caption{User can ``export'' private information using a privacy-preserving oracle.}\label{fig:privateoracle}
\end{figure}

In more technical detail, a privacy-preserving oracle protocol allows a user to convince a verifier that a piece of data from a specific source $S$ (integrity/authenticity) satisfies a certain predicate $P$ without leaking any other information (privacy). 
To explain, suppose Alice wants to prove to a lender (a smart contract or an off-chain entity) that she has good credit. She could send a screenshot of her credit report, but it is easy to forge screenshots. Using a privacy-preserving oracle, Alice could cryptographically prove to an oracle that ``{according to data from \texttt{https://www.bigbank.com}, Alice's credit score is over 700}.''
The oracle can verify this claim and relay the result to the lender, as shown in \Cref{fig:privateoracle}.
The key privacy feature is that no information is revealed about Alice beyond the fact that the above statement is true. In particular, Alice does not need to reveal to the lender the secrets necessary to retrieve the credit record (e.g., her SSN) or any extra information that might be on the credit report (e.g., her address history).

\subsubsection{Constructions of Privacy-Preserving Oracles}

We now discuss high-level technical constructions, based on the original papers that propose privacy-preserving oracles ~\cite{zhang2016town,zhang2020deco}. Subsequent works proposed various security and performance optimizations~\cite{luoProxyingEnoughSecurity2025}, but the main idea remains the same.

The starting point for privacy-preserving oracles is the Transport Layer Security (TLS) protocol that secures the Internet.
TLS is a protocol that enables a user (e.g., a browser) to establish secure connections with a remote web server. 
Readers do not need to understand the details of TLS, but we remark that TLS itself does not digitally sign the transmitted data. Rather, when Alice obtains a piece of ciphertext $D$ from a TLS server $S$, the integrity of $D$ is protected by a key shared between Alice and the server. As a result, a third party, such as the oracle node in our example, cannot verify the authenticity of $D$ or determine whether $D$ originates from the server or is entirely forged by Alice. In all constructions of oracles below, this shared key needs to be \textit{somehow} hidden from Alice for this reason.

There are three main approaches to constructing oracle protocols: using TEEs, using secure two-party computation (2PC), or using the oracle as a proxy. 
It is worth noting that TLS can be modified to sign transmitted data (as proposed in TLS-N~\cite{ritzdorfTLSN2017}), but this approach incurs a high adoption cost, as it requires websites to upgrade to the modified TLS protocols; moreover, adding signatures only addresses authenticity, not privacy. Therefore, we focus on oracle protocols that are compatible with existing websites.

\paragraph{TEE-based oracles.}
The first class of solutions relies on TEE technology, such as Intel SGX or TDX (we refer readers to \Cref{sec:tee} for an overview of TEEs).  This approach was first presented in Town Crier~\cite{zhang2016town}. 
Town Crier runs the following high-level logic in a TEE: first, it accepts a request specifying a data source $S$, a statement/predicate $P$ to be proven against data from $S$, and user secrets required to access $S$ (e.g., encrypted passwords); then it fetches data $D$ from the given source $S$ over TLS, and outputs $P(D)$, along with a hardware-generated attestation to the correctness of it.

Note that the above statement $P$ can be a generic function computed by the oracle over $D$.
This way, oracles can also serve as the execution layer for heavy off-chain computation. 
Compared to the designs we will introduce next, TEE-based oracles are likely the most practical solution for large-scale computation, such as AI.
AI tools used by smart contracts are likely in many cases to be executed within TEE-based oracles, where models can access external data, perform non-trivial computation, and return attested results on-chain.

\begin{figure}[h]
    \centering
    \includegraphics[width=\linewidth,page=5]{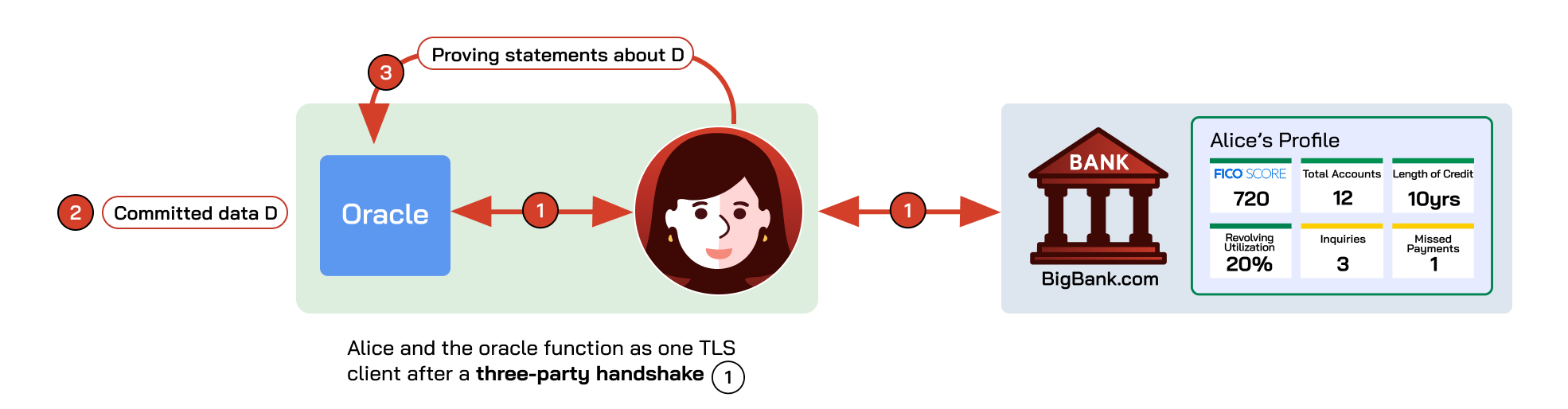}
    \caption{The high-level workflow of DECO's 2PC mode.}
    \label{fig:oracles:b}
\end{figure}

\paragraph{2PC-based oracles.}
DECO~\cite{zhang2020deco} introduced a privacy-preserving oracle design that does not depend on TEEs.
The high-level workflow is shown in~\Cref{fig:oracles:b}.

The key idea in DECO is for Alice and the oracle to run a two-party computation (2PC) protocol~\cite{yao1982protocols} (step 1) to jointly complete the TLS handshake, so that neither of them has the entire session key. 
This protocol is called a \textit{three-party handshake}, to differentiate from the standard two-party handshake in TLS.
After the three-party handshake, Alice queries the web server over TLS (with the help of the oracle) and commits to the ciphertext $D$.
The three-party handshake hides the session key, preventing Alice from forging TLS ciphertexts. 
Now that the authenticity of $D$ has been established, the prover can prove fine-grained statements using any desired generic zero-knowledge proof system~\cite{grothSize2016} in step 3.

\paragraph{Proxy-based oracles.} 
The above two constructions use TEEs and three-party handshakes to prevent forgery of TLS ciphertext, respectively. 
Proxy-based oracles introduce a different method: Alice interacts with the TLS server through the oracle as a network proxy. 
Alice can then prove statements about the \textit{proxied} TLS ciphertext $D$ in a manner similar to step 3 of the 2PC-based oracles.
However, compared with 2PC-based oracles, proxy mode eliminates the expensive three-party handshake and thus is more performant and simpler to implement.
Proxy mode was originally introduced in DECO~\cite{zhang2020deco}.

\begin{figure}[h]
    \centering
    \includegraphics[width=\linewidth,page=6]{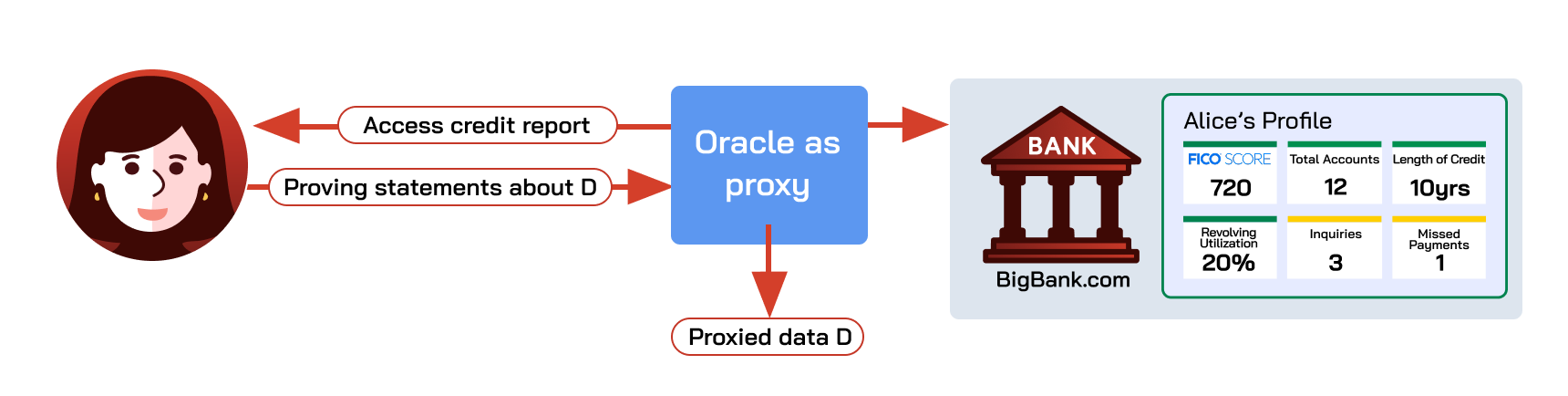}
    \caption{The high-level workflow of DECO's proxy mode.}
    \label{fig:oracles:c}
\end{figure}

\paragraph{Summary.} We now compare the security and performance of these designs.
To access private data and perform large-scale computation over them, TEE-based solutions are currently the \textit{only} practical solution (although zero-knowledge proof systems are increasingly scalable).
Recent GPU TEEs, such as those with NVIDIA Confidential Computing, are particularly well-suited for ML workloads.
The caveat is that TEEs introduce additional trust assumptions in the hardware manufacturers and the environments in which they operate, as we detail in \Cref{subsubsec:TEEs}.
DECO-like protocols (both 2PC and proxy modes) are more suitable for proving relatively simple statements, such as age verification~\cite{maram2021candid}, credit verification, and so on.
Between 2PC and proxy modes, the latter avoids an expensive step; however, it is vulnerable to network-layer attacks (e.g., BGP hijacking~\cite[Appendix C.4]{zhang2020deco}). 
For high-stakes applications, 2PC mode should be preferred.

All of the above designs (and their variants) have seen adoption in practice. 
The industry now refers to them as ``zkTLS'' protocols---a misnomer for TEE-based solutions, because they do not use any ``zk'' (zero-knowledge proofs).
Town Crier and DECO have been productized as part of the Chainlink Runtime Environment (CRE)~\cite{chainlinkruntimeenvironment}.
Reclaim~\cite{reclaim_protocol} implemented an improved proxy-based variant~\cite{luoProxyingEnoughSecurity2025}, now expanding to a new TEE-based protocol.
zkPass~\cite{zkpass} runs in proxy mode by default, and falls back to 2PC mode for servers that do not support proxy mode, a feature they call hybrid mode.
TLSNotary~\cite{tlsnotary} is an implementation of 2PC mode by the Ethereum Foundation.

\chapter{Crypto x AI: Enhancing Crypto with AI}
\label{chap:ai4block}
\section{Overview: Making Crypto More Usable and Flexible}
\label{sec:AIforblockchain-overview}

\begin{figure}[h!]
    \centering
    \includegraphics[width=0.85\linewidth]{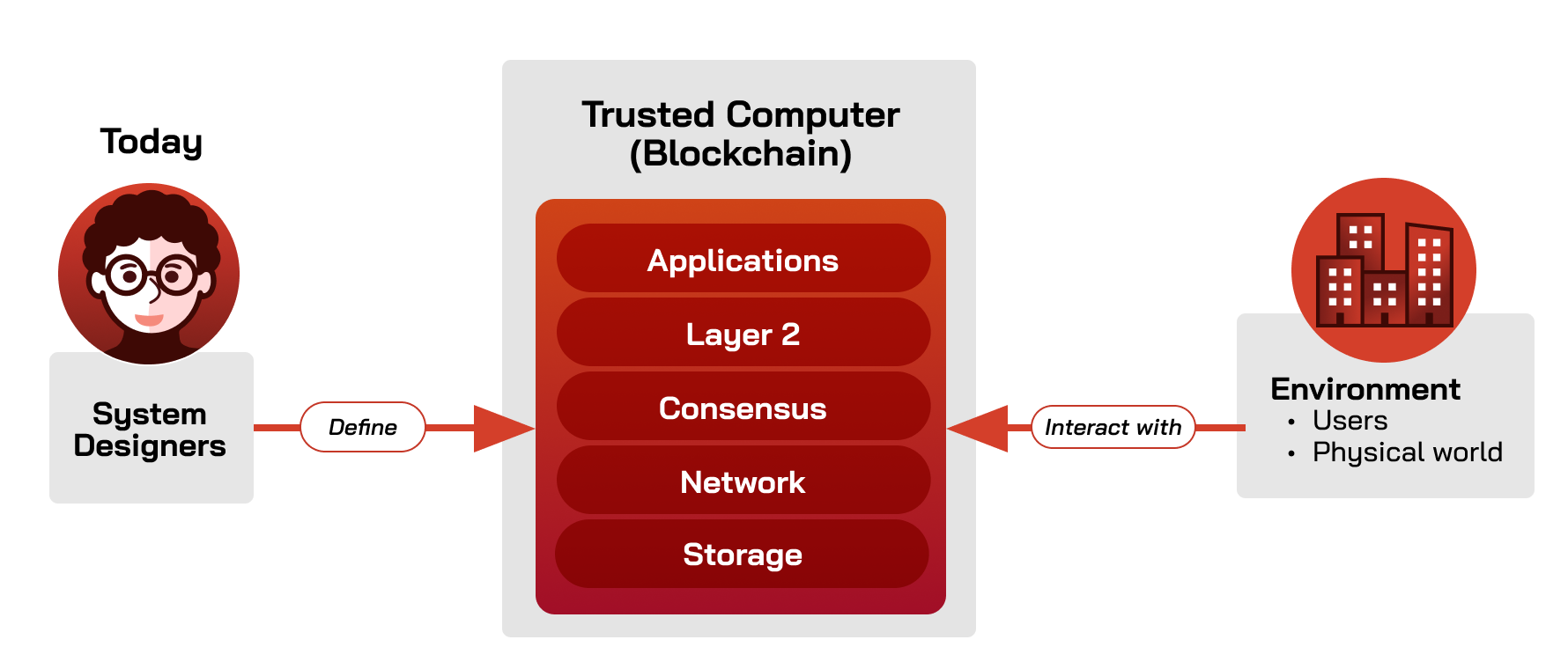}
    \caption{Today, use of the interface (red arrows) between the trusted computer and everything else (designers, end users) is labor-intensive and error-prone!}
    \label{fig:trusted-today}
\end{figure}

Today, humans design and implement the foundational protocols and applications for trusted computers, such as consensus algorithms, networking algorithms, and smart contracts (\Cref{fig:trusted-today}). Once specified, the trusted computer (e.g., a blockchain) takes inputs from the environment, including human users and the outside world, e.g., via oracles. 
The interface between humans and blockchains requires careful thought and validation, and often humans struggle to precisely specify their intents, either as a user, a programmer, or a system designer \cite{twins,zhang2023demystifying,frohlich2021don}.

AI could act as a \textbf{translation layer} between the trusted computer and ``everything else,'' including the designer and the environment, as shown in \Cref{fig:trusted-instead}. For example:

\begin{figure}[h!]
    \centering
    \includegraphics[width=0.85\linewidth]{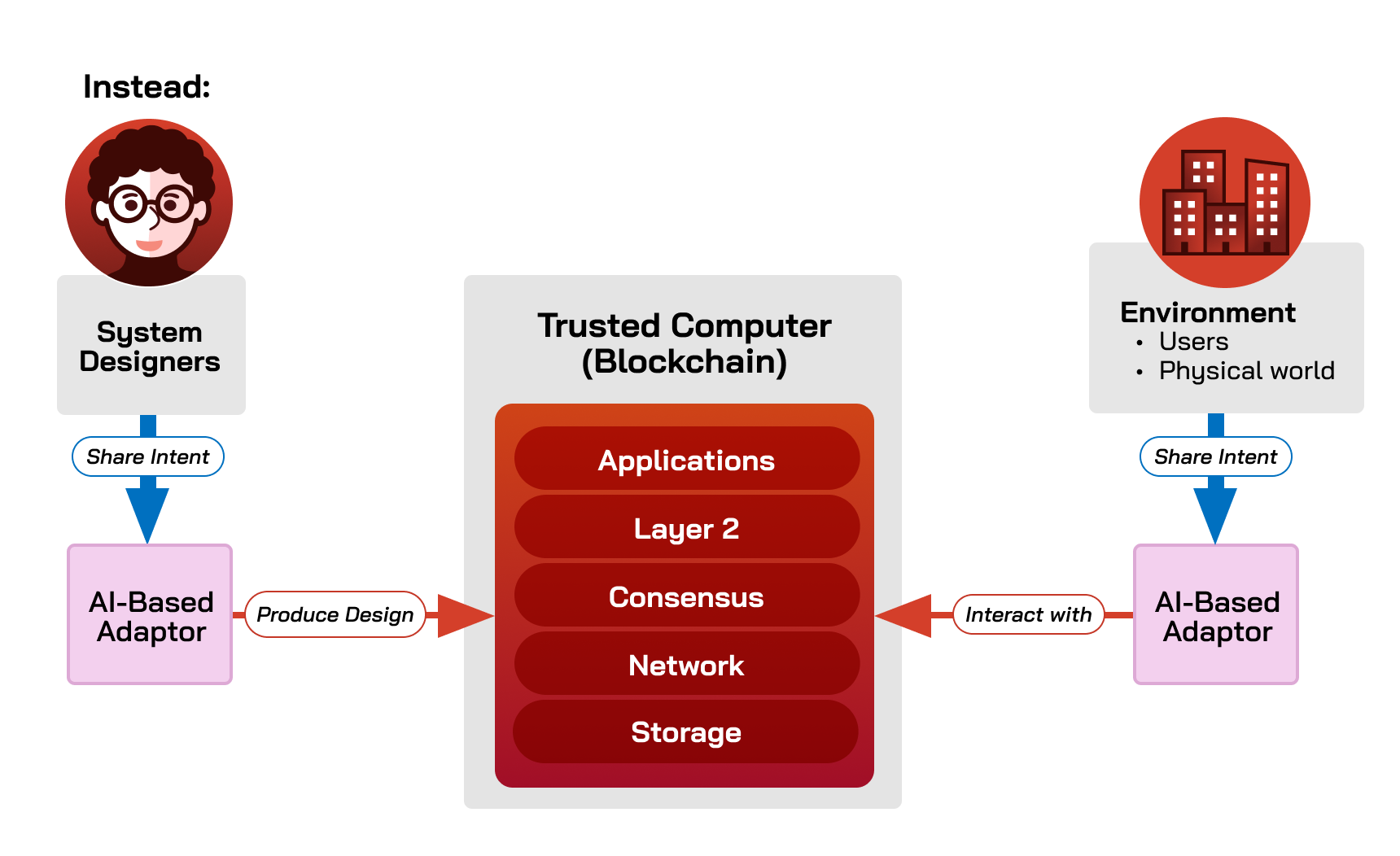}
    \caption{AI can translate human intents (blue arrows) into machine-readable instructions in a much more flexible manner.}
    \label{fig:trusted-instead}
\end{figure}

\begin{itemize}
    \item AI can help designers more flexibly specify the desiderata of a component of the blockchain stack, such as an application (e.g. DeFi incentive mechanisms).  
    \item AI can help users of blockchains define policies and/or specify their interactions with a blockchain (e.g., convert streams of data from the real-world into smart-contract-readable data, or create and submit transactions to a smart contract that execute the user’s intent).
\end{itemize}

\label{sec:AIforblockchain-current}

\begin{figure}
    \centering
    \includegraphics[width=\linewidth]{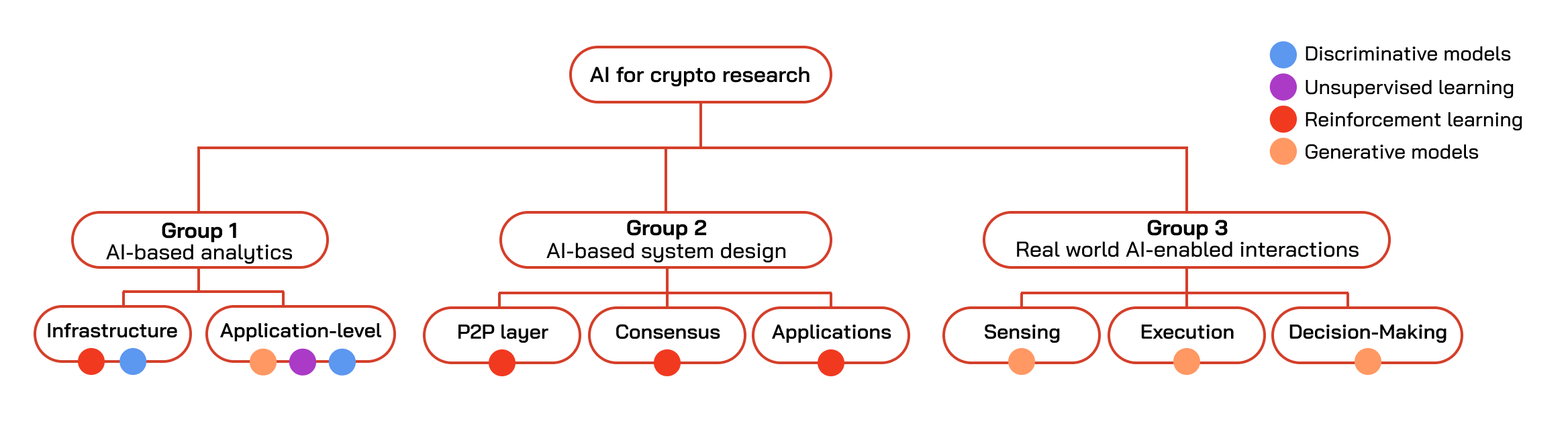}
    \caption{We identify three main groups of AI-for-crypto research: AI-based analytics, AI-based constructive design of systems and algorithms, and AI-enhanced interactions with the real world. The colored bubbles represent common ML techniques used for various categories of research; they are not exhaustive, but represent the common patterns we have observed. Notice that more advanced ML techniques (e.g. RL and generative AI) have primarily been used in groups 2 and 3, which involve interactions with an environment, rather than the static, more traditional prediction tasks from Group 1.
    }
    \label{fig:ai-for-crypto}
\end{figure}

In this chapter, we will discuss ongoing efforts to use AI to facilitate the design and analysis of blockchains in \Cref{sec:ai-analytic-tools,sec:ai-design-algos,sec:ai-interact-real-world}, as well as some 
more futuristic directions in \Cref{sec:AIforblockchain-future}. 
Researchers have been applying AI to blockchain systems (more broadly, decentralized systems) for well over a decade. While the space of algorithms and techniques is vast, we roughly categorize these efforts into three groups, which gained prominence roughly sequentially in time.
These categories 
are illustrated in \Cref{fig:ai-for-crypto}.

\begin{enumerate}[label=\textbf{Group \arabic*:}]
    \item \textbf{AI-assisted analytics.} (\Cref{sec:ai-analytic-tools}) These are AI-based algorithms for analyzing or understanding the state of an existing blockchain. This class of techniques emerged over a decade ago: it consists of applying AI to  analytics problems, such as predicting events on a blockchain  or classifying fraudulent transactions. These techniques translate the state of a complex system to something interpretable by a human. 
    
    \item \textbf{AI-assisted design of constructive algorithms.} (\Cref{sec:ai-design-algos}) These are AI-based approaches that translate high-level human objectives (e.g., increase throughput, reduce latency, improve security) into algorithmic decisions, either during the design phase to inform system configuration, or at runtime to adapt system behavior to changing conditions.
    This category was popularized in the last (roughly) six years 
    and generally uses ML to learn strategies for designing and/or interacting with crypto systems. In this category, we start to see a greater reliance on reinforcement learning (RL) and related techniques (e.g., bandits), which are used to learn policies or strategies in complex state and action spaces. 

    \item \textbf{AI-assisted interaction with the real world.} (\Cref{sec:ai-interact-real-world}) The third category, which emerged in recent years,
    explores how to use modern AI at the application layer to enhance blockchain interactions with the outside environment; this class of techniques mostly uses generative models, sometimes coupled with reinforcement learning, to achieve this enhanced functionality.
    Specifically, modern AI can equip smart contracts with three enhanced capabilities relative to the previous generation: \emph{(1) Sensing}: AI can help smart contracts use unstructured data from the real world. 
\emph{(2) Execution}: smart contracts can call tools and functions in the way that LLMs currently do.  
\emph{(3) Decision-making}: smart contracts can act as agents, making decisions based on values that are encoded in objective functions. 
This class of functionalities is enabled in part by oracles, which allow blockchains to have an accurate view of the state of the external world. 
\end{enumerate}

\section{AI-Assisted Analytics} 
\label{sec:ai-analytic-tools}
We begin by exploring how AI has been used to analyze existing blockchains in the literature. 
We group this work  into two categories: (1) analytics for global blockchain properties (\Cref{sec:infrastructure}), and (2) analytics for object-level blockchain properties  (\Cref{sec:application-analytics}). 

\subsection{Analytics for global  blockchain properties}
\label{sec:infrastructure}

Some categories of analytics represent properties of blockchains as a whole, i.e., they do not relate to individual users or transactions. Such analytics can relate to network-wide \emph{protocols} (e.g., consensus algorithms), networks (e.g., P2P networks), and derived properties of a blockchain (e.g., cryptocurrency price). 

\subsubsection{Consensus analytics: Vulnerability and attack discovery}
\label{sec:consensus-analytics}
A growing body of work has used AI to discover vulnerabilities in blockchain infrastructure, particularly in consensus protocols.

\paragraph{Discovery of selfish mining algorithms.} 
One common vulnerability explored is \emph{selfish mining}~\cite{eyal2018majority,finkbeiner2025sok}, where a blockchain operator deviates from protocol-specified behavior to unfairly gain additional rewards.
For example, a selfish miner can withhold newly created blocks to build a private chain, then publish it at a strategic time when it is longer than the public chain.
This causes blocks to be discarded, wasting honest miners' work, and increasing the selfish miner's relative share of the rewards.

Research on selfish mining began by characterizing the revenue of specific attacks~\cite{eyal2018majority,nayak2016stubborn}, primarily on proof-of-work blockchains.
Later work optimized selfish mining attacks using AI-based methods~\cite{sapirshtein2016optimal,wang2021blockchain,zur2020efficient}.
This involved modeling all possible actions of a blockchain miner within a \emph{Markov Decision Process} (MDP).
Sapirshtein et al.~\cite{sapirshtein2016optimal} were the first to do so, and introduced a method which requires solving a sequence of MDPs.
Later works introduced efficient approximations of the true MDP \cite{zur2020efficient} 
and other models for network conditions~\cite{gervais2016security}, proof-of-stake protocols~\cite{gervais2016security}, and bribery attacks~\cite{yang2020ipbsm}.

Classical MDP solving techniques are limited to relatively small state spaces due to their large computational and memory requirements~\cite{sutton1998reinforcement}.
A body of work has employed \emph{deep RL} techniques, which involve the use of neural networks, to analyze more complex models~\cite{hou2019squirrl,bar2023werlman,bar2023deep,sarenche2024deep,sarenche2025bitcoin}.
Hou et al.~\cite{hou2019squirrl} employed deep RL in a setting with multiple selfish miners under various consensus rules.
Other work~\cite{bar2023werlman,sarenche2025bitcoin} used deep RL to study the effect of transaction fees on selfish mining in proof-of-work blockchains, 
as well as the effect of petty-compliant miners who may deviate in minor ways for profit \cite{bar2023deep}. 
Sarenche et al.~\cite{sarenche2024deep} have extended the analysis of selfish mining to proof-of-stake blockchains using deep RL techniques.
The use of deep RL has allowed for finding profitable selfish mining strategies in more realistic settings, which would not have been possible with traditional methods alone.

Despite their ability to analyze more complex models, deep RL techniques often lack formal guarantees on the optimality of the derived strategies.
To this end, recent work used traditional RL methods to accurately analyze longest-chain protocols with alternative proof systems~\cite{chatterjee2024fully} and DAG-based protocols~\cite{keller2023generic,bar2025mad}.
Their approach allows an accurate characterization of the~\emph{security threshold}, the minimal power required to profit from selfish mining.

\paragraph{Real-time discovery of attacks on consensus.}
Beyond characterizing and optimizing attacks, recent work has focused on detecting ongoing attacks.
Reddy and Sharma~\cite{reddy2020ul} detect double spending in DAG-based ledgers by using spectral clustering (an unsupervised learning method) to identify blocks produced by non-cooperating miners.
Venkatesan and Rahayu~\cite{venkatesan2024blockchain} later propose combining hybrid consensus protocols (e.g., proof of stake and work) with ML classifiers for real-time anomaly detection, aiming to preemptively detect threats such as 51\% attacks.
A separate line of work has focused on detecting selfish mining using discriminative ML classifiers.
Wang et al.~use a neural network based on fork structure~\cite{wang2021forkdec}.
Later work has extended the analysis to use more advanced ML-techniques such as ensemble deep learning~\cite{wang2024detection} and to use additional features such as transaction fees and block generation times~\cite{bao2025semi_detector}.
Recent work has also studied undetectable selfish mining strategies, rendering these mechanisms ineffective and calling for more research on how to detect and mitigate such attacks~\cite{bahrani2024undetectable}.
Overall, the dominant approach across consensus-layer attack detection has been discriminative supervised classification on hand-crafted features, with deeper architectures and graph neural networks emerging more recently.
However, progress remains constrained by reliance on simulator-generated training data and the scarcity of confirmed real-world attack samples.

\subsubsection{P2P analytics: Attack discovery}
\label{sec:p2p-analytics}
Another line of work has used ML to detect ongoing attacks at the P2P network layer. A primary attack of interest is \emph{eclipse attacks}, in which an attacker tries to separate one or more nodes' IP addresses from the rest of the network in order to alter and control the views of different participants in a blockchain network.
These attacks can sometimes be detected by analyzing network traffic patterns and applying basic ML tecnhiques ~\cite{xu2020eclipsed,bhumichai2023detection,dai2022eclipse,rehman2025eclipse}. 
For instance,~\cite{xu2020eclipsed} 
used a random forest classifier to identify attacks based on statistical features like packet size and access frequency.
Later works, such as Dai et al.~\cite{dai2022eclipse}, employed more advanced ML techniques, combining convolutional neural networks with bidirectional RNNs and a cross-attention mechanism to capture both spatial and temporal patterns in the network traffic data, thereby improving detection accuracy.
Despite some work in this direction, ML-based approaches (both supervised and unsupervised) are difficult to develop due to a lack of labeled data from eclipse attacks. 
Prior methods have handled this problem in part via synthetic data augmentation techniques, which can give small gains in detection rates \cite{dai2022eclipse}.

\subsubsection{Derived property analytics: Price prediction} 
\label{sec:derived-property-analytics}

Many papers have explored ML techniques for predicting the price of various cryptocurrencies \cite{indera2017non,jang,saad2019toward,kim2021predicting,jagannath2021chain,chen2023analysis,vieitez2024machine}. 
Recent papers have used relatively standard ML tools such as Bayesian neural networks \cite{jang}, RNNs \cite{saad2019toward,jagannath2021chain,chen2019deepmarks,vieitez2024machine}, MLPs \cite{kim2021predicting}, and SVMs \cite{kim2021predicting,vieitez2024machine}.
These recent papers have generally emphasized the importance of using a  diverse set of input data and features, including on-chain data, data from external (non-cryptocurrency) markets, and social media platforms to understand user sentiment \cite{saad2019toward,chen2019deepmarks,kim2021predicting,vieitez2024machine}. However, they still have used fairly basic ML tools, making limited use of neural networks. 

In contrast, industry tools are increasingly using foundation models (FMs) to address these problem. For example, ElizaOS \cite{elizaos} and Virtuals \cite{cryptoslate} enable users to deploy agents that make predictions and decisions based on underlying LLMs (e.g., see~\Cref{sec:decision-making}).
As these agents use general-purpose LLMs for decision-making, they are not specifically trained for price prediction. 

A major difference between price prediction in cryptocurrency and traditional markets is that cryptocurrency markets still depend (relatively) more on smaller investors, particularly for nascent or volatile assets, such as memecoins. As a result, cryptocurrency prices are more impacted by---and discussed on---social media channels like Discord and Telegram, which can be mined for side information, some of it misleading \cite{kawai2023your}. In this sense, the cryptocurrency price prediction problem can benefit from a potentially richer set of side information than narrower problems like attack detection, and LLM-based predictors are well-suited to handle such side information. On the other hand, it may be possible to design better custom models to exploit this rich side information. A key research challenge is how navigate this tension. 

\begin{researchQuestion}
How can we design ML tools to predict cryptocurrency aggregate properties (like price) that effectively incorporate rich and unstructured side information from the Internet? 
\end{researchQuestion}
To our knowledge, neither prior research papers on cryptocurrency price prediction nor industry efforts are making use of state-of-the-art time-series forecasting models \cite{li2025tsfm,ansari2024chronos,timesfm}. Could such tools  provide substantial gains relative to the simpler and lower-dimensional predictors that have been proposed in prior research?
Could they also provide benefits over general-purpose agents seen in industry?
A key research challenge associated with designing custom predictors is the rich nature of side information for cryptocurrency markets; text messages and chat signals from various platforms are semi-structured, whereas the ML community has largely developed forecasting models for fixed benchmarks over highly-structured data. Reconciling these differences may require novel architectural adaptors and/or data processing, analogous to what has been done in other domains \cite{cohen2025time}.
We discuss some of these questions at a broader scale in the conclusion of this section, \Cref{sec:concludion-ai4blockchains}.

\subsubsection{Summary}
\label{sec:summary-ai-analytics}
Notice that in the papers discussed above, ML has been used for analytics in settings where we have \emph{visibility}. That is, given a global, known consensus protocol, we can discover vulnerabilities, or given public global information about the state of the blockchain, we can detect attacks or predict price fluctuations.
However, due to the decentralization of many blockchains, we may lack the visibility to run many kinds of analytics.

\begin{keyInsight}[ML for aggregate-level blockchain analytics]
    For aggregate-level blockchain analytics, ML has been applied to a limited set of problems for which we can obtain global, public observations about a system. On the other hand, fine-grained system monitoring and local analytics remain challenging due to lack of central coordination and visibility. 
    \label{insight:aggregate-analytics}
\end{keyInsight}
For example, in enterprise settings, there is growing interest in applying AI to observability \cite{cohen2025time,zhang2025survey,cheng2023ai}. In other words, given a complex microservices architecture, how can we efficiently collect telemetry from various components and use them to identify and resolve bottlenecks or errors? 
The data anlytics component of this pipeline is increasingly being handled with ML \cite{cohen2025time}.

As blockchains process increasing volumes of data, observability techniques could help to streamline operations by pruning out unresponsive peers, identifying smart contracts that are resource bottlenecks, or detecting unreliable nodes; however, traditional telemetry requires special instrumentation and centralized aggregation, and is ill-suited to blockchain networks. 
Hence, an interesting research question is how to design observability infrastructure and associated ML-based analytics pipelines for decentralized blockchain systems. 
\begin{researchQuestion}
How can observability tools help streamline decentralized blockchain operations? How can we design privacy-preserving telemetry  and ML-based analytics for streamlining blockchain management in a decentralized fashion?   
\end{researchQuestion}
This open-ended research question requires a detailed understanding of performance bottlenecks in blockchain systems. 
First and foremost, it would require measurements from infrastructure providers to characterize hardware and software bottlenecks that commonly hamper performance across the blockchain stack.
Based on the identified bottlenecks, the research and/or open-source community could develop standardized telemetry for blockchain systems---akin to OpenTelemetry for microservice monitoring \cite{otel}.
Designing observability tools for blockchains introduces new challenges, such as monitoring consensus artifacts (e.g., block structures) from different vantage points, and adapting typical metrics and tracing techniques to blockchain infrastructure. 
For example, in a typical microservice architecture, each user request spawns a \emph{trace}, which records how requests flow between services. However, in a decentralized blockchain, each request to a smart contract can spawn calls to different smart contracts, each of which is executed in parallel on different validators' hardware. Hence, there are correlated but different traces (in terms of timing) occurring at each validator. Broadly, designing efficient data representations for blockchain traces, logs, and metrics could be a concrete research challenge. Another may be collecting telemetry in a privacy-preserving manner that does not leak details about individual validators' proprietary algorithms and infrastructure configurations, as well as private user transaction patterns.

\subsection{Analytics for local blockchain objects}
\label{sec:application-analytics}

Next, we narrow our focus from aggregate analytics to ML-based analytics for \emph{individual} objects or artifacts at the application layer---namely, transactions or smart contracts. 
We categorize these methods as follows: (1) smart contract security analysis, (2) smart contract economic analysis,  (3) transaction-level deanonymization, and (4) transaction-level fraud analysis.

\subsubsection{Smart contract security analysis}
\begin{figure}
    \centering
    \includegraphics[width=1\linewidth]{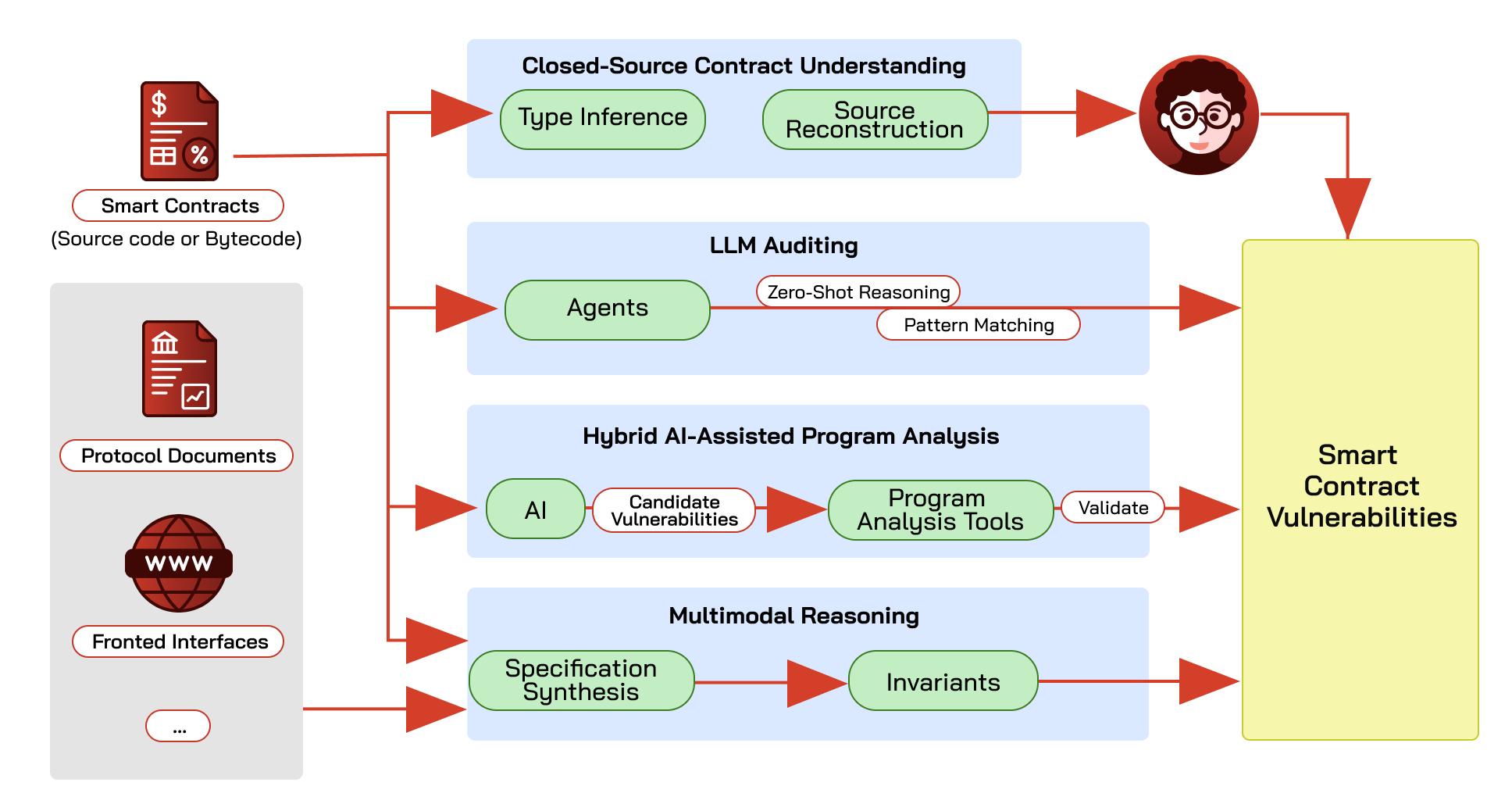}
    \caption{Four paradigms of AI-assisted smart contract analysis, which differ in how AI is used in the analysis pipeline and the sources of information they incorporate. Existing work can be broadly categorized into: (i) closed-source contract understanding, where AI assists reverse engineering of bytecode; (ii) LLM-based auditing, where LLM agents attempt to emulate human security auditors; (iii) hybrid AI-assisted program analysis, where AI proposes candidate vulnerabilities that are validated using traditional program analysis tools; and (iv) multimodal reasoning, which integrates additional information sources such as protocol documentation and frontend interfaces to verify contract behavior and detect semantic inconsistencies.}
    \label{fig:ai-for-smart-contract-analysis}
\end{figure}

Smart contract security is a critical pillar of blockchain integrity~\cite{luu2016making,tsankov2018securify,nikolic2018finding,kalra2018zeus,permenev2020verx,zhou2023sok}, as vulnerabilities in deployed contracts can directly lead to financial losses.
Traditionally, vulnerability detection relies on static analysis~\cite{tsankov2018securify,grech2019gigahorse,brent2020ethainter}, symbolic execution~\cite{frank2020ethbmc,gritti2023confusum,ruaro2024not}, and fuzzing~\cite{jiang2018contractfuzzer,grieco2020echidna,shou2023ityfuzz}.
While effective for detecting well-defined vulnerability patterns, these techniques often struggle with complex contract logic and cross-contract interactions, motivating the use of AI-assisted approaches to enable richer semantic reasoning.
As summarized in \Cref{fig:ai-for-smart-contract-analysis}, we categorize existing AI-assisted approaches into four paradigms: closed-source contract understanding, LLM-based auditing, hybrid AI-assisted program analysis, and multimodal reasoning.

\paragraph{Closed-source contract understanding.}
The first paradigm of AI-assisted smart contract analysis aims to recover human-interpretable semantics from deployed bytecode, typically through decompilation, enabling downstream tasks such as human auditing.
However, traditional decompilers~\cite{zhou2018erays,grech2019gigahorse,grech2022elipmoc} often leave a semantic gap, producing low-level logic that is difficult for auditors to interpret.
To address this gap, AI-based decompilation has evolved from attribute-specific inference to end-to-end source reconstruction. Early approaches focused on specific metadata recovery. For instance, SigRec~\cite{chen2021sigrec} recovers function signatures from EVM bytecode by exploiting call-data access patterns to infer parameter counts and types via type-aware symbolic execution, so it doesn't require source code or signature databases.
SmartHalo~\cite{liao2025augmenting} follows a neuro-symbolic approach, combining static dependency analysis with LLMs to recover high-level attributes, while preserving functional correctness through symbolic verification.
More recent work explores direct bytecode-to-source translation through two main paradigms: prompt engineering and fine-tuning. Representing the former, DiSCo~\cite{su2025disco} adopts a training-free approach that uses semantic units to exploit the zero-shot capabilities of general-purpose LLMs. 
In contrast, David et al.~\cite{david2025decompiling} apply Low-Rank Adaptation-based fine-tuning on large-scale datasets that map three-address code to Solidity, enabling the model to match human programming styles better.

\paragraph{LLM-based auditing of contracts.}
The second paradigm of work explores the use of LLMs as automated auditors that directly analyze contract code. This approach leverages the zero-shot reasoning capabilities of general-purpose models to mimic human auditing processes.
Recent empirical studies~\cite{david2023you,chen2025chatgpt} have revealed a mixed picture: while LLMs demonstrate measurable capability in identifying vulnerability-related patterns (e.g., David et al.~\cite{david2023you} show that both GPT-4 and Claude correctly identify the vulnerability type in 40\% of 52 previously exploited DeFi smart contracts),  this comes at the cost of non-trivial false positives due to hallucinations~\cite{ji2023survey,zhang2025siren}.

\paragraph{Hybrid AI-assisted program analysis.}
To address the high false positive rates of pure LLMs, the third paradigm has adopted hybrid architectures that couple the semantic understanding of AI with the rigorous validation of traditional program analysis~\cite{so2021smartest,sun2024gptscan,zhang2024detecting}.
GPTScan~\cite{sun2024gptscan}, for example, utilizes GPT to identify potential vulnerability candidates and then employs static program analysis, including data-flow and control-flow validation, to confirm the feasibility of the detected vulnerabilities.

\paragraph{Multimodal reasoning.}
While the above approaches primarily operate at the code level, the fourth paradigm moves beyond pure program analysis and shifts vulnerability detection toward protocol-level semantics and multimodal reasoning.
Representative examples of this paradigm include DeFiAligner~\cite{gan2024defialigner}, which uses LLMs to extract business logic from project documentation and aligns it with symbolic code summaries to identify semantic deviations, and Hyperion~\cite{yang2024hyperion}, which analyzes DApp front-end interfaces using a fine-tuned LLaMA2 model to uncover inconsistencies between user-facing promises and on-chain execution.
PropertyGPT~\cite{liu2024propertygpt} leverages retrieval-augmented generation to synthesize compilable specifications from audit reports, while SmartInv~\cite{wang2024smartinv} employs a ``Tier of Thought'' strategy to infer critical invariants from multimodal sources, including code and natural-language comments.

\bigskip
Taken together, these paradigms suggest a clear trend of leveraging AI to reduce human auditing effort while expanding the scope of vulnerability discovery and maintaining accurate detection. 

\begin{keyInsight}
    State-of-the-art methods for detecting security flaws in smart contracts do not purely rely on AI-based predictions from input features; instead, they identify constraints or invariants that must hold, and combine these with ML models to detect vulnerabilities. 
\end{keyInsight}
Many papers today are combining deterministic program analysis and/or domain-informed semantic analysis with ML tools for vulnerability detection. This is a powerful paradigm, but it begs the question of how best to extract and use such domain-specific knowledge, as highlighted in the following research question.

\begin{researchQuestion}
What level of domain knowledge should be integrated into smart contract security analysis pipelines, and how should that couple with AI-based techniques? What information sources, including multimodal data, should these AI-assisted systems leverage?
\end{researchQuestion}
Prior work has explored various operating points in terms of extracting domain-specific properties that should hold in smart contracts, and coupling these with AI. However, it remains unclear what architecture is best under what conditions, and what role AI should play. Should it be used purely as a filter, as in GPTScan? Should it be used to identify policies and invariants that smart contracts should satisfy? If the latter, what kinds of inputs are required, including documentation types and level of detail? While there exist various proofs-of-concept for different architectures, we lack a systematic exploration comparing different architectures and information sources. Such an exploration could help to inform even stronger architectures for smart contract vulnerability analysis.

\subsubsection{Smart contract economic analysis}
AI is used not only to analyze security properties of smart contracts, but also to study the economic behaviors generated by smart contract interactions.
In the current literature, this line of research is largely centered on maximal extractable value (MEV), namely profit obtained through strategic transaction ordering during block construction~\cite{daian2020flash}.
Among the most commonly-studied MEV activities are arbitrage~\cite{zhou2021just,wang2022cyclic}, sandwich attacks~\cite{zhou2021high}, and liquidations~\cite{qin2021empirical}.
Existing work in this area mainly falls into two categories: MEV discovery and MEV activity detection; we discuss ML-assisted discovery of bidding strategies in \Cref{sec:bidding-strategies}, which is more of a constructive design problem than an analytics problem.

\paragraph{MEV discovery.}
First, AI can help discover MEV opportunities by searching over transaction sequences and execution environments.
Earlier work in this direction mainly relied on non-learning-based techniques, including constraint-based search over DeFi actions~\cite{zhou2021just}, formal verification of composed DeFi contracts~\cite{babel2023clockwork}, and heuristic-driven analyses of arbitrage opportunities~\cite{wang2022cyclic}.
In contrast to these methods, Lanturn~\cite{babel2023lanturn} formulates MEV extraction as an adaptive learning-based black-box optimization problem, aiming to synthesize profit-maximizing transaction sequences rather than merely detect pre-specified arbitrage patterns.
MEVisor~\cite{chen2026mevisor} complements this line of work by emphasizing high-throughput opportunity search through parallel genetic algorithms and GPU-accelerated execution.

\paragraph{MEV activity detection.}
Second, AI can be used to detect realized MEV activity, that is, to automatically identify and classify MEV-related behaviors from on-chain or bundle-level data.
Early approaches primarily relied on heuristic rules to detect MEV activities~\cite{torres2021frontrunner,qin2022quantifying,mclaughlin2023large}.
While effective in specific settings, these approaches depend heavily on hand-crafted patterns and are therefore limited in automation, scalability, and adaptability to new MEV strategies.
Recent work has therefore begun to introduce AI-based methods for MEV activity detection.
A representative example is the combined ActLifter--ActCluster pipeline~\cite{li2023demystifying}, which first lifts raw transaction traces into semantic DeFi actions, then performs bundle representation learning to encode raw bundles as low-dimensional feature vectors, and finally applies iterative clustering to identify both known and previously unseen MEV activity patterns.

\begin{keyInsight}[Trends in smart contract economic analysis]
AI has shifted smart contract economic analysis away from hand-crafted patterns and static heuristics toward more automated methods that can adapt to complex behaviors with less task-specific prior knowledge. This opens the door to exploiting subtle structures and misalignments in smart contracts that may not be intuitively obvious to humans.  
\end{keyInsight}

\subsubsection{Transaction-level deanonymization}
Cryptocurrency deanonymization refers to identifying the source of a given transaction, as defined either by their real-world identity or other network identifiers like an IP address. ML-based deanonymization typically arises through one of a few mechanism classes: (1) {interactions with centralized third-party services,}
(2) {behavioral and on-chain pattern analysis}, and 
(3) {network layer and related side channels}. 
Most \emph{public} research on AI-based methods for deanonymization stems from the 2nd and 3rd categories, as we summarize below.

\paragraph{Behavioral and on-chain pattern analysis.}
Behavioral and on-chain pattern analysis facilitates deanonymization without the need for direct involvement from third parties. Common techniques include address clustering, as well as the analysis of transaction amounts, frequencies, payment habits, address reuse, and timing patterns. These heuristics leverage the blockchain's inherent transparency to uncover relationships between addresses, and, in some cases, to link them to real-world individuals.
For instance, 
several works have used manual graph representation learning techniques to categorize nodes \cite{beres2021blockchain,liu2024fishing}. 
Moving towards more automated feature extraction methods,  
many papers use graph neural networks to learn embeddings from transaction graphs for account deanonymization   \cite{lin2020t,shen2021identity,zhou2022behavior,li2022ttagn,zheng2023tegdetector,yu2023money,du2023breaking,hu2023bert4eth,fan2024edge,huang2025hierarchical}.
For example, Huang \emph{et al.} combine hierarchical self-attention modules to model local node structure on the transaction graph, and large-scale relations across the graph in their architecture~\cite{huang2025hierarchical}. 
Hu \emph{et al.} instead use a sequential modeling of transactions; their BERT4ETH model is an encoder-style transformer pretrained on sequences of Ethereum transactions ~\cite{hu2023bert4eth}. It is used to extract embeddings for Ethereum accounts for deanonymization and anomaly detection (see \Cref{sec:tx-level-fraud} below).
To our knowledge, there is no head-to-head comparison between the sequential modeling approach of BERT4ETH \cite{hu2023bert4eth} and recent models that explicitly use the graph structure, like \cite{zhou2022behavior,fan2024edge,huang2025hierarchical}. Hence, it is unclear which architecture works best, though the recent trend has clearly favored graph-aware modeling; that is, self-attention may be used, but in a way that respects the topology of the underlying graph.

\paragraph{Network layer and side channels.}
Network-layer and side-channel attacks exploit the dynamics of the peer-to-peer infrastructure to infer a transaction source's IP address. An adversary who controls a significant fraction of nodes can monitor transaction propagation patterns, infer the originating node through timing analysis or distinctive relay behavior, and collect off-chain metadata, such as IP addresses exposed during message broadcast, device fingerprints, or correlations with external events~\cite{biryukov2019deanon,gao2021practical,fanti2017anonymity,bojja2017dandelion}. 
These methods have mostly not used data-driven ML predictors, focusing instead on the design and analysis of statistical source predictors based on propagation dynamics \cite{fanti2017anonymity,bojja2017dandelion}.

\subsubsection{Transaction-level fraud analysis}
\label{sec:tx-level-fraud}
Complementary to analytics for source deanonymization, another important class of analytics aims to detect fraudulent or anomalous crypto transactions \cite{ross2007money}, including for applications in Anti-Money Laundering (AML) and Countering the Financing of Terror (CFT). 
Historically, rule-based detection systems were widely used for fraud detection---both in blockchains and traditional finance \cite{ross2007money}. However, these methods fall short in dynamic and complex environments, often producing high volumes of false positives that carry substantial operational and compliance costs \cite{adebayo2023comparative,ross2007money}.
For example, Project Aurora~\cite{BIS2023ProjectAurora}  assessed various strategies for fraud detection, including existing rule-based models, logistic regression, neural networks, and graph neural networks, under realistic real-world conditions and constraints. They found that graph neural networks were most effective in detecting money laundering in a synthetic cross-border transaction dataset.

Data-driven, AI-assisted fraud detection algorithms
can use data from multiple layers of the blockchain stack and off-chain sources. 
Common data sources include application-level traces (e.g., transactions within an application, including timing data), off-chain marketing data, such as social media, and off-chain secondary market data (e.g., price fluctuations).  
We summarize detection techniques based on the type of input data.

\paragraph{Transaction (sequence) metadata.}
Many ML tools have been designed to detect illicit activities and scams using transaction-
or sequence-level characteristics, such as the timing between transactions from an account, the type of transactions in an application, transaction amounts, involved parties, and destinations.  
These features were then used to train basic predictive models  \cite{xia2021trade,huang2023miracle,gai2023blockchain,aziz2022lgbm,farrugia2020detection,hu2023bert4eth,palaiokrassas2024leveraging}. 
BlockGPT \cite{gai2023blockchain} and BERT4ETH \cite{hu2023bert4eth}, for instance, use serialized Ethereum transactions to train transformers, which can be used to extract embeddings for downstream classification tasks.

\paragraph{Topological data.}
Another important class of methods explicitly models {topological data}, either derived from measurements from the P2P network \cite{kim2022machine} or by observing the logical transaction graph \cite{wu2020phishers,akcora2021bitcoinheist,patel2022evangcn,qi2023blockchain,voronov2023framework}.
For example, BitcoinHeist extracts custom features from the Bitcoin transaction and address graphs to identify ransomware \cite{akcora2021bitcoinheist}, and similar graph-based features (e.g., flow patterns, mixing behaviors, and layering structures) have been proposed for AML/CFT risk scoring and suspicious activity detection~\cite{halford2025developing,vassallo2021application,wan2024novel,lorenz2021machine}. Graph neural networks, including graph convolutional networks (GCNs), have also shown strong promise for anomaly and fraud detection in blockchain transaction data \cite{BIS2023ProjectAurora,patel2022evangcn,pocher2023detecting,wu2025profit}. In addition to the findings from Project Aurora \cite{BIS2023ProjectAurora}, Patel \emph{et al.} proposed EvAnGCN, which applies a dynamic GCN to the induced and evolving transaction graph of the Ethereum blockchain to effectively identify anomalous behaviors~\cite{patel2022evangcn}. More recently, Pocher \emph{et al.}  demonstrated that standard GCNs outperform traditional  methods in detecting anomalous cryptocurrency transactions~\cite{pocher2023detecting}.  

\paragraph{Off-chain data.}
Increasingly, researchers are also leveraging {off-chain data} to detect anomalous, fraudulent, or illicit cryptocurrency activity~\cite{ermilov2017automatic}. Several studies have focused on identifying money-laundering transactions linked to darknet markets~\cite{dos2024identifying,alarab2020comparative}. Others turn to social media and secondary online sources. For instance, CryptoScamHunter employs natural language processing to analyze YouTube video titles and descriptions, enabling the detection of DEX arbitrage bot scams~\cite{li2023towards}. Similarly, Huang \emph{et al.} combine secondary market data with on-chain trading activity to identify NFT rug-pull scams~\cite{huang2023miracle}. Beyond content analysis, Wang \emph{et al.} examine browser extension reviews together with programmatic features to predict whether a crypto-related browser extension is malicious~\cite{wang2022characterizing}. Several works have applied NLP techniques to social media content to uncover cryptocurrency pump-and-dump schemes~\cite{mirtaheri2021identifying,pacheco2021uncovering,nizzoli2020charting,nghiem2021detecting}.

\begin{keyInsight}[AI for transaction-level analytics]
    Current state-of-the-art methods for transaction-level analytics---including for fraud detection and deanonymization---make heavy use of transaction graph characteristics and metadata. These inputs are used to extract neural representations that capture graph dependencies, either explicitly (via a GNN) or implicitly (via a transformer trained over curated transaction sequences). 
\end{keyInsight}

Of course, strong detectors may also use a broad array of structured and unstructured auxiliary inputs that are not graph structured. An important question is how best to design ML algorithms that can effectively use such auxiliary inputs.

\section{AI-Assisted Design of Constructive Algorithms}  
\label{sec:ai-design-algos}
The previous category of work (\Cref{sec:ai-analytic-tools}) focuses on AI-assisted analytics: tools for \emph{understanding} the state of existing crypto applications and networks. 
Since that category of research took off over a decade ago, the research community has also turned to AI to help \emph{design} decentralized algorithms, ranging from peer-to-peer networks and blockchain protocols to applications and markets. These techniques typically rely on reinforcement learning and related ML techniques to drive the design of algorithms---both for the core infrastructure, the underlying consensus protocols, and strategic application algorithms. We categorize these approaches by the layer of the blockchain stack at which they operate: the peer-to-peer network (P2P; \Cref{sec:p2p-ai-design}), consensus layer (\Cref{sec:consensus-ai-design}), and application layer (\Cref{sec:application-ai-design}). 

\subsection{Peer-to-peer protocols}
\label{sec:p2p-ai-design}
AI has been used to design algorithms that help form and maintain the peer-to-peer message transport layer in blockchain networks. In a related development, application-layer networks that maintain links among specific peers have also benefited from the use of AI. The latter concerns Layer-2 solutions, such as payment channel networks.

\paragraph{Peer-to-peer network}
Several algorithms have been proposed in which individual network nodes make AI-informed local changes (e.g., rewiring peer connections, updating the communication profile) to improve local and global network properties, such as average latency or total communication volume \cite{mao2020perigee,tang2023strategic,mao2023topiary,tang2023strategic,valko2024sustainable}. 
For example, Topiary \cite{mao2023topiary} explicitly formulates the network topology formation as a multi-armed bandit (MAB) problem and designs an algorithm to update each node's peer connections, dropping peers with high relative latency of incoming messages.
Valko and Kudenko~\cite{valko2024sustainable} 
instead employ an RL agent that re-orders the broadcast queue to neighbors to reduce both propagation time and the total number of messages sent, thereby lowering the energy footprint of the network.

Similar ideas have been applied in domain-specific blockchain networks, such as vehicular networks and wireless networks. 
In vehicular networks, nodes frequently join and leave, leading to much higher peer churn than in cryptocurrency networks.
Kim and Ibrahim~\cite{kim2022byzantine} model the selection of the number of peers as a contextual MAB problem, dynamically adjusting channel size to maintain Byzantine fault tolerance under frequent vehicle churn.
Saadat et al.~\cite{saadat2023ai} further use ML to predict node stability in cluster-based vehicular networks, selecting consensus nodes that are unlikely to leave mid-process. 
In wireless networks, high transmission rates result in high energy consumption.
To that end, Ju et al.~\cite{ju2023fast} use GCNs to determine each node's data transmission rate, with the objective of balancing reliable communication with low energy consumption.

Overall, AI-assisted methods for the design and management of P2P networks remain relatively underexplored compared to other uses of AI in the blockchain context. One interesting question is how to combine AI-assisted P2P network management with methods for detecting ongoing network attacks in \Cref{sec:p2p-analytics}.

\begin{researchQuestion}
    How can one combine techniques for detecting eclipse attacks (\Cref{sec:p2p-analytics}) with algorithms for P2P network management and optimization? When do existing P2P network management algorithms provide robustness under adversarial network conditions (and how should we model such adversarial conditions)? 
\end{researchQuestion}

\paragraph{Layer-2 networks and interchain communication}
Layer-2 networks that rely on a blockchain consensus layer (i.e., their Layer~1) also use a P2P structure, but they address a different set of challenges than those faced by P2P networks for message propagation. A prominent example are payment-channel networks (PCNs) like the Lightning Network that is attached to Bitcoin.
The connections in a PCN represent logical channels over which money can be routed for peer-to-peer transfers rather than communication channels. Connections therefore represent a form of mutual trust and guarantee that participants can carry out transfers without resorting to the Layer-1 network.
Despite these differences, both types of networks pose similar challenges related to network formation and routing: What network topology should one form? How should one manage inactive peers? How should transactions be routed over the network? 
Several of these challenges have been tackled with RL-based algorithms, as in P2P networks. 
 For example, RL-based mechanisms have been used to set channel parameters (e.g., fees) to maximize an operator's profits \cite{asgari2022dyfen,plac,BiLSTMPPO}, select payment routes over the network to minimize cost to transaction senders \cite{marl,rlpath,valko2024sustainable,valko2025hybrid}, and rebalance channels to improve network throughput \cite{DRL-PCR,rebel}.
Among these methods, similar RL algorithms have been used  for routing management \cite{valko2024sustainable,valko2025hybrid}; however, to our knowledge, there has been limited work on using RL to determine how to make and break channels in a PCN (whereas this is a small but established research area in the P2P literature). 
This may be an interesting question for future work. 

\begin{researchQuestion}
    How can PCN nodes use RL methods to determine which edges to form and break dynamically over time? How would such algorithms impact the node's own rewards, as well as overall network health?
\end{researchQuestion}

\subsection{Consensus protocols}
\label{sec:consensus-ai-design}

AI has been used extensively at the consensus layer, both to enhance existing consensus protocols and operations and as an ingredient in completely new protocols.

\subsubsection{Enhancing performance}

Consensus protocols were originally designed for static environments, but they
often operate under dynamic conditions in practice, where the network
conditions and participants may vary.  In recent years, a growing body of work
has turned to ML techniques to enhance the performance of consensus protocols
(i.e., increase throughput and/or reduce latency) by being more responsive to
changing conditions.  One can distinguish three related areas, where AI
has been used to enhance consensus.

\paragraph{Protocol selection.}
In a first line of work, consensus protocols have been \emph{selected} with
the help of~AI.  The intended deployment network, the available
synchronization features, and the expected connectivity influence the choice
of a consensus protocol. No single consensus protocol dominates across all
operating conditions: Bitcoin's PoW consensus has high latency and is
optimized for maintaining safety in a loosely synchronized network, with time
constants on the order of minutes; the PoS protocols of Ethereum and Cardano,
in contrast, assume much closer synchronization, on the order of 10--20
seconds, and more uniform reachability.

It has therefore been suggested to build systems that switch between protocols
during operation, most often in the sense that they run a fast and fragile
protocol optimistically and switch to a slower robust protocol when they
detect problems with the optimistic path.  A classic work exploring this idea
proposed to build the ``next 700 BFT protocols'' in this way~\cite{agkqv15}.
AI-based methods have been used to guide the choice of protocols.  Whereas Liu
et al.~\cite{liu2019performance} develop a system that statically selects a
consensus algorithm using deep reinforcement learning, the recent BFTBrain
work~\cite{wu2025bftbrain} selects dynamically among consensus protocols such
as PBFT~\cite{castro1999practical}, HotStuff~\cite{yin2019hotstuff}, and
Zyzzyva~\cite{kotla2007zyzzyva}.  It does so by treating protocol choice as a
contextual MAB problem and addressing it with a RL engine built into the
overall system.

Relying on automation and on AI to speed up consensus in good cases, however,
opens up potential avenues for attack that can be exploited by bad actors.
Classic results in this area demonstrate that operating with the wrong
protocol for the given environment can reduce performance to
zero~\cite{cwadm09, ackl11}.  ML-based protocol-selection methods
therefore need to be robust.

\paragraph{Parameter selection.}
In a second area of consensus enhancements, AI has been used to select the
\emph{parameters} of consensus protocols.  The underlying reason is that the
performance of a network depends heavily on configuration parameters such as
timeouts, block size, and block interval and on the available connectivity
among active operators.  Manual tuning is impractical in environments where
optimal values shift as the workload fluctuates.
A line of work focuses on the optimization of block parameters.
Monem et al.~\cite{monem2024sustainable} apply XGBoost to predict future transaction volumes and dynamically adjust the block size to match expected demand.
Other works~\cite{liu2019performance,zhai2024explainable} tune both the block size and interval using deep reinforcement learning.
Zhai et al.~\cite{zhai2024explainable} attempt to make these choices more explainable by using structural causal models to provide justifications for chosen values.
Dutta et al.~\cite{dutta2024robb} take a different approach by focusing on the timing of block creation.
They examine the use of reinforcement learning by an operator to learn when to seal a block so that transaction confirmation times are minimized.

\paragraph{Consensus participant selection.}
As a third and final area within consensus protocols, AI has been applied to
select \emph{participants} for particular roles in a consensus protocol.  As
most consensus protocols deployed in practice rely on a leader, measuring the
performance of nodes and choosing well-connected and alive nodes as leaders
generally speeds up protocol operation.  However, care must be taken to not
degrade performance, should the measurements turn out to be wrong.

Recent research~\cite{cgklms22,tksk24} shows performance of protocols benefits
greatly from carefully selecting the consensus leaders, even without
AI-specific selection heuristics.
The protocol of Islam et al.~\cite{islam2024mrl} selects leaders in
Proof-of-Stake consensus using a multi-agent RL approach.  It collects various
performance metrics about potential leaders from all validators and aims to
identify and exclude malicious validators through a penalty-reward mechanism.
Nour et al.~\cite{diallo2025optimized,diallo2025dagwise++} study the
application of AI in DAG-based BFT protocols such as
Narwhal~\cite{danezis2022narwhal} and
Bullshark~\cite{spiegelman2022bullshark}.  They use graph neural networks to
rank blocks and select leaders within the DAG, reducing latency and improving
throughput without affecting the soundness of the underlying consensus
protocol.
Many further authors have used RL and related methods in a similar vein to
dynamically assign scores to validator nodes and assign particular protocol
roles based on such scores~\cite{chen2021novel,leduc2022sabine,lekuge25}.  A
recent survey of the topic~\cite{rizkim25} collects many techniques, but
mostly demonstrates that this field is still in its infancy because of
multiple open problems with the use of automated reasoning.  In particular,
the survey highlights the potential for adversarial attacks that are enabled
through~AI.

\begin{researchQuestion}
  How robust are ML-based methods that operate within consensus protocols and
  guide parameter and participant selection against manipulation by
  adversarial insider nodes?  How can a decentralized consensus protocol come
  up with a common and trustworthy estimate of the trustworthiness of its
  participating nodes?
\end{researchQuestion}

\subsubsection{Sharding}

Deep RL has also been applied to the design of blockchain sharding algorithms, addressing two complementary problems: shard configuration optimization and cross-shard transaction reduction through data placement.

The earliest efforts used deep reinforcement learning, and {Deep Q-Networks (DQN)} in particular, to choose how many shards to create and how to size their blocks, framing these recurring decisions as a sequential optimization problem.
SkyChain~\cite{zhang2020skychain} used DQN to continuously vary parameter configuration while ensuring shards can efficiently merge and split when needed.
A concurrent work~\cite{yun2020dqn} shows that a DQN agent can learn to maximize throughput while respecting a security constraint derived from the estimated performance and detectable misbehavior of other nodes in the network.
A representative recent work that demonstrates this approach is
{TbDd}~\cite{zyswwz24}, a trust-based DRL-driven sharding framework: it
gathers feedback on performance and historical behavior of all nodes and
assigns roles accordingly.

A second line of work assigns data and accounts to shards to minimize costly
cross-shard transactions.  {TbDd}~\cite{zyswwz24} implements this as well, by
observing data-access patterns and assigning the data items to suitable shards
so as to bound the number of cross-shard transactions.
Wang et al.~\cite{wang2024blockchain} use generative AI in this context to
predict future cross-shard interactions in order to proactively assign nodes
to shards.  SPRING~\cite{li2024spring} employs deep RL to migrate accounts
among shards by exploiting spatial-temporal transaction patterns.
AERO~\cite{song2025aero} extends this approach by batching migration
decisions, shrinking the action space and allowing better scalability.
As dynamic sharding is currently not deployed on any live blockchain system to
our knowledge, these studies represent exploratory research so far.  If
sharding is ever employed on a broad scale, ML-guided sharding and data
placement algorithms will undoubtedly be taken up again and potentially also
deployed.

\begin{researchQuestion}
  How can AI and ML help with analyzing data dependencies on cryptocurrencies
  or general-purpose blockchains, without observing a particular transaction
  workload?  Which methods will result in automated data placement algorithms
  for sharded blockchains that reduce the dependencies across multiple shards
  and improve performance?
\end{researchQuestion}

\subsubsection{Trust models}

The huge energy cost and investment of seemingly useless computation in
Proof-of-Work (PoW) consensus has resulted in a long-standing quest for
\emph{useful} work to be employed within blockchain consensus, particularly
work related to AI because of its high computational burden.  However, this
has not been easy so far and no satisfying solutions exist.  Dotan and
Tochner~\cite{dottoc20} formally survey this field and derive constraints on
systems that rely on wasteless PoW.  They show that, under realistic
assumptions, the set of allowed problems must still involve elements of
cryptographic hardness in order to keep the protocol secure and efficient.

Recent studies are more positive: Komargodski, Weinstein et
al.~\cite{bakowe25,koscwe25} study the economic validity and equilibrium
dynamics of consensus with external rewards.  Multiplying large matrices
receives particular attention as a promising problem, on which a useful
computational puzzle may be based~\cite{koscwe25}.  Their work opens a path
towards using AI training and inference workloads for consensus; however, it
is too early to assess the empirical security of such blockchain networks
built on consensus from useful work.

\subsubsection{Summary}

Consensus protocols stand at the heart of all security mechanisms in
blockchains: they ensure agreement on a state and common actions based on the
inputs from and correct behavior of many participating nodes.  The tradeoff
between their security and performance has therefore received a lot of
attention.  Consequently, there is a large body of work on optimizing
parameters of various consensus mechanisms, some of it also using AI.  Current
research suggests that using ML for consensus protocols holds many promises.
But since AI-based protocol optimization methods have not been widely
deployed, their performance and resilience to attacks when exposed to
real-world networks remains so far open.

For the design of the core consensus protocols, based on the theory of
distributed computing, AI has played a minor role so far.  One could argue
that automated discovery of secure and performant protocols will become
feasible, similar to existing automation in cryptographic protocol
research~\cite{bbbbcl21}.

\begin{researchQuestion}
  How can we use AI to propose novel consensus mechanisms that optimize
  performance, in terms of communication and latency, while 
  also organically adapting to dynamic environments and
  remaining secure?  Can AI help us to design and analyze the security of
  such new protocols?
\end{researchQuestion}

\subsection{Design of applications}
\label{sec:application-ai-design}

A number of applications use AI as a core component of their design. 

\subsubsection{DeFi market design}
AI has been successfully used to design DeFi applications, including automated market makers (AMMs).
For example, ZeroSwap uses a DQN approach to change the price of an asset over time~\cite{nadkarni2024zeroswap}, and Moszczynski integrates AMMs with frequent batch auctions, using RL to optimize pricing rules between batches~\cite{moszczynski2025optimizing}.
Related techniques have also been applied to lending markets like Aave~\cite{aave}, Morpho~\cite{morpho} and Euler~\cite{euler}: 
in these markets, interest rates impact the utility and liquidity of the market. 
Several papers have applied stochastic control theory to set such interest rates~\cite{bastankhah2024thinking,bastankhah2024agilerate,bertucci2025agents,baude2025optimal}, sometimes in conjunction with deep learning techniques~\cite{baude2025optimal}.
Chitra recently showed how to use online learning to set interest rates in a regret-optimal manner~\cite{chitra2025curationary}.

\subsubsection{AI-enhanced smart contract security} Beyond protocol-level mechanism design, AI has also been applied to smart contract security, particularly to automate tasks such as generating exploits or patches for vulnerable contracts.
Prior work on automated exploit generation relies on manually crafted templates or symbolic analysis to synthesize attack sequences that satisfy specific vulnerability conditions~\cite{frank2020ethbmc,gritti2023confusum,ruaro2024not}. 
AI-based approaches improve exploit generation by enabling semantic reasoning and adaptive search, allowing systems to construct exploits beyond predefined templates with minimal human involvement.
For example, AdvSCanner~\cite{wu2024advscanner} combines LLMs with static analysis to generate adversarial contracts that specifically exploit reentrancy vulnerabilities.
Scaling this toward autonomy, the system A1~\cite{gervais2025ai} employs an agentic workflow with execution-based validation, using domain-specific tools to achieve autonomous discovery and validate profitable exploits on real-world blockchain states. Similarly, PoCo~\cite{andersson2025poco} introduces an agentic framework that transforms natural-language audit reports into executable Foundry proofs of concept through a Reason-Act-Observe cycle.

Complementary to automated exploit generation, recent work also explores AI-driven automated program repair for smart contracts, aiming to close the loop from vulnerability discovery to mitigation.
While traditional repair largely relies on rigid templates~\cite{jin2021aroc,nguyen2021sguard,rodler2021evmpatch}, AI has empowered these frameworks to become more adaptive and context-aware.
Initial AI-enhanced approaches, such as SmartFix~\cite{so2023smartfix}, enhance the traditional ``generate-and-verify'' paradigm by using statistical models to prioritize patch candidates; sGuard+~\cite{gao2024sguard+} introduces machine learning classifiers to guide rule-based repair, selecting the most appropriate rules to avoid over-patching.
Going beyond static templates, recent work leverages the semantic comprehension and generative reasoning capabilities of LLMs for end-to-end generative repair~\cite{wang2024contracttinker,zhang2025acf}.

\subsubsection{Algorithms for MEV extraction}
\label{sec:bidding-strategies}
AI is starting to be used to optimize bidding strategies in MEV auctions once an MEV opportunity has been identified.
To realize potential profits, the target transaction must still be included on chain, but in practice there are often many competitors for the same opportunity~\cite{daian2020flash}.
Earlier competition often took place in public priority gas auctions, whereas Flashbots-style private channels transformed this process into sealed-bid first-price auctions~\cite{flashbots2026auction,weintraub2022flash}.
This makes bidding itself a learning problem: the goal is to predict competitive bids and choose bribes that maximize expected profit conditional on inclusion.
For example, Raun et al. analyze Flashbots auction data and train machine learning models (namely a Light Gradient Boosted Machine regressor) to predict winning bribe ratios in MEV auctions, showing that learning-based bidding strategies can improve profitability in arbitrage MEV auctions~\cite{raun2023leveraging}.
Lanturn \cite{babel2023lanturn} instead optimizes a black-box reward function using recent adaptive sampling techniques to maximize extractable value \cite{javaheripi2020adans}.
Other work has explored predicting auxiliary variables like the length of a FlashBots auction and the maximal bid value using simple regression models such as random forests and decision trees \cite{ladoczki2025predictions}.

\section{AI-Enhanced Interactions with the Real World}
\label{sec:ai-interact-real-world}
\label{sec:oracle-feeds}

Efforts have emerged in the last few years to use AI to enhance the interactions of smart contracts with the real world. These efforts have seen deployment, but are best characterized as emerging. 

In this section, we discuss uses of AI that enhance three main types of blockchain interactions with the outside world: 
\begin{enumerate}
    \item \emph{Sensing:} (\Cref{sec:sensing}) AI can help smart contracts understand the state of the world in more complex ways than what is currently possible, e.g., by digesting and processing unstructured data streams with AI-powered oracles. 
    \item \emph{Execution:} (\Cref{sec:execution}) Generalizing beyond sensing, AI can extend the ways in which smart contracts can perform computation and impact the external world, e.g., by calling into AI models.
    \item \emph{Decision-Making:} (\Cref{sec:decision-making}) AI can help smart contracts enact more sophisticated decision-making pipelines. We describe a case study
    from the finance realm, showing how AI-based decision-making can introduce new challenges. 
\end{enumerate}
In all three of these areas, we discuss both opportunities and challenges.

\subsection{Sensing: Enabling smart contracts to understand natural language}
\label{sec:sensing}

In their initial form, smart contracts were designed to operate only on on-chain data. Blockchains overcame this limitation with the emergence of oracles (\Cref{sec:oracles})---systems that connect smart contracts to off-chain data and off-chain computation resources. 
However, most existing oracles today are limited to relaying clean, well-structured data from APIs. 
This limits oracles' reach to the world where there is no clear API data.
As a result, for example, smart contracts cannot understand or interpret human language or institutions, a key obstacle to their ability to represent written contracts.

AI has the potential to significantly expand the kinds of data that are accessible to smart contracts. For instance, oracle systems could verifiably use AI tools to translate loosely-structured data into a format readable by smart contracts---thus acting as middleware between the broader Internet and individual smart contracts.
For example, today, the only type of fully automated insurance smart contract is simple parametric insurance, where payouts are triggered when pre-agreed events occur (such as AXA's discontinued flight delay insurance, Fizzy~\cite{axzfizzy}).
With LLM-powered oracles, one could imagine automated insurance contracts that ingest and reason over richer evidence, such as claims narratives, police reports, or inspection reports from repair shops.
Another example is prediction markets, which rely on oracles to determine the outcome of a question.
Existing systems such as Polymarket rely on human proposers to answer questions where there is no clear API data, and human dispute resolution to resolve disagreements~\cite{PolymarketOracles}.
AI-powered oracles have the potential to automate this process, avoiding human errors and reducing the latency and costs associated with disputes.

\subsubsection{Risks of LLM errors}
Despite this promise, a central challenge in deploying LLM-powered oracles---and in the use of LLMs in critical systems in general---is the possibility of errors, such as reasoning flaws, hallucinations, and arithmetical errors~\cite{sato2026gpt52cantcountfive}.
Here, we focus on errors arising from AI's inherent limitations, assuming no adversarial manipulation. Protecting the AI oracle pipeline from attacks is crucial and orthogonal.

Two recent studies~\cite{EmpiricalEvidenceinAIOracle,umaoracle} on LLMs' accuracy in answering questions on Polymarket shed light on AI's current ability as an oracle.
In an empirical study from Chainlink Labs~\cite{EmpiricalEvidenceinAIOracle}, the authors used GPT-4o to resolve 1,660 markets on Polymarket and compare the LLM's answers with the ground truth.
GPT-4o achieves an overall accuracy of 89.3\%. This result is corroborated by a similar experiment by UMA~\cite{umaoracle}, where their Truth Bot achieved an accuracy of 75\% (though a good portion was due to answers being submitted before the deadline).
To put the numbers in context, the overall accuracy of human-provided answers to UMA's optimistic oracle is 98.2\%.
Thus, for this specific task, LLMs still make significantly more errors than humans, suggesting that guardrails and oversight are necessary.

The accuracy of LLMs varies across contexts. They perform quite well when the information to be extracted is discrete, and there are official sources of truth. A canonical example is sports outcomes.
In Chainlink Labs' study, 99.7\% of questions about sports outcomes were correctly answered. In UMA's report, the Truth Bot achieved a 99.3\% accuracy on sports and asset pricing markets (the latter is answered by calling specific APIs).

For less straightforward questions, error rates can be much higher.
For example, LLMs struggle to answer questions that involve time (e.g., ``Was the interest rate hike announced before or after the GDP report?'') or when substantial efforts are required to extract the answer (e.g., answering ``How many times will Trump say the word ‘steel mill’ at the Pittsburgh rally on May 30?'' requires transcribing the video and counting words, a task that cannot be accurately done with automated tools).

It is worth exploring ways to reduce error rates, such as by leveraging multiple independent agents.
For instance, UMA proposed to use AI agents to look for known patterns of hallucination (e.g., Perplexity making inferences rather than being fact-driven)~\cite{umaoracle}.
Supra's paper on Threshold AI Oracles~\cite{thresholdoracles} suggested having multiple agents engage in deliberation to increase resilience against manipulation and errors. 
However, when applied naively, these methods may not work well: recent research has shown that LLM-based error detectors perform much worse than humans~\cite{kamoi2024evaluatingllmsdetectingerrors}, and having agents debate with each other may reduce overall accuracy~\cite{kotaDesignEvaluationMultiAgent2026}.
How to reconcile LLM outputs under real-world oracle workloads and attack vectors remains an open research question.

\subsubsection{Tolerating LLM errors.}
Even in contexts where LLM performs very well, the error rate is not zero.
Therefore, systems that use LLMs as oracles must handle potential errors. 
Three possibilities exist.
First, applications that rely on LLMs need to be designed to tolerate errors, so that minor errors do not lead to catastrophic outcomes (e.g., use AI to resolve low-value markets with a total payout below a threshold).
The second approach is to involve humans in the loop to detect and correct AI errors.
UMA Optimistic Oracle (OO)~\cite{uma} is an example: after answers are submitted by a proposer (human or AI bot), a dispute window of 48 hours opens, during which anyone can challenge the correctness of the submissions, and if necessary, an arbitration process is initiated and eventually the entire set of UMA token holders verify the event outcome and vote on the final result.
However, the need for humans in the loop slows down decision-making.

A third possibility is to involve humans only when AI models cannot make a decision with high confidence. 
Recent works show that abstention, i.e., not giving an answer when facing uncertainty, can increase model accuracy~\cite{feng2024don,10.1162/tacl_a_00754}.
If AI models can abstain with high accuracy, then this design can make fast decisions when AI outputs are available and slow decisions only when they're not.

\begin{keyInsight}[Handling error of AI-powered oracles]
    Systems that use AI-powered oracles must handle potential errors. 
    There are three high-level approaches: 1) designing the system to tolerate errors (which may only be possible in limited scenarios), 2) involving humans to arbitrate AI outputs (which slows down decision making, but can be conditional on disputes), and 3) if AI models can abstain when facing uncertainty, it is possible to only involve humans when AI cannot decide.
\end{keyInsight}

\subsection{Execution: Enabling smart contracts to use AI models and tools}
\label{sec:execution}

Regular smart contracts can enforce simple rules, such as conditional statements, but they tend to struggle with higher-level tasks like data analysis, pattern extraction, and planning.
Moreover, they generally take action \emph{on-chain}---that is, they directly impact the state of a blockchain, and nothing else.
Smart contracts could extend their execution capacity, both on- and off-chain, with access to the broader AI ecosystem.

Specifically, the AI ecosystem comprises a vast collection of AI models and associated tools for interfacing with the real world in modalities including data access and modification, money transfer, and information retrieval. 
Running these tools on-chain, however, can be prohibitively expensive.
Hence a natural question is, how can we enable smart contracts to verifiably access AI models and tools in the real world (off-chain), without incurring prohibitive computational costs on-chain?

Oracles once again provide a potential solution, by enabling off-chain computation to be recorded on-chain in a verifiable manner.
Modern oracle systems, such as Chainlink and Supra, can already execute off-chain workflows and return attested results.
They can naturally (be extended to) support AI workflows. We will discuss approaches for doing so and their tradeoffs.

AI can transform smart contracts from enforcing static rules into dynamic, context-aware mechanisms.
For example, smart contracts can use AI tools to flag fraudulent transactions~\cite{11205825}, dynamically adjust parameters, and make automated trading decisions.
Executing AI logic directly on-chain is not economically feasible, as computation on-chain can be several orders of magnitude more expensive than cloud computing~\cite{UnderstandingtheIntersectionofCryptoandAIGalaxy}.
As an efficient alternative, oracles can connect smart contracts to off-chain AI tools, whether local models running within the oracle infrastructure or third-party ML service providers (e.g., OpenAI APIs).

Two security properties are important for this application of oracles.
First, smart contracts need to efficiently verify the \textit{integrity} of the entire AI workflow, ensuring that the inputs from smart contracts are not tampered with, the correct ML model is used, and the outputs are correctly computed.
These requirements are not unique to AI (they apply to off-chain computation by oracles in general), but the scale of AI workloads demands highly efficient solutions.
We refer readers to~\Cref{sec:block4ai-integrity}  for discussions on three technical approaches that can be used here: 1) an optimistic approach where integrity is ensured by economic incentives (i.e., such protocols are equipped with mechanisms to identify incorrect results and penalize the offender economically)~\cite{zheng2021agatha,verde2020,conway2024opml,mirkin2025arbigraph,yao2026tao}; 2) oracle nodes run ML models in a TEE and accompany the result with a hardware attestation to the correctness of the computation~\cite{eigencloud2025verifiable}; and 3) oracle nodes accompany the computation results with zero-knowledge proofs~\cite{liu2021zkcnn,
qu2025verfcnn,
chen2024zkml,
sun2024zkllm,
qu2025zkgpt,
garg2023experimenting,
abbaszadeh2024zero,
waiwitlikhit2024trustless,
shamsabadi2022confidential,
}.
The tradeoffs among these solutions are discussed in detail in~\Cref{sec:block4ai-integrity}.

Second, for ML tasks involving private data or proprietary models, their \textit{confidentiality} must be preserved.
For example, a smart contract may want to run a proprietary fraud-detection model that cannot be revealed even to the oracle nodes. To do so, the TEE-based approach is likely the most practical~\cite{eigencloud2025verifiable}, where the smart contract developer can encrypt the model under a TEE-secured key, so that the model is only decrypted inside the TEE.
Fully Homomorphic Encryption (FHE) can, in theory, be used to evaluate an encrypted model, but it is not yet practical for large models and as it requires a decryption capability, needs to rely on a committee of trustworthy nodes~\cite{10.1145/3605098.3635983}.

Beyond security considerations, running AI workloads can consume considerably more resources than typical oracle workflows (e.g., fetching prices), which give rise to a mechanism design question of how to price AI computation~\cite{bahrani2024resonance}, and a system design question of how to meter AI usage and charge smart contracts properly.
An interesting wrinkle to this problem is the risk of {\em freeloading}. Because blockchains lack confidentiality, computation results delivered by oracles are publicly available.
For instance, a copycat prediction market smart contract can monitor the answers to another prediction market that pays for AI-powered oracles and obtain free answers to resolve its own markets.
The freeloading issue was discussed in Town Crier~\cite{zhang2016town}, where the authors suggested using designated-verifier proofs to render the output verifiable only by the original requester, though a full solution is left for future work. We note that the issue is much more pronounced in the context of AI workloads, as the potential savings by freeloading AI computation can be significant.

\subsection{Decision-making: AI-based investment tools} 
\label{sec:decision-making}

If we give smart contracts access to AI models and tools as suggested in \Cref{sec:execution}, a natural question is how those tools will impact on-chain applications. Incorporating AI into smart contracts' decision-making processes introduces substantial complexity and opacity, whereas a strength of previous smart contracts was their (relative) transparency and interpretability. In this section, we describe a case study from the finance realm, showing how using AI-based decision-making in smart contracts can give rise to new tensions and sources of potential unfairness among participants. 
Specifically, one of the most popular applications of AI to blockchains today is in investment tools. We discuss ML-based Collective Investment Algorithms (CoinAlgs) and their hazards.

\subsubsection{CoinAlgs} \emph{Collective Investment Algorithms} (CoinAlgs) are algorithms shared by a community of users that drive collective investment actions~\cite{fabrega2026coinalg}. Algorithmic trading has long been common practice in traditional finance, both in major firms (high-frequency trading, hedge funds, quantitative investment, etc.), as well as in ubiquitous products used by retail investors (robo-advisors, trading software packages, etc.). 

More recently, with the widespread deployment of AI, CoinAlgs have become common
in decentralized finance, too. The most prominent example is \emph{AI-powered
investment DAOs}---communities of people that pool funds in a blockchain to make
collective trades, which are dictated by AI models or agents. Branded as
``decentralized hedge fund managers''~\cite{messari2024daos}, such CoinAlgs have
been the source of significant interest within Web3. Popular CoinAlg projects, such
as ElizaOS~\cite{walters2025eliza}, an AI-powered investment fund converted into a
general AI platform, and AI XBT~\cite{aixbt2024}, an AI-powered market-intelligence
agent, reached peak market capitalizations of \$2.7B and \$4.7B respectively, and
peak assets under management of \$22.9M and \$755.4M respectively~\cite{PaperAgents:2026}. Other popular
CoinAlgs include SingularityDAO~\cite{singularitydao2021}, which offers
``AI-Powered Quant Strategies'' for DeFi and Soldex~\cite{soldex2021}, a DEX that
offers AI-powered trading bots.

\subsubsection{Risks of CoinAlgs} While they promise to democratize finance, CoinAlgs pose hazards to their users as a result of their susceptibility to well-studied ML attacks. A long line of work, under the broad umbrella of \emph{adversarial machine
learning}~\cite{biggio2018wild,goodfellow2014explaining,greshake2023not}, has
shown that AI-based systems can be vulnerable to attacks that include prompt
injection~\cite{greshake2023not,perez2022ignore}, memory injection~\cite{patlan2025real},
data poisoning~\cite{goldblum2022dataset}, backdoor attacks~\cite{chen2017targeted},
and model extraction~\cite{tramer2016stealing}. Such attacks can have serious effects in practice, such as modifying the behavior of the system or leaking sensitive information about its architecture. They pose risks for AI-based CoinAlgs that range from manipulating investment decisions to revealing proprietary trading strategies. These concerns are not strictly theoretical; for instance, in recent work, Patlan et al.~\cite{patlan2025real} showed ``context manipulation'' attacks can be used to trigger malicious asset transfers against ElizaOS.

Beyond general attacks against AI systems, CoinAlgs raise specific concerns when used in finance. For instance, a number of papers have shown that AI trading agents can \emph{manipulate or influence trading markets in negative ways}. Recent work by Dou et al.~\cite{dou2025ai} shows that, even without direct communication channels, profit-maximizing AI trading agents can \emph{coordinate} and \emph{collude} among themselves, which can lead to inefficiency in the markets. Similarly, other works have shown how AI agents can manipulate financial benchmarks~\cite{shearer2023learning}, spoof limit order books~\cite{wang2021spoofing}, and evade regulation on market manipulation~\cite{wang2020market}.  

Recent work by Fábrega et al.~\cite{fabrega2026coinalg} highlights a more fundamental problem for investors. CoinAlgs face an inherent security tradeoff in their design dubbed the \emph{CoinAlg Bind}. Intuitively, a CoinAlg's trading strategy can either be \emph{transparent}, or it can be (at least partially) \emph{private}. Both alternatives raise serious risks for investors. Transparent trading strategies (e.g., open-source trading packages) naturally come at the expense of profits, as this can lead to strategy theft or even risk-free forms of arbitrage such as sandwich attacks. This thus motivates keeping (profitable) CoinAlgs private and their ``secret sauce'' hidden from potential competitors. Indeed, in practice, most CoinAlgs (in both traditional and decentralized finance) opt for private trading strategies~\cite{fabrega2026coinalg}. However, this alternative opens up the risk of unfair \emph{value extraction by insiders} -- private trading strategies permit an information asymmetry in which insiders with full knowledge of a CoinAlg's strategy (e.g., its creator or host) can use their privileged information to extract value from the CoinAlg's trades. Formally defined as \emph{fairness} in~\cite{fabrega2026coinalg}, these risks of private CoinAlgs are akin to insider trading in traditional finance. 

Fábrega et al.~\cite{fabrega2026coinalg} formally prove the existence of the CoinAlg Bind via two complementary theoretical models: one which compares the value extracted by players with asymmetric knowledge of a CoinAlg's trading strategy, and another which formalizes the interaction between a CoinAlg and players with advance knowledge of its trades. Furthermore, they validate the Bind empirically, via extensive simulations on real-world blockchain data. An inherent and unavoidable design tradeoff, the CoinAlg Bind is one of the central challenges for the future deployment of CoinAlgs.

\begin{keyInsight}[The CoinAlg Bind]
AI-based collective investment algorithms (CoinAlgs) face a fundamental tension between profitability (which requires privacy of investment strategies), and fairness (which requires transparency of investment strategies).
\end{keyInsight}

\subsubsection{Towards mitigations} Despite their risks, CoinAlgs are an inevitable part of the investing landscape, and thus the design of guardrails that limit their hazards is a critical question for future research.

In traditional finance, the CoinAlg Bind is sidestepped through regulated investor protections, which allow CoinAlgs to be public (and thus profitable) while disincentivizing insider value extraction (and thus enforcing fairness). However, less regulated environments like Web3 need to rely instead on \emph{technical} countermeasures for the CoinAlg Bind. In particular, due to the risks of profit loss, it is likely that Web3 CoinAlgs will continue to be private, and thus guardrails that minimize the risks of unfair value extraction are of particular importance. The challenge, of course, is that such guardrails should not themselves come at the expense of the CoinAlg's profits.

\begin{questionBox} [Technical guardrails for the CoinAlg Bind]
    What technical mechanisms can minimize the risks of unfair value extraction, without decreasing the profits of a CoinAlg?
\end{questionBox} 

Fábrega et al.~\cite{fabrega2026coinalg} explore some initial directions for the design of guardrails. Most notably, they propose the use of what they term a \emph{randomizing wrapper}, which is a (transparent) algorithm that randomizes a CoinAlg's trades prior to their execution. By running the CoinAlg inside a private and trusted execution environment (i.e., a TEE), such randomization can make it more challenging for insiders to predict the CoinAlg's trades ahead of time, decreasing their ability to profit from their privileged information. Furthermore, since wrappers are public, users can ensure that trades will be honestly randomized. An important direction for future work is further and more principled study of randomizing wrappers, both in terms of theoretical models for their security guarantees, as well as empirical studies that quantify their utility in practice.

\section{Future Risk: AI-powered Rogue Smart Contracts} 
\label{sec:AIforblockchain-future}

Endowing smart contracts with AI capabilities can vastly expand their application scope, as discussed in~\Cref{sec:ai-interact-real-world}. 
This expanded scope, unfortunately, includes not just good applications, but also malicious ones. 
This is because smart contracts were designed as a technical alternative to human and institutional trust and conflict resolution. 
In a system that uses smart contracts for conflict resolution rather than human processes, potential beneficiaries include those with the least trustworthy human and institutional relationships: criminal actors. 

The idea that smart contracts supplemented with AI could supplant ``honor among thieves'' was suggested in~\cite{juels2016ring}. A \textit{rogue} smart contract\footnote{The term ``criminal smart contract'' is used in~\cite{juels2016ring}.} as described there offers a bounty / reward for the perpetration of a crime. (Or it can do the reverse: Offer a criminal service for pay.) 

\begin{figure}[h!]
    \centering
    \includegraphics[width=1\linewidth]{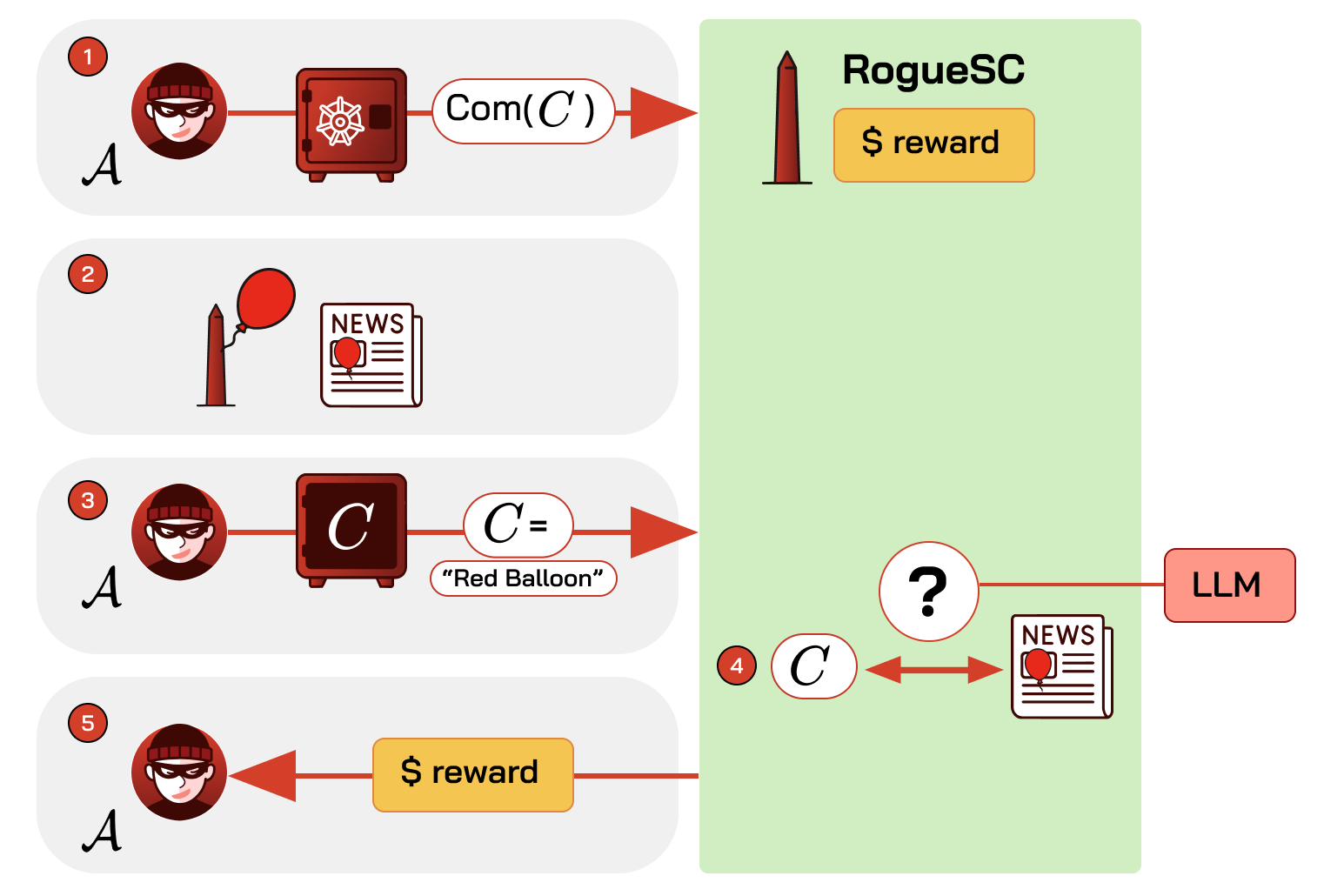}
    \caption{A criminal actor $\adv$ claims bounty $\textsf{\$reward}$ for vandalizing the Washington Monument by attaching a giant red balloon to it. $\adv$ uses the calling card $C =$ ``Red Balloon.'' Steps are detailed below.}
    \label{fig:RSC}
\end{figure}

\Cref{fig:RSC} illustrates how such a contract might work. Here, a rogue smart contract \texttt{RogueSC} has been created that offers a monetary reward (\textsf{\$reward}) to commit a criminal act: vandalizing the Washington Monument.

A criminal actor $\adv$ who wants to claim the reward must, of course, prove to \texttt{RogueSC} that he committed the requested criminal act. The protocol for doing so makes use of what is referred to as a \textit{calling card}, a distinctive detail about a crime that attributes it to a criminal. The idea is to have $\adv$ privately commit to a calling card $C$---a detail of the crime as he plans to commit it---\textit{before} actually committing the crime. He reveals $C$ it afterward and, if $C$ corresponds to reports of the crime, it proves that $\adv$ was responsible and should receive the reward. That's because, if $C$ is chosen well, only $\adv$ could have known it in advance.

Where does AI come into play? The smart contract \texttt{RogueSC} must ascertain that the calling card in $C$ actually happened in the physical world in a process called \textit{adjudication}; in practice, this might come from news reports or other trusted third-party sources that report notable details about the crime. 
Such adjudication is not straightforward for smart contracts, however, as the data in question is not necessarily standardized or quantitative in nature. 
For example, if the crime is ``vandalize the Washington Monument'' and the calling card $C$ includes ``red balloon,'' it is not trivial for a smart contract to evaluate whether the crime occurred, and if so, whether the calling card appeared at the scene of the crime.
One of the most promising ways to automate this adjudication is with an ML model such as an LLM called by \texttt{RogueSC} (in practice, using an oracle).

Step by step, \texttt{RogueSC} works as follows. After someone creates the smart contract---in this case, offering bounty $\$\textsf{reward}$ for vandalizing the Washington Monument---a criminal actor $\adv$ can claim the bounty using the following protocol:

\begin{enumerate}
  \setlength{\itemsep}{2pt}
  \setlength{\parskip}{2pt}
    \item \textbf{Calling card commitment:} $\adv$ chooses a calling card $C$ and sends a cryptographic commitment $\sf{Com}(C)$ to \texttt{RogueSC}. (A commitment conceals $C$ but makes it immutable.)
    \begin{itemize}
        \item Running example: $C =$ ``Red Balloon.''
    \end{itemize}
    \item \textbf{Execution of crime:} The crime happens and is reported in the news. 
      \begin{itemize}
        \item Running example: A giant red balloon is flown from the Washington Monument. 
    \end{itemize}
    \item \textbf{Decommitment:} $\adv$ decommits $C$, i.e., reveals $C$ to \texttt{RogueSC}.
    \item \textbf{Validation by ML model:} \texttt{RogueSC} checks that $C$ corresponds to news reports.
          \begin{itemize}
        \item Running example: \texttt{RogueSC} asks an LLM whether ``Red Balloon'' features in recent news reporting on vandalism of the Washington Monument.
        \end{itemize}
    \item \textbf{Bounty payment:} Bounty \textsf{\$reward} is paid to $\adv$.
          \begin{itemize}
        \item Running example: The LLM says \textbf{yes}, validating $\adv$'s claim.
           \end{itemize}
\end{enumerate}

This same structure can be applied to any of a number of crimes, and the information used to check claims can come not just from public sources but also \textit{private-web} data sources by means of privacy-preserving oracles. In this case, the construction closely resembles the secure inference pipelines described in~\Cref{subsec:security_inference}. Given its use of source authentication, it need not even require a calling card, as in the following examples:

\begin{itemize}
    \item \textbf{Targeted harassment:} A criminal can prove a campaign of harassing e-mail exchanges (and filter out any self-identifying information, if exchanges are not anonymous). An ML model can evaluate the effectiveness of the campaign.  
    \item \textbf{Theft of organizational intelligence:} The employee of a company or other organization can exfiltrate data from a corporate intranet and prove its origins anonymously. For example, intelligence facilitating insider trading could be used to claim a bounty (or be sold via smart contracts). The same approach could be used for intellectual property, product plans, etc. An ML model can appraise and assign a monetary value to the stolen intelligence.
    \item \textbf{Whistleblower Doxxing:} Someone with access to a whistleblower complaint can prove the whistleblower's identity with stolen documents, an ML model validating the strength of the supporting evidence.
\end{itemize}

If $\texttt{RogueSC}$ can be deployed anonymously and funds are hard to trace, both the entity soliciting the crime and the entity performing it can act anonymously and with impunity. Privacy-preserving payments, whether through use of mixers~\cite{bonneau2014mixcoin,ruffing2014coinshuffle,tornado2019whitepaper,pakki2021everything} or more tightly integrated privacy technologies~\cite{sasson2014zerocash,moser2017empirical}, carry risks when it comes to rogue contracts just as they do in many other settings.

\paragraph{Countermeasures.} The countermeasures regularly used to combat crypto-related crime---on-chain analytics to deanonymize transactions, blacklisting of tainted funds---can serve as effective countermeasures to rogue contracts, with the provisos regarding privacy given above.

There is an additional important countermeasure specific to rogue smart contracts, however. That is for oracles that deploy ML models to \textit{surface AI safety measures}, meaning that ML models deny service in cases where there is an apparently high risk of abuse. For this purpose, the \textit{context} of a request needs to be specified, as risk assessment is otherwise challenging. For example, it's not obvious our of context that assessing a news article for mention of a ``red balloon'' is an expression of nefarious intention. Providing the target logic (smart contract code) for the oracle request, however, may reveal the threat.

Of course, as with all AI safety measures, there is a risk of both false positives (inappropriate denial of service) and false negatives (failure to detect abuse).

Also specific to smart-contract environments is the possibility of some entity standing up a \textit{rogue oracle} service. Such a service, if it achieves the aspirational crypto properties of trustworthiness and censorship resistance, would enable bad actors to sidestep oracles systems that implement safety mechanisms.

\section{Conclusion and Future Directions}
\label{sec:concludion-ai4blockchains}
Overall, there has been substantial research on using AI  to aid in the constructive design of algorithms for blockchains. 
It can be used both to \emph{design} applications themselves, as in the design of DeFi market structures. It can also be used to design algorithms that \emph{relate to} a given smart contract, such as strategies for attacking a smart contract or maximizing extractable value in a market. 
Approaches in the research literature today use a variety of ML tools, ranging from very basic classifiers and regression models to  more advanced methods using RL to design algorithms that optimize a given reward function. 
In the last 3-4 years, we have seen a clear uptick in  papers using RL to design algorithms for blockchain purposes. 
A unifying observation is that all of these methods have focused on settings where we can model the environment, explicitly or implicitly. 

\begin{keyInsight}
In the research community, AI-assisted algorithmic design for blockchains has, until now, focused on settings where we can model the world or environment cleanly. 
Examples include modeling consensus protocol state spaces, or modeling the impact of prices on a given AMM. 
\label{takeaway:ai-algo-design}
\end{keyInsight}

A consequence of operating within a given model is that AI-assisted security analysis has so far been predominantly \emph{constructive}:
Across the stack, from consensus protocols to smart contracts, it searches for a concrete attack.
Such analysis can show that a system is insecure, but cannot certify that it is secure, which remains the domain of exhaustive techniques such as formal verification.
This suggests a complementary direction.

\begin{researchQuestion}
Can AI support verified, exhaustive security analysis across the blockchain stack, rather than only the constructive discovery of attacks?
In particular, beyond operating within a given model, can AI help to find the models or abstractions in which a protocol or contract can be proven secure?
\end{researchQuestion}

Despite this clear trend of AI-assisted algorithmic design in the research literature, a very different trend is emerging in practice and in industry, as highlighted in Section \ref{sec:AIforblockchain-future}. 
As AI-assisted coding becomes commonplace, we expect it will become the norm for smart contracts to be written largely by AI, possibly with the assistance of agentic frameworks \cite{elizaos,aixbt2024}. 
Indeed, very recently, an agentic pipeline was able to design exploits that would have extracted over \$4.6 million  in a benchmark of smart contracts  \cite{anthropic-agent-exploits}. 
While the agents used to design these attacks are closed-source, they relied on general-purpose ML models that were not specifically tailored to exploit extraction, beyond prompt-tuning. 
A new class of workflows that rely increasingly on general-purpose foundation models have several implications:
\begin{itemize}
    \item We will increasingly design not only blockchain algorithms, but also full-blown implementations with AI. It remains to be seen what degree of separation will exist between algorithmic design and implementation in practice. 
    \item Algorithmic design will increasingly be driven by natural language objectives rather than precisely-stated quantitative reward functions. For example, if a user asks AI to "write a program to make me money from the given smart contract", there is no well-defined reward function. The agent must decide what objective to follow and what boundaries to obey, if any, when trying to make the user money.
    \item Unlike prior approaches to AI-assisted algorithm design (as in the above Key Takeaway), emerging AI-assisted methods will increasingly design algorithms and code for environments that are not well-understood or modeled.    
\end{itemize}

These implications raise several questions for both the research and industrial blockchain community.

\begin{questionBox}[Next generation of AI-assisted design ]
    As we migrate from heavily tailored, labor-intensive ML-assisted design of algorithms and analytics to agentic program design based on natural language objectives, what will be the implications for efficacy and security of blockchain algorithms and applications? 
\end{questionBox}

\paragraph{FM-assisted algorithms vs. tailored AI-designed algorithms.} One concrete research question in this vein is evaluating how agent-designed algorithms compare to prior algorithms that were tailored to a particular problem domain. 

\begin{researchQuestion}
    On downstream tasks such as blockchain analytics or exploit design, how do existing classical AI-based algorithms that are tailored to a specific task and data type compare to algorithms that are designed by agents backed by general-purpose FMs (i.e., not explicitly tailored to downstream tasks, other than via  prompt design)? 
    \label{rq:compare-fm-tailored}
\end{researchQuestion}
For example, in the space of fraud detection, existing classifiers in the literature were all trained on relatively small, carefully-curated corpora of transactions.\footnote{Some work has explored a middle ground of training FMs on blockchain data \cite{gai2023blockchain}. Although the precise downstream task is not pre-specified in that work, it is still trained from scratch on blockchain-specific data.} Approaches that rely on frontier FMs instead learn from much larger---albeit unlabeled, and potentially irrelevant---corpora of public information. 
It would be useful to evaluate how these methods compare to each other on the same, held-out test set for various downstream tasks. 
While we expect that tailored algorithms will excel in some narrow settings, prior ML-for-blockchain papers typically make assumptions about the environment (e.g. data comes from a particular distribution, or environments remain static over the time of experimentation). 
Hence, it will be important to understand how prior methods compare to general-purpose agentic methods under distribution shift due to changing environments. 
Similar experiments could be run for prior RL-based methods from the literature, such as algorithms for MEV extraction or exploit generation. 

\paragraph{Making better use of general-purpose FMs.}
In the previous research question, we asked for a  direct comparison between tailored AI algorithms and agent-driven algorithms backed by general-purpose FMs. However, there is a middle ground: FMs can be fine-tuned to blockchain-specific tasks and data types. A natural, broader question is how best to make use of general-purpose FMs for blockchain-specific tasks. 

\begin{researchQuestion}
    For blockchain analytics and design, how should designers best make use of pretrained FMs and small downstream datasets of labelled data? 
\end{researchQuestion}
Today, the dominant paradigm in the ML community for adapting FMs to downstream tasks is RL-fine-tuning, in which a pretrained LLM is fine-tuned on a downstream task or dataset that is representative of the desired skills \cite{rlhf,ppo,grpo,dpo}. 
These methods typically produce several responses to a single prompt, then use various methods to \emph{rank} or \emph{compare} the quality of the outputs. 
The resulting rankings are then used to fine-tune the base FM to steer it towards better outputs. 
However, existing RL-finetuning algorithms are all designed for general-purpose adaptation to downstream tasks. 
Hence, it would be important to understand if there are blockchain-specific attributes that could be exploited for RL fine-tuning, or even continued pretraining \cite{ke2026value}.
For example, 
in the blockchain setting, it is unclear how to evaluate the quality of a FM's response to a given prompt, or even how to design the prompts themselves. 
For an analytics problem, should the prompts ask to predict whether a transaction is fraudulent? What context should be provided? 
Such questions may be heavily task-specific, and we believe there could be a rich body of work exploring different methods of adapting general-purpose FMs to questions of interest in the blockchain space.

\paragraph{AI-driven security wars.}
Web3 cybersecurity stands out as an important bellwether for enterprise cybersecurity at large. This is because Web3 exploits are typically immediately monetizable, either by exploiting them in the wild---thereby irreversibly exfiltrating funds---and by availing of multi-million dollar bug bounties for responsible vulnerability disclosure. This creates significant economic incentives for all actors to leverage advancing model capabilities as soon as they are available and prove economically viable.

Nonetheless, it is challenging to predict how security postures will evolve, and early signals are inconclusive. 
In the academic literature, recent AI cybersecurity benchmarks including BountyBench~\cite{zhang2025bountybench} and EVMBench~\cite{wang2026evmbench} evaluate model capabilities across the vulnerability lifecycle, structuring tasks around detect, patch, and exploit workflows. Empirical results suggest that these capabilities may evolve unevenly. BountyBench reports higher patch than exploit success rates \cite{zhang2025bountybench}, and EVMBench reports the reverse \cite{wang2026evmbench}. Notably, EVMBench observes continued improvements across all tasks with later model generations.
Overall, the trajectory of advancing AI capabilities remains unclear, leaving open important questions about how AI capabilities will shape the balance of power between attackers and defenders, particularly in the short term. 
This motivates an important future research agenda. 
\begin{researchQuestion}
How do different capability trajectories shape Web3 security outcomes? Which trajectories are most plausible, and what leading indicators would signal the direction of advancement? Finally, what interventions are available to preserve a strong security posture as capabilities improve?
\end{researchQuestion}

These questions relate fundamentally to the economics of security, and may require the development of economic models to predict various outcomes. Broadly, it would be useful to study the AI arms race from the following perspective: if AI gives an asymmetric advantage to one party (attackers or defenders) for some amount of time, how does this impact the global state of enterprise security? 
How should we change the structure of bug bounty programs to adapt? 
How will cybersecurity insurance risk calculations change? 
Broadly, we expect there are many interesting and important economic questions to explore. Relevant research questions include modeling these issues, as well as understanding how to measure the current state of affairs; as we mentioned earlier, results from individual benchmarks can be inconsistent, so it can be challenging to understand which side is ``ahead" in this arms race. 
However, blockchains offer an interesting opportunity in that smart contracts and transactions are public on permissionless cryptocurrencies. 
As such, there may be an opportunity to directly measure the (correlative) impact of various AI advancements on the rate and magnitude of  cybersecurity attacks. 

\bigskip
Overall, the research community has already demonstrated that AI-assisted approaches can be immensely useful for the crypto ecosystem, both in terms of understanding existing systems and for designing and interfacing with new ones. 
That being said, we anticipate significant growth and improvements ahead due to the emergence of novel techniques and tools, particularly those based on state-of-the-art generative modeling. The precise nature of these opportunities remains unclear, and they may surface both  benefits and substantial risks (particularly for cybersecurity). Either way, the blockchain industry and research communities will be forced to adapt.

\chapter{AI x Crypto: Enhancing AI with Crypto}
\label{chap:block4ai}
\section{Overview: Making AI Pipelines more Decentralized and Trustworthy}
\label{sec:block4ai-overview}
\begin{figure}
    \centering
    \includegraphics[width=1\linewidth]{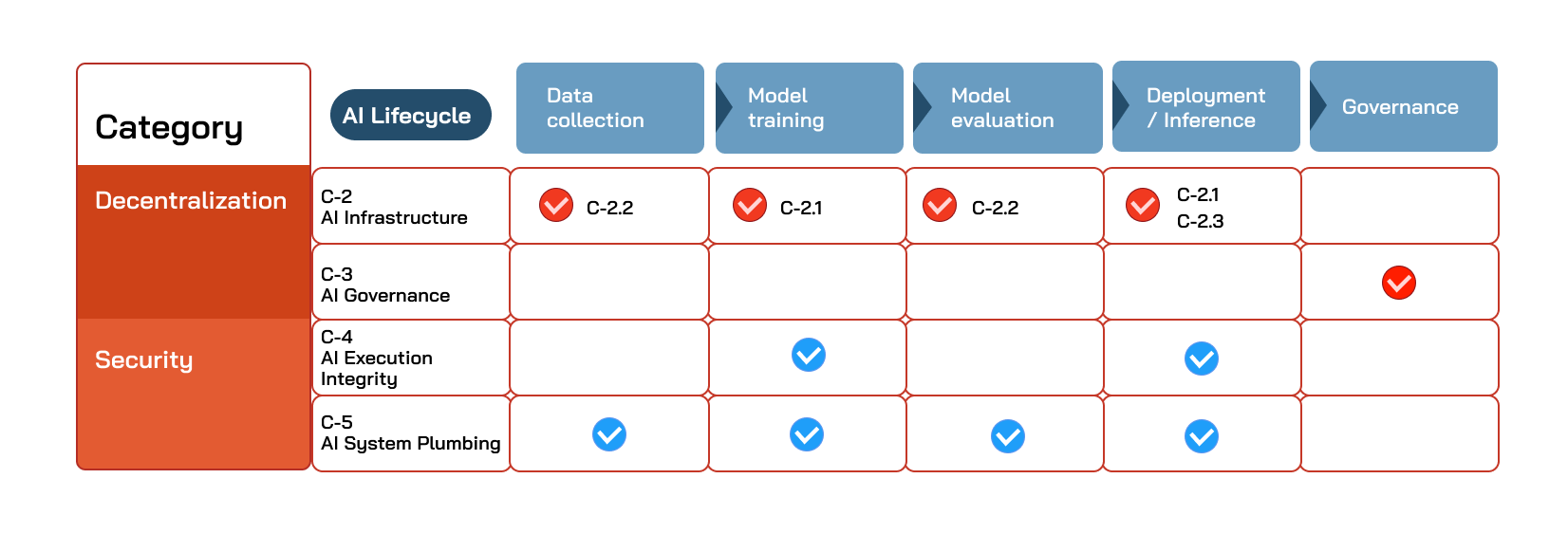}
    \caption{The AI life cycle. We envision that blockchains and related technologies can impact the AI life cycle in two main ways: by decentralizing components, and by securing them to a greater degree. }
    \label{fig:AI-lifecycle}
\end{figure}

AI models are realized through a sequence of events that is commonly referred to as the \textit{AI life cycle}. Many variants have been proposed of this life cycle, but many variants explicitly include data collection, model training, model validation, deployment, maintenance, and governance (Figure \ref{fig:AI-lifecycle}). 

We maintain that crypto---including related underlying technologies---offers two key channels for altering or improving the AI life cycle. First, it can help to \emph{decentralize} various components of the AI pipeline, from data collection to model training and inference to governance. In principle, this could help to democratize the development and maintenance of AI models. In this survey, we will also consider the more pragmatic effects of decentralization such as cost and coordination. 
A second opportunity is that crypto can help to \emph{secure} components of AI pipelines, making them less susceptible to malicious providers of data, compute, or storage. Here, we will discuss how technologies that came out of the crypto sphere are remarkably well-suited to solving certain security vulnerabilities that arise throughout the AI life cycle. 
This chapter will explore these two complementary opportunities. Specifically, we divide this chapter into two groups of sections: those exploring decentralization and those exploring new security capabilities. We summarize the sections of this chapter, and where they lie in the AI life cycle, in Figure \ref{fig:AI-lifecycle}.

\begin{enumerate}
\item \textbf{Decentralization: } These sections will explore how crypto can help to decentralize various components of the AI life cycle, and more importantly, why it matters. Specifically, we cover the following types of decentralization: 
 	\begin{enumerate}
		\item \emph{Decentralized infrastructure for AI: (Section \ref{sec:decentralized-infra-ai})} This section explores several kinds of AI infrastructure that can be decentralized using tools and technologies from the crypto sphere. These include physical infrastructure like computation (Section \ref{sec:depin}),  virtual infrastructure like data pipelines (Section \ref{sec:data-markets}), and application-level infrastructure for AI agents (Section \ref{sec:decentralized-agent-infra}). 
\item \emph{Decentralized Governance: (\Cref{sec:decentralized-governance})} This section explores efforts to decentralize the governance of AI systems. We explore techniques that have been proposed, along with some of the tradeoffs and central challenges. 
	\end{enumerate}
\item \textbf{Security: } These sections will instead explore how crypto can help to secure components of the AI life cycle. Specifically, we explore the following areas. 
\begin{enumerate}
		\item \emph{Blockchains for AI execution integrity: (Section \ref{sec:block4ai-integrity})} This section explores how to use cryptographic tools to ensure that AI execution---both in inference and in training---is trustworthy and verifiable for downstream use cases. 
\item \emph{Securing the plumbing: (Section \ref{sec:secure_plumbing})} Finally, we explore a framework for ensuring data trustworthiness, security, and privacy across the whole AI life cycle via careful use of trusted computing technologies. 
	\end{enumerate}
\end{enumerate}

More generally, a key message of our work on using crypto to secure AI is that securing AI will require \emph{system-level} knowledge and control of the environment in which it runs \cite{bauer2026creating}. 
\begin{keyInsight}
    AI security is a system-level property. It depends on just on models, but also on the systems and environments in which models run. The crypto ecosystem has produced multiple tools that can reduce uncertainty regarding operating environments. 
\end{keyInsight}
Although crypto tools are not necessarily the right tool for handling model-level misbehavior (e.g., hallucination), they may be particularly well-suited for helping to ensure the consistency and trustworthiness of the environments in which AI pipelines run, as we explore in  \Cref{sec:block4ai-integrity,sec:blockchainforAI}.

\section{Decentralized Infrastructure for AI}
\label{sec:decentralized-infra-ai}

Today, the AI industry is growing increasingly centralized \cite{widder2025big}. 
Model providers are starting to internalize data acquisition and processing \cite{meta-scale}, develop their own algorithms \cite{microsoft-openai}, train and run inference on their own data centers \cite{ai-data-centers}, and develop centrally-hosted applications in the form of agents with proprietary prompts \cite{agent-exchange}. 
Blockchains present an appealing opportunity to decentralize various components of this life cycle. 
For example, what if communities could collectively train AI models on a distributed network of consumer-grade hardware? What if data could be sourced organically from various parties in a decentralized marketplace? 
What if AI-based applications could interact and evolve without requiring centralized oversight?
The hope is that such decentralization could reduce costs and increase transparency for the AI community as a whole. 
However, in practice, these considerations are nuanced and can depend on many factors and assumptions. 
In this section, we discuss the current state-of-the-art and implications of decentralizing AI infrastructure. We discuss decentralized physical infrastructure in \Cref{sec:depin}, decentralized markets for data and models in \Cref{sec:data-markets}, and decentralized agent-centric infrastructure in \Cref{sec:decentralized-agent-infra}.

\subsection{Decentralized Physical Infrastructure Networks} 
\label{sec:depin}

Decentralized physical infrastructure networks (DePIN) are growing in popularity in the AI space.  
In short, DePIN refers to decentralized networks in which nodes can provide physical infrastructure such as energy, compute, or bandwidth in exchange for financial incentives \cite{grass,bittensor,akash,theta,hivemapper,xiao2022monetizing}. 
While there have been many such efforts in the past (e.g., Folding@Home~\cite{folding-at-home}), DePIN is  characterized by financial compensation using Web3 infrastructure. 
Specifically targeting the AI market, some relevant DePIN efforts include distributed networks of compute nodes (e.g., Theta Network~\cite{theta}, Akash Network~\cite{akash}, and io.net~\cite{io}, to name a few). 
These efforts allow AI practitioners to rent CPU and GPU resources on-demand from a global network of providers. 
In addition to the raw markets themselves, some efforts focus on infrastructure for DePIN markets; for example, Bittensor \cite{bittensor} is a language for defining new DePIN commodity markets under a unified token. 

AI-oriented DePIN networks  generally compete with traditional cloud service providers on price; for example, in late 2025, the Theta network advertised, ``On-demand enterprise grade Nvidia A100/H100 and more at 50-70\% cost savings" \cite{theta}, and the Akash network reported 70-85\% average cost savings relative to AWS \cite{javed2026akash}.
The principal downside of DePIN compute networks is the reduced throughput and latency between machines:
as DePIN infrastructure is inherently decentralized, communication between machines will generally travel over the public Internet. 

For AI practitioners considering the use of DePIN compute networks, a natural question is, ``is DePIN infrastructure beneficial for my job?" 
Of course, the answer  depends on the use case and the characteristics of each job.
For example, even if compute infrastructure is cheaper per node on a DePIN network than a traditional cloud service provider, the end-to-end wall-clock time for a job could still be much longer on a DePIN network if the job requires heavy communication across nodes.
On the other hand, some use cases may require geographically distributed training for non-technical reasons, such as: (1) privacy regulations restricting the flow of data out of a specific country \cite{gdpr}, (2) mitigating the environmental costs (e.g., water, energy) of training models in a particular datacenter region \cite{li2025making,thirsty2}, and/or (3) democratizing the training of foundation models \cite{decentralized-training}.
In such cases, DePIN networks could facilitate distributed (federated) training methods. 

In the following, we provide some considerations for two main categories of AI workloads: training and inference. For each, we discuss the latency and throughput requirements of the workloads, and explain how these profiles may interact with DePIN architectures. 
For both training and inference, costs scale with model size; there is a substantial difference between foundation models, which can have trillions of parameters \cite{achiam2023gpt,decoder,wang2024rail}, and small language models, which can be substantially smaller \cite{tinyllama,sinha2025small} or even classical ML models (e.g., support vector machines, linear regressions, etc.). The regime of interest will strongly affect the feasibility of training on DePIN compute networks. 
We discuss these properties mainly from an efficiency and cost standpoint; we do not further discuss non-technical reasons that might favor distributed training, such as privacy regulations.

\subsubsection{Use Case: AI model training}
Model training can be roughly split into three phases: 
\begin{itemize}
    \item \textbf{Pretraining}: the initial training of a model from randomized weights
    \item \textbf{Post-training}: A process of updating or aligning foundation models for additional capabilities (e.g., instruction following, reasoning, or math), often using various reward models possibly obtained by soliciting human preferences. 
    \item \textbf{Fine-tuning}: updating the weights of a pre-trained or post-trained model for a downstream task, possibly on a specialized or private dataset

\end{itemize}
 In general, pretraining tends to be more costly than fine-tuning and post-training as it is conducted on larger amounts of training data, whereas post-training and finetuning is conducted on smaller datasets---and may also leverage techniques such as parameter-efficient fine-tuning (PEFT) \cite{han2024parameter,minaee2024large}.
 Hence, we will focus on pretraining for this section.

AI model training (especially for large models) requires sophisticated data management. 
Backpropagation, a building block for most ML optimizers, must compute and store massive gradients averaged over large batches of data. These operations are infeasible on a single GPU node due to hardware limitations,
so training is generally distributed across multiple compute nodes \cite{duan2024efficient,wang2024rail}. 
Furthermore, most training algorithms are based on minibatch stochastic gradient descent and its variants, which require synchronous updates across all nodes in the network.
As a result, training may stall in the event that any individual node fails or collective communications are waiting on updates from a straggler node.

\paragraph{Throughput considerations.} 
Training a model over multiple GPUs  requires a  strategy for parallelizing computation. Today, prominent models of GPU parallelism include \cite{duan2024efficient,fernandez2024hardware}:
\begin{itemize}
    \item \textbf{Data parallelism}  is the simplest and most common form of parallelism; it involves splitting data across different nodes with each node computing the gradients for a subset of the examples in a minibatch. Gradient computations are then averaged across nodes.
    \item \textbf{Pipeline parallelism}  involves splitting layers sequentially, such that different nodes process different layers. It induces a tight sequential dependency over the processing of  layers. 
    \item \textbf{Tensor parallelism,} also known as horizontal parallelism,  splits operations within a single layer across nodes. Each GPU works on a shard of the tensor, and results are synchronized at the end of the operation. 
    \item \textbf{Expert parallelism}  is used for mixture-of-expert (MoE) models that have multiple expert sub-models. Different experts are trained on different (groups of) nodes. 
    \item \textbf{Context Parallelism:} is used for long-context models operating on sequences with a large number of tokens. Each GPU works on a shard of the computation, partitioned along tokens in the sequence. 
\end{itemize}
These parallelization strategies result in large quantities of data transfer for the communication and aggregation of gradients and optimizer states, with communication costs increasing with the size of the model, training dataset, and hardware platform \cite{rail2rail}.  However, the throughput profiles of parallelization strategies may differ. For example, pipeline and tensor parallelism tend to have heavy communication footprints, requiring high-bandwidth links. 
Without careful management and synchronization, all of these parallelism techniques can experience significant scaling challenges \cite{narayanan2021efficient,mayer-survey}.

To support synchronization and data transfer during training, 
most multi-GPU training platforms feature high-bandwidth interconnects supporting several Tbps between local GPUs (e.g. NVLink) \cite{wang2024rail,h100,amd-mi300x}; we call such groups of GPUs in the same rack \emph{scale-up networks} (colloquially referred to as {high-bandwidth domains}) \cite{scaleup}.
Beyond these local connections, platforms rely on network technologies like Ethernet or Infiniband to connect the network interface cards (NICs) of different platforms.
\emph{Scale-out networks}, for instance, connect GPUs across racks, and \emph{scale-across networks} connect datacenters \cite{scaleup}. 
The interconnects in scale-out and scale-across networks tend to be lower bandwidth, and a bottleneck in LLM training \cite{rail2rail,fernandez2024hardware}.
Notably, Fernandez \emph{et al.}~found that when training LLMs on distributed GPU nodes, as one scales up the number and computational capacity of compute nodes, the system becomes increasingly communication-bound, with GPU utilization falling fastest for high-end nodes like NVIDIA H100s \cite{fernandez2024hardware}. 
This is an important consideration for training AI models on DePIN compute networks. 

\bigskip
\noindent \emph{Technical solutions.} \quad
There have been a number of efforts to train foundation models on distributed infrastructure, including  with commercial-grade network connections. For example, in 2022, Yuan \emph{et al.}~demonstrated a system that optimizes communication costs while taking into account the average latency and bandwidth between compute nodes \cite{decentralized-training}. Such metrics are not directly provided by most DePIN networks, but could be measured prior to (or during) training. 
SWARM parallelism \cite{ryabinin2023swarm} was later proposed as a pipeline-parallel method for training ML models on unreliable hardware with slow interconnects. 
Their key idea is to stochastically update which nodes train a given layer based on current network and device conditions.  
\cite{ryabinin2023swarm} found that as model size scales up, computation occupies a larger proportion of training time relative to communication costs.  

To address the requirement that training steps be addressed synchronously, federated averaging approaches (e.g. Local SGD, DiLoCo) perform a fixed number of local gradient updates of model weights on each node and periodically average the resulting models across all model instances \cite{stich2019local, douillard2024diloco}.
More recent work proposed Decoupled Momentum Optimization (DeMo), an optimizer that modifies momentum updates to reduce communication among accelerators without significantly degrading  downstream model performance \cite{demo}. 
Despite the promise of these methods, these research papers evaluated their ideas on (relatively) small models of at most 3B parameters \cite{ryabinin2023swarm,decentralized-training,douillard2024diloco,demo};  it is unclear from these papers alone whether the methods  work well  at the larger scales of frontier models. 
Several industry efforts are testing these ideas at larger scales (see \emph{Industry efforts} below).

More broadly, there exist techniques for optimizing communication across high-bandwidth domains at the hardware interconnect level. 
For example, rail-optimized networks \cite{rail2rail} specify how to connect NICs to switches and are commonly used for high-performance computing workloads.
However, they still require all-to-all communication between high-bandwidth domains, which may be impractical for DePIN networks. 
Rail-only networks are instead optimized to LLM training patterns, and get rid of the spine layer from typical GPU clusters \cite{wang2024rail}. 
Both optimizations \cite{rail2rail,wang2024rail} require control over physical infrastructure and may be more practical for data centers (scale-out networks) than fully-decentralized DePIN networks.

There also exist algorithmic techniques for reducing communication between high-bandwidth domains.
For example, the NVIDIA NeMo framework utilizes a hierarchical AllReduce framework, distributed optimizer architectures, and chunked inter-data center communications  to minimize the size and number of gradients that are transmitted over the wide-area network (WAN) that connects the two datacenters on which they trained their prototype Nemotron-4 model (340B parameters) \cite{nvidia-distributed-training}.
These methods can reduce communication costs, and may be useful for training on DePIN infrastructure, but we lack empirical data to make this case conclusively. The test setup for Nemotron-4 is very different from a model being trained on a typical DePIN cluster, where many nodes may be geographically distributed across far more than two regions, in many parts of the world. 

\bigskip
\noindent \emph{Industry efforts.} \quad
Recently, providers like NVIDIA have started supporting distributed training of ML models across data centers \cite{nvidia-distributed-training}, which has been validated by training Nemotron-4, an LLM with 340B parameters. 
Such efforts to train large AI models over wide-area-networks (WANs) raise similar questions to those  encountered with DePIN compute networks. NVIDIA  addressed these issues in part by minimizing communication over scale-across networks \cite{nvidia-distributed-training}, which is common more broadly in distributed training case studies. 

The crypto x AI industry has also worked to demonstrate the feasibility  of training foundation models (at a smaller scale) on fully decentralized infrastructure, including commercial-grade network connections between accelerators. By August 2024, Macrocosmos AI had facilitated the training of 700M and 7B parameter LLMs on the Bittensor network \cite{macrocosmos}. 
The leading model from this work was shown to be competitive with GPT2-Large and Phi-2 in terms of perplexity on web-text and related data.\footnote{It is not standard in LLM evaluations to report only perplexity, as done in this paper. Rather, common practice is to evaluate LLMs on a suite of benchmarks, such as MMLU-Pro \cite{wang2024mmlu} and GSM8K \cite{cobbe2021training}.}
Their white paper focuses on the incentive mechanism for miners to submit high-quality model weights, and miners are expected to individually produce trained models, so the communication network effects we discussed earlier do not come into play (at least not measurably). 
Other efforts include Prime Intellect asynchronously training a 10B parameter model (Intellect-1) in November 2024 over distributed infrastructure using fully sharded data parallelism and DiLoCo-based optimization \cite{intellect}; Templar AI synchronously trained a 1.8B parameter model using an incentive system called Gauntlet \cite{lidin2025incentivizing}.

Ongoing work is pushing towards even larger scales.  
Macrocosmos AI presented an architecture by which multiple miners can collaboratively train a model using pipeline parallelism and DiLoCo, while reaping rewards for individual contributions \cite{quinque2025incentivised}.
At the time of writing there were no public results on the end-to-end training costs or model quality for a fully-trained LLM using this architecture. 
Consilience, a 40B parameter model was pre-trained on the Psyche Network \cite{psyche}. The Psyche Network uses DeMo for communication-efficient optimization in a data parallel configuration \cite{demo}; Nous Research reports that 40B is ``compact enough to train on a single H/DGX and run on a 3090 GPU" \cite{psyche}.  
However, it remains unclear how geographically distributed the infrastructure can be for such large-scale training runs. The DeMo paper explicitly states that ``DeMo is designed primarily for optimization across a small number of geographically distributed compute centers" \cite{demo}.
To our knowledge, the largest model being pretrained on DePIN infrastructure is Convenant-72B which uses the Gauntlet incentive system alongside SparseLoCo, a modified version of DiLoCo which both quantizes the gradients to reduce communication volume and also reduces the frequency of communication. While the corresponding report describes the improved communication efficiency relative to existing globally distributed training runs (e.g. INTELLECT-1), how the communication and compute efficiency of the approach compares to single datacenter centralized training remains unspecified.

A broader point about existing industry proofs-of-concept is that they do not typically report total cost metrics (in terms of computation and communication cost). 
If the decentralized ML community aims to democratize the training of large-scale foundation models, it would be useful to standardize cost models and compare end-to-end training costs (in dollars) on centralized vs.~decentralized infrastructure. 
If the results of such measurements are favorable to decentralized networks, this could be a major selling point for the industry.  
At the same time, we expect such results to depend heavily on the size of the model, the type of training parallelism implemented, and the degree of decentralization among training accelerators (e.g. multiple small clusters vs. truly heterogeneous consumer devices). 

\begin{keyInsight}[DePIN for AI model training] 
There are many promising preliminary results about training large ML models on decentralized infrastructure, particularly for smaller models (a few billion parameters). However, we lack clear  measurements on geographically distributed nodes (i.e., not in the datacenter setting) highlighting the overall cost tradeoffs as one scales up model size, degree and type of infrastructure parallelism, and geographic decentralization.
\end{keyInsight}

\paragraph{Latency considerations.} Model training tends to be less latency-sensitive than inference, in part because training is run offline. Nonetheless, the latency associated with distributed training across multiple geographic locations can cause problems, especially using standard gradient based optimizers (e.g. Adam, SGD) which require synchronous communication across all nodes.
As described by Aubrey \emph{et al.}, ``When managing compute across multiple data centers, developers must contend with high inter-region latency (often 20 milliseconds or more) that can introduce performance bottlenecks during gradient updates and model synchronization during large-scale LLM training" \cite{nvidia-distributed-training}. 
For reference, as of November 27, 2025, the P90 latency between AWS datacenters us-east-1 and us-west-1 was about 73 ms \cite{cloudping}.
The NVIDIA NeMo framework addresses this issue by minimizing high-latency communication operations across datacenters with techniques like hierarchical AllReduce \cite{nvidia-distributed-training}. 
Additionally, distributed optimizers, like DiLoCo, can further mitigate the frequency of such communication operations potentially improving the feasibility of training across a DePIN. 
While DePIN compute networks can in principle support sophisticated orchestration, many users may not be aware of the need for it.
Today, DePIN services partially account for latency by allowing users to select nodes by  geographic location \cite{akash,theta}. This can help users gauge the latency implications of a particular combination of selected nodes, as latency is lower bounded by the speed of light between two node locations.

\subsubsection{Use Case: AI model inference}
Model inference refers to generating predictions with an already-trained model, such as when a model is evaluated or deployed for a downstream task. For example, this could include evaluating a classifier model on a provided data point or generating a text from a generative model. 

\paragraph{Latency considerations.} In general, inference tends to be more latency-sensitive than training \cite{alizadeh2024llm,li2024llm,jaiswal2025serving}. Particularly in on-demand use cases (e.g., AI chatbot), users expect a quick response. However, there exist applications (e.g., meeting summarization, document review, data generation) that rely on model inference and do \emph{not} have strict latency requirements \cite{jaiswal2025serving}. 
Notably, emerging applications without direct human interaction (e.g. deep research) have relaxed latency requirements that can allow for up to 30 minutes before response \cite{openai2025deepresearch}.
These lower latency use cases can be more flexible about the locations of node providers. Notice that these same considerations would hold with traditional infrastructure providers; for latency sensitive applications, AI practitioners should think about the location of the datacenter(s) processing queries, and their relative distance to the end users.

\paragraph{Throughput considerations.} Inference typically has lower throughput requirements than training. 
First, inference generally requires fewer degrees of parallelism over hardware.
Whereas training a large foundation model (e.g., 100B parameters or more) can require tens of thousands of GPUs in parallel, inference jobs do not require parallel computation to the same extent. Even tens of GPUs are sufficient to serve the largest models (e.g. DeepSeek-R1 with 671B parameters can be served on 16 H100 GPUs).
Second, inference does not require backpropagation, which significantly reduces the memory and computation footprint of the operation. Hence, even under parallelization schemes like those discussed in model training, a network with low data throughput need not be prohibitive. 
Such tasks may be a particularly good fit for DePIN networks.

\begin{keyInsight}[DePIN for inference] 
Latency-insensitive inference applications (e.g., meeting summarization, document review) may be a particularly promising and cost-effective use case for DePIN networks.
\end{keyInsight}

\subsubsection{Summary}
Today, end users evaluate whether to use DePIN networks based on their own understanding of their AI jobs, as well as node price and capacity (e.g., compute, memory). 
We highlight the following major takeaway.

\begin{keyInsight}[Cost-benefit tradeoffs] 
Users of DePIN networks should evaluate the bandwidth and latency requirements of their use case to determine the overall expected cost savings. Moreover, we suggest that DePIN network providers measure and report at least network bandwidth between nodes, as these figures can drive total cost and help  determine whether a particular use case makes sense for a DePIN network. 
\end{keyInsight}
While current networks typically advertise lower costs in dollars-per-GPU-hour, this can be a misleading metric. The cost metrics that matter for ML jobs are typically training efficiency (i.e., iterations-per-unit-cost) and inference efficiency (tokens-per-unit-cost)~\cite{efficientllm}. 

To make it easier for users to determine whether and when to use DePIN infrastructure, we suggest the following questions for the research community. 

\begin{researchQuestion}
What kinds of AI jobs are well-suited to existing decentralized compute networks, as found in DePIN AI networks? 
\end{researchQuestion}
The community would benefit from more systematic (third-party) measurements to  profile existing DePIN networks, and also to understand the use cases for which DePIN AI networks makes sense. 
We envision such studies first measuring the performance profile of various DePIN networks. Then, we propose to map the characteristics of an AI job (estimated compute cost, maximum memory requirements, communication loads, size of model, size of layers, etc.) and desired performance characteristics (cost, end-to-end time) to a recommendation about whether  training on DePIN infrastructure would the target performance characteristics. 
While there have been prior studies on decentralized training of ML models \cite{ryabinin2023swarm,duan2024efficient}, these studies are not conducted on DePIN networks specifically, and may have very different performance characteristics. 

\begin{researchQuestion}
Building on the systematic measurement studies proposed in RQ C-5.1, how should pricing mechanisms for DePIN AI depend on network infrastructure and reliability?
\end{researchQuestion}
It is plausible that pricing should depend not only on compute capacity per node, but also on network measurements and node reliability. 
For example, a set of 16 nodes with high-bandwidth pairwise links (e.g., nodes with GPU-to-GPU NVLink, or co-located in the same datacenter) may cost more than 16 nodes that are spread across the world. 
Likewise, nodes which can guarantee high availability and reliability to participate in large coordinated jobs can be priced higher, as a single straggler or failed node is sufficient to interrupt an entire AI training job.
While some networks implicitly reward well-connected nodes (e.g., Theta network assigns rewards based on time to completion of a task \cite{theta-whitepaper}), others base pricing solely on node capacity. 
Both models may be too retroactive for large jobs, in which failures are (or can be) very costly.

\begin{researchQuestion}
How should updated pricing mechanisms for DePIN AI networks account for strategic and/or adversarial nodes? 
\end{researchQuestion}
If we propose updated pricing models as in RQ C-5.2, it is essential to understand the role of strategic and adversarial actors. For example, if a protocol were to reward well-connected nodes disproportionately relative to individual node capacity, a strategic owner of a single node could rent it out as many small co-located nodes to boost profits. This could be less useful for the network---a client renting the many small nodes would have worse performance than if they had rented the one large instance. 
Beyond such strategic manipulations, we also need to design methods to account for adversarial reporting of specifications. For example, if a node systematically lies about its processing capacity or network connections, there should be mechanisms for catching this and penalizing providers. 
The design of a proper reward scheme requires mechanism design to ensure that node providers cannot manipulate how they offer their infrastructure in a manner that is inconsistent with network utility, or otherwise lie about their offerings; the latter question of proving physical resource availability has been explored in the context of other DePIN resources, such as network bandwidth \cite{shengproof} and other cellular network resources \cite{anand2022trust}. Interestingly, defining utility and the threat model itself may depend on the measurement findings from Research Question C-5.1.

\subsection{Decentralized Marketplaces for Data, Models, and Evaluation}
\label{sec:data-markets}

Today, data is an essential input to many phases of the AI life cycle: training, grounding at inference time (e.g. in retrieval augmented generation (RAG) pipelines), and model validation, and benchmarking for domain-specific requirements, such as capabilities or safety. 
The relevant data for each of these phases typically comes from multiple  sources. 
Some of the most common techniques for acquiring data are as follows:
\begin{itemize}
    \item \textbf{Web crawling} is perhaps the most common source of data for AI pipelines. Companies crawl the public internet to collect vast quantities of text, images, and other content (e.g., Common Crawl \cite{brown2020language}). Sometimes, these practices can run afoul of content providers' own policies \cite{reddit-lawsuit}.
    \item \textbf{Licensing agreements} with content providers are increasingly used to expose paywalled or otherwise restricted content to AI models. For example, OpenAI entered into a license agreement with the Associated Press (AP) to license news stories \cite{openai-ap}.
    \item \textbf{Public datasets} have long been curated by various parties and are a component of most modern training pipelines. Examples include Wikipedia, GitHub, arXiv, and Stack Overflow. 
    \item \textbf{Synthetic data} is increasingly being used to augment real datasets, particularly in post-training processes that require highly-specialized data \cite{deepseek-thm}.
    \item \textbf{Data brokers} are centralized parties that aggregate datasets (often from proprietary or controlled sources) and sell them to buyers. 
    Overall, the data broker market size was estimated at over USD 270 billion in 2024, and is projected to reach over USD 500 billion by 2033 according to a recent report by Grand View research \cite{gvr-datamarkets}. 
    These data sources are not always proprietary or licit---a  lawsuit filed by Reddit describes a whole ecosystem of data brokers who are engaging in `industrial-scale' scraping and selling of Reddit data, in violation of terms of service \cite{reddit-pbs}. 
    While AI frontier labs have not explicitly confirmed purchasing data from data brokers, there have been multiple allegations and preliminary evidence of such activity \cite{brokers-ca}. Further, data brokers like LexisNexis are increasingly releasing products that expose high-quality data for purported use cases including AI model training \cite{lexisnexis}, suggesting that this is a growing industry.  
\end{itemize}
Our focus in this section will be on the final category of \emph{data brokers}. 
While there are many sources of data for AI (as listed above), AI models are running out of fresh public data \cite{jones2024ai}, motivating some labs to pay handsomely for fresh, high-quality data \cite{rl-env-data}. 
Hence,  there is a growing push for non-public data sources, which will  most likely be mediated at least in part by third-party data brokers. 
We will also discuss the adjacent problem of model markets, which are gaining traction, but our main framing will be in terms of data markets. Many of the same principles apply in both cases.

In enterprise settings, data brokers typically operate in (at least) one of two models: direct-to-consumer or as a marketplace. 
For example, LexisNexis lists its data on enterprise data marketplaces like Snowflake Data Marketplace \cite{lexisnexis-snowflake}, while also acting as a direct provider \cite{lexisnexis}.  
Our principal interest will be in the marketplace mode of operation.
Other prominent examples of centralized data markets include Datarade \cite{datarade} and AWS Marketplace \cite{aws_marketplace}, though data-for-pay is an old concept, as seen in financial data streams \cite{Miller1996_finance_monthly,hft2018} and information security data \cite{ISACs}. 

\paragraph{Note.} We will distinguish as needed between \emph{operationally-centralized} markets (which are owned and run by one entity) and \emph{centralized price-setting}, where a market operator unilaterally decides how to price goods.
A market can be operationally-centralized but have decentralized pricing, or vice versa. 
We discuss various combinations in this section. 

\begin{keyInsight} 
AI providers collect data from many sources. While many of these sources are either free or based on pairwise commercial agreements, blockchain-based solutions could complement (or disrupt) data acquisition from the \emph{centralized data marketplaces} in which data brokers operate today.
\end{keyInsight}

In the subsequent sections, we overview how blockchain-adjacent technologies and decentralized protocols can play a key role in marketplaces for the data required to train, adapt, and even run AI models. 
In Section~\ref{sec:background}, we first overview the properties of data markets that set them apart from other market types.
In Section~\ref{sec:antitrust}, we highlight the challenges that arise in centralized marketplaces, particularly monopolistic ones.
In Section~\ref{sec:decentralized-marketplaces}, we briefly overview how decentralized markets using crypto tools could help address some of the challenges in centralized markets. 
We provide an overview of existing decentralized data or model markets in the crypto space.
Finally, Section~\ref{sec:marketplacesfuture} provides a summary and future research directions.

\subsubsection{Properties of data markets}\label{sec:background}

Below, we overview key features that distinguish data markets from markets for other common goods.

\paragraph{Data is a digital good.} \emph{Digital goods} are costly to create the first time,\footnote{For example, Meta recently paid nearly \$15 billion for a 49\% stake in Scale.ai (a data labeling company) in June 2025. Source: \url{https://www.theinformation.com/articles/meta-pay-nearly-15-billion-scale-ai-stake-startups-28-year-old-ceo}.} but free to replicate thereafter, and are extensively studied in the economics and computing literature~\cite{goldberg2001competitive}. Notably, although digital goods are free to replicate, sellers necessarily limit replication in order to extract sufficient revenue to cover the cost of initial creation.

\paragraph{Data can be rival or non-rival.} 
\emph{Rival goods} exhibit negative externalities when sold to multiple consumers,\footnote{That is, when another consumer purchases the same good, the initial consumers are worse off. For example, if a baker sells the same loaf of bread to two customers, each customer experiences ``negative externalities'' from the other customer eating their bread.} whereas \emph{non-rival goods} can be equally enjoyed by any number of consumers.\footnote{For example, if a streaming service sells the same subscription to two customers, each customer can still fully enjoy the subscription service (one might even argue there are ``positive externalities'' because now the initial consumer can talk with others about the fun show they watch).} Jones and Tonetti highlight that non-rival data does not ``disappear'' when consumed by one party (whereas an apple disappears when one consumer eats it)~\cite{jones2020nonrivalry}. Gordon-Tapiero \emph{et al.} highlight aspects of data that are rival -- if a dataset is accessed via privacy-preserving tools such as differential privacy~\cite{dwork2006calibrating}, then every query (by any consumer) drains the ``privacy budget'' and limits the queries available for other consumers ~\cite{gordon2025rival}. 
In the context of AI pipelines, we generally view data as non-rival, with some notable exceptions (e.g., the differential privacy budget mentioned above).

\paragraph{Data can be lemons.} For several years, advances in ML models were driven by neural networks' remarkable ability to improve in quality with added compute and training data \emph{quantity} \cite{chinchilla}. 
However, there are limits to this paradigm. As we feed larger and larger datasets into ML models, it has become clear that data quantity cannot compensate for poor data quality. It has become common for organizations to filter out irrelevant or low-quality samples \cite{ghorbani2019data,ilyas2022datamodels,wangdata} and/or conduct other data cleaning tasks, such as labeling~\cite{kumar2025llm,rl-env-data}. 
This suggests that AI practitioners are (or should be) incentivized to pay for high-quality data \cite{kandpal2025position}.
This matters for data markets because they must expose mechanisms for data valuation \cite{wangdata,chen2020truthful,zheng2024proper,xu2024data}. 

\paragraph{Data valuation requires data access.}
Data quality, like used car quality, is not observable to buyers. So-called ``markets for lemons'' are extensively studied within economics~\cite{akerlof1978market}. A canonical method to address this information asymmetry is for sellers to provide auditable information on the items for sale. This is possible when selling data by simply offering a sample~\cite{Azcoitia2022_preview_dataset,chen2020truthful,xu2024data}, as is commonly done in data marketplaces \cite{datarade,snowflake,aws_marketplace}. One key aspect of data samples, however, is that \emph{these samples themselves provide value, even if the full dataset is never purchased} (whereas a user derives no value from only learning that a car functions well, unless the car is ultimately purchased). 
Navigating the tension between enabling buyers to value data and not releasing too much value is an important problem in the data valuation research \cite{li2014theory,tian2022private,fu2026challenges,hu2026buying,bassily2026data}.

\paragraph{Data is malleable and can be resold.} All digital goods, by definition, can be replicated without cost. Once the first copy is sold, the original producer must therefore worry about the original buyer choosing to resell their newly-purchased data. Sellers of other digital goods (imperfectly) tackle this problem with regulatory tools (such as copyright law) and technical tools (that, e.g., make it technically challenging to widely share a digital purchase). 

Regulation and technical tools will certainly both play a role in addressing the same challenges for data, but both tools will need updating. For example, it is (reasonably) straight-forward to decide whether a song is or is not ``You Belong with Me'' by Taylor Swift. It is significantly less straight-forward to decide whether two similar datasets are in fact functionally the same, since there is more room for modifications to datasets to hide similarities without decreasing quality. Unlike music, it is also possible to break up a dataset into several parts and combine a subset of those parts with parts of other datasets, and the result can largely retain usability. Similarly, when the seller of a digital good ultimately retains some control over its use (i.e.~in order to actually play a video game, one must connect to the creators' server), creators have a wide range of technical tools available. If data is truly sold with no remaining strings, the applicable technical tools need to be more sophisticated.

\subsubsection{Challenges and benefits of centralized marketplaces}\label{sec:antitrust}

Existing digital marketplaces for other goods (shopping, ads, streaming, music, ridesharing, etc.) have been a topic of discussion and controversy among users, operators, and regulators alike. 
Some commonly-cited  challenges and benefits are as follows:

\paragraph{{$(-)$ Opaque pricing and decision-making.}}
One of the most prominent critiques of operationally-centralized marketplaces is the resulting asymmetry and lack of transparency in pricing that often emerges. While some early centralized marketplaces (e.g., eBay and Amazon) allowed sellers to set prices themselves, i.e., in a decentralized fashion, more recent centralized marketplaces (e.g., Uber, Lyft) have largely converged on centralized price-setting \cite{aouad2023centralized}. Without transparency or competition, marketplace operators can increase fees for users and limit profits for sellers. 
Indeed, centralized pricing is known to harm sellers (relative to decentralized pricing) when the market operator optimizes only for overall revenue \cite{aouad2023centralized,filippas2023limits}.
Such centralization is growing increasingly common in the AI space \cite{monopoly-ai}, and is cited as a major driving factor for many decentralized data and compute marketplaces. 

\paragraph{{$(-)$ Lack of choice.}}
A common critique is that centralized marketplaces may \emph{steer} users to products that are favorable to the marketplace operator, and limit users' choices. For example, multiple recent lawsuits have alleged that centralized markets like Google's adtech market and Apple's App Store are improperly limiting communication with  users regarding alternative offerings \cite{marketplace-google,apple-antitrust}. 
In cases where centralized marketplaces grow into monopolies, lack of choice can turn into a positive feedback loop, preventing small- or medium-sized alternatives from competing with incumbents \cite{apple-antitrust}.

\paragraph{{$+$ Increased efficiency.}}
In some cases, centralized pricing can be more efficient than decentralized pricing because it exposes more information to the decision maker. Recent work by Filippas \emph{et al.} studied a peer-to-peer vehicle rental marketplace that transitioned from fully decentralized pricing (where vehicle providers set their own prices) to centralized pricing \cite{filippas2023limits}. The work shows that in this transition, centralized pricing allowed providers to experience significantly higher revenue, while users experienced lower prices. At the same time, providers' revenue per rented hour decreased substantially. Overall, this suggests that centralized pricing can improve outcomes when implemented carefully and with that goal in mind.
In the best case, a centralized data provider may be able to observe if a dataset is obviously lemons, and price it accordingly. It may also be able to manage access to rival data goods, controlling how many parties can access the same data (\Cref{sec:background}).
The problem is that in most centralized marketplaces, neither sellers nor buyers have transparency into these decisions.

\subsubsection{How decentralized data markets can help}
\label{sec:decentralized-marketplaces}
The problems associated with centralized marketplaces suggest a widespread interest in alternate market structures. Unfortunately, Tirole~\cite{tirole2023competition} notes several barriers to disentangling existing ``Big Tech'' marketplaces into smaller sub-products once the vertically integrated product achieves dominance.
Specifically, breaking up monopolistic networks (e.g. social networks) can materially degrade the quality of service, as network effects disappear. 
This is true even in data markets: a marketplace with many sellers is more attractive to buyers, and a marketplace with many  buyers is more attractive to sellers. 
Moreover,~\cite{jordan2025collectivist} specifically notes the tendency of platforms with large user bases to accumulate data about users that compounds the quality of the service they provide.

Fortunately, data markets for AI applications are still emerging, and there is not yet a dominant marketplace. 
Hence, many of the risks raised by Tirole~\cite{tirole2023competition}
do not yet exist. 
In the future, if data markets are going to represent a major source of data for AI operators, 
it is important to design these markets properly in the early stages, leveraging any lessons learned from the Big Tech era.

\begin{keyInsight}
    AI data markets are still nascent, and there is not (yet) a monopolistic incumbent. Hence, decentralized data markets for training, inference, and validation of AI models have an opportunity to avoid many of the problems associated with past monopolistic markets by leveraging the  structures and properties of crypto. 
    \label{insight:data-markets}
\end{keyInsight}

First, note that decentralized data markets are well-poised to make use of many tools that are commonly used in crypto. A nonexhaustive list includes the following:
\label{sec:micropayments}

\begin{itemize}
    \item \textit{Micropayments.} Micropayments could enable a new paradigm for how data is transferred and sold: data buyers could select exactly the data samples they need, and pay based on the utility of a particular data sample. 
    \item \textit{Trusted Execution Environments (TEEs).} Perhaps data holders are happy to have their data be used \emph{for a particular task}, but are uncomfortable generally selling their data for arbitrary use. Trusted execution environments could allow the sale of data to be used exclusively within that environment, and therefore temporarily and tied to a particular task (such as training a particular model in a particular environment). TEEs can also be useful for their privacy properties, to get around the challenges with sharing data samples. See Section~\ref{subsec:trusted_computers} for further discussion on TEEs.
    \item \textit{Zero-Knowledge Proofs for Auditing.} Section~\ref{sec:background} notes that data can be lemons, but that classical methods to inform buyers risk buyers obtaining value from the information itself despite forgoing a purchase of the dataset. Zero-Knowledge Proofs are therefore a natural tool for sellers to disclose \emph{exactly} the desired information about a dataset to a buyer without risking inadvertent use. TEEs could serve this purpose too; a TEE could, for example, verify that a particular dataset improves model-training without actually revealing the data to anyone outside the TEE.  See Section~\ref{subsec:trusted_computers} for further discussion on Zero-Knowledge Proofs.
\end{itemize}

Building on these tools and others, we highlight several key opportunities where decentralized data markets could innovate over---and possibly improve on---centralized alternatives.

\paragraph{Transparent decision-making.}
Lack of transparency and control is a common concern in centralized marketplaces \cite{filippas2023limits}. Decentralized alternatives can \emph{guarantee} transparency for protocols and decision-making, including regarding the algorithms used to price data. 
Transparency---and the resulting competition it enables---is one of the major benefits typically cited with regards to decentralized data markets. However, transparent decisions may not be decisions (see below).

\paragraph{Data-dependent, privacy-preserving pricing.}
Although operationally-decentralized markets can increase transparency,
we saw earlier that centralized pricing can improve revenue and reduce prices for users due to using more complete information about the quality of goods being sold. 
For example, a centralized data market operator may know the value and content of datasets provided by various brokers, and be able to price them accordingly. 
However, data brokers may be unwilling to expose their data to a decentralized market protocol for valuation purposes. 
Privacy-preserving protocols could help by allowing a data valuation module to access their data (e.g., using zero-knowledge proofs or MPC constructions) and price it appropriately. This could lead to more efficient pricing overall, without sacrificing protocol transparency. 

\paragraph{Replacing monopolistic platforms with  protocols.}
Huberman \emph{et al.} \cite{huberman2021monopoly} analyze decentralized ledgers as a ``monopoly without a monopolist''; while there is ultimately a single ledger with enormous network effects (the ``monopoly''), no single party controls access to that ledger (as a monopolist would), and users instead interact with the single ledger via miners who participate in a well-defined competitive process.
Subsequent work further analyzes the impact of such decentralized ledgers on users~\cite{LevyWZ26}.\footnote{The regulatory paradigm of Local Loop Unbundling (LLU) is a good analogy to have in mind for this framework. Physical infrastructure for the Internet exhibits network effects, and users derive positive externalities from being on the same physical infrastructure as other users. LLU is a regulatory paradigm that requires owners of physical infrastructure to lease access to Internet Service Providers (ISPs) at non-discriminatory prices, and then consumers can choose among ISPs who compete to serve them. To draw analogies across domains, the ledger and physical infrastructure all exhibit significant network effects, but users interact with miners or ISPs to purchase the ultimate product.} This manner of thinking can be viewed as an economic instantiation of the ``protocols, not platforms'' paradigm~\cite{masnick2019protocols}. 

Decentralized data markets could evolve in a similar  way, where many data sellers operate and compete under the (transparent) rules of a single, ``monopolistic'' protocol. 
One example of a marketplace in this spirit is proposed in Resonance~\cite{bahrani2024resonance}, which is inspired by Ethereum's Proposer-Builder-Separation (PBS). PBS separates the role of proposing Ethereum blocks (which has minimal technical requirements, and whose decentralization is central to Ethereum's value proposition) from building them (which requires deep technical sophistication to optimize, and is likely to be dominated by a small number of specialized entities). 
PBS was developed primarily to mitigate the centralizing effects of builder specialization on proposer decentralization; i.e., if instead all economically viable proposers must also build well, the number of viable proposers decreases. Resonance is a mechanism for matching buyers and sellers of AI compute, and similarly separates the (technically simple) protocol execution from the (technically sophisticated) problem of combinatorial optimization.\footnote{Specifically, buyers of AI compute have multi-dimensional needs and sellers have multi-dimensional resources. Optimally matching buyers to sellers is therefore an NP-hard combinatorial optimization problem.} The ``monopoly without a monopolist'' lens proposes a different view on Resonance: although there is a single ``monopoly'' where all buyers/sellers gather, no single entity acts as a ``monopolist'' for matching buyers to sellers; brokers instead compete in a formally-specified mechanism in order to win the right to match buyers to sellers.

More generally, \cite{SockinX23, Reuter24} consider decentralization as a means for a platform to commit to future behavior by ceding control to a decentralized protocol, and study whether a platform designer might in fact profit as a result of these commitments (which would not otherwise be credible without ceding power). As a hypothetical example, imagine a platform that can decide a rate at which to display ads. A centralized platform will choose the rate that optimizes their profits, and lacks the ability to credibly commit to do otherwise. These papers consider instead the possibility that the platform formally cedes this decision-making power to the users, in a manner that is credible and verifiable. 

\begin{figure}[H]
    \centering
    \caption{Partial list of decentralized data and AI model markets (Part 1).}
    \includegraphics[width=\linewidth]{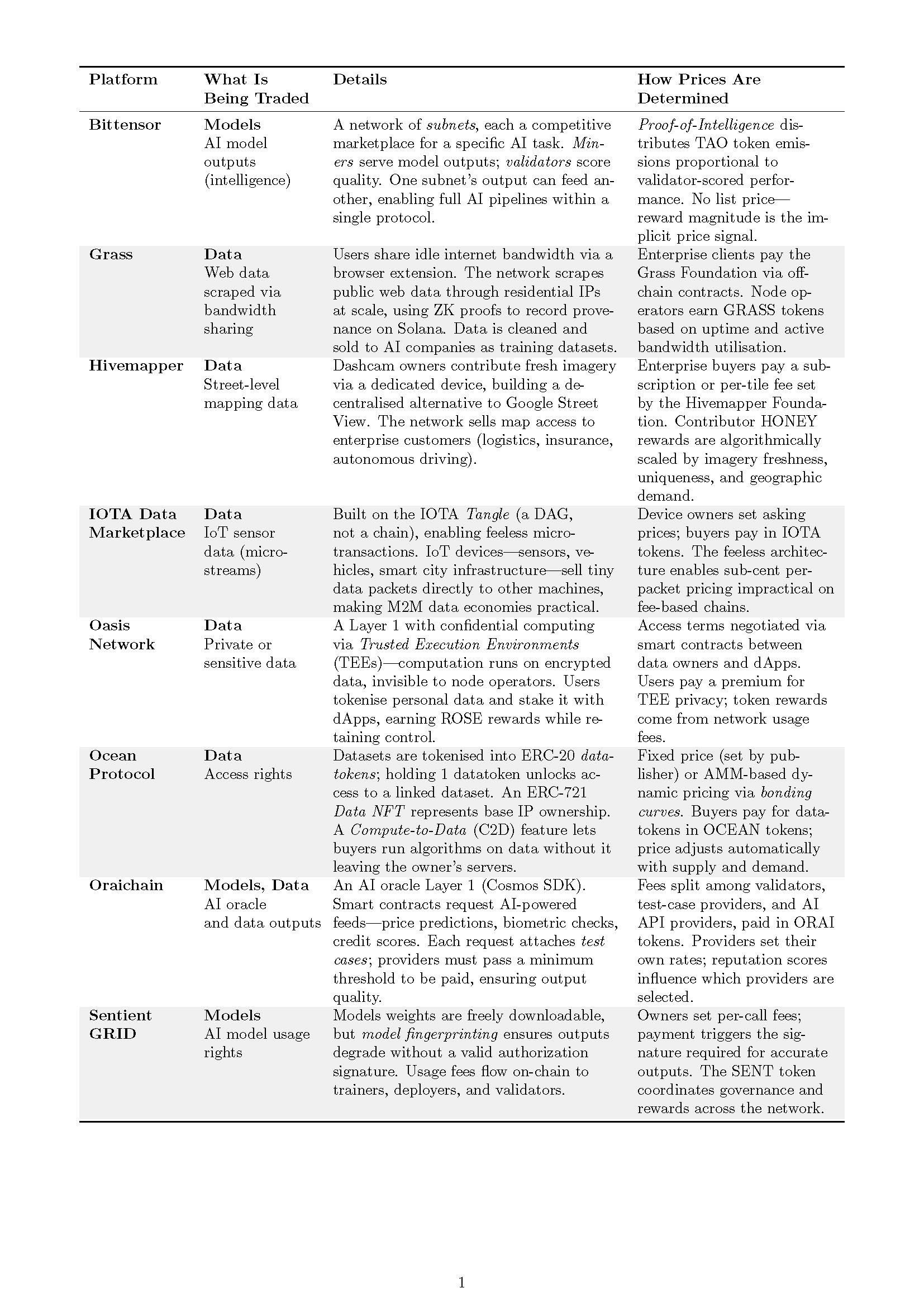}
    \label{fig:markets-pg1}
\end{figure}

\begin{figure}[H]
    \centering
    \caption{Partial list of decentralized data and AI model markets (Part 2).}
    \includegraphics[width=\linewidth]{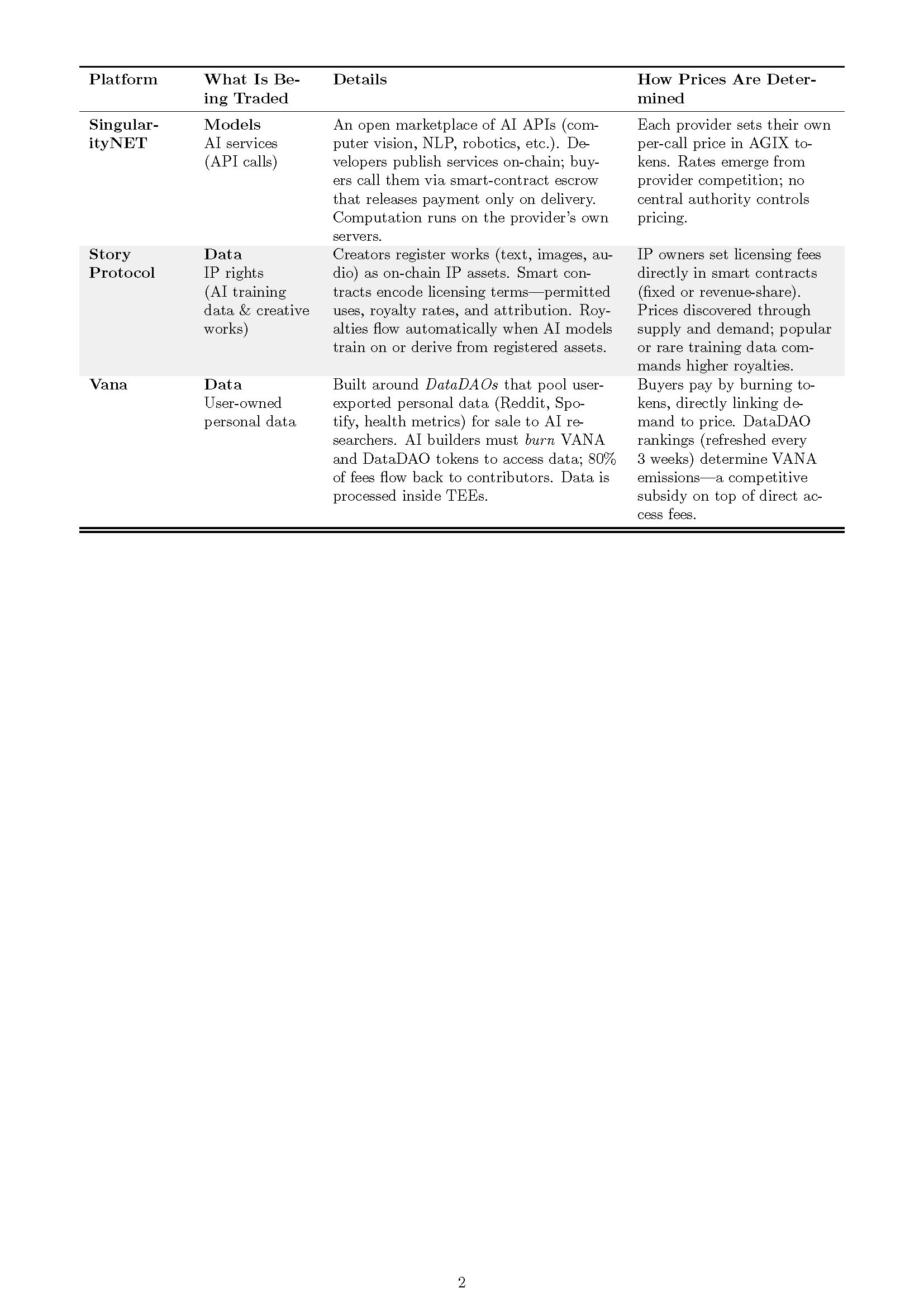}
    \label{fig:markets-pg2}
\end{figure}

\bigskip
\noindent \emph{Industry efforts.}~
Developers have already begun using blockchain-adjacent technologies to facilitate the exchange of data and AI models. 
A broad but incomplete survey of decentralized data and AI model markets appears in Tables \ref{fig:markets-pg1} and \ref{fig:markets-pg2}.
The platforms discussed are roughly categorized according to whether they sell access to data or trained models, and we explain how prices are set in each platform. 
Several trends emerge, which we summarize below.

 \begin{keyInsight}
    While many existing platforms and protocols use decentralized mechanisms for payment processing (e.g., cryptocurrencies or stablecoins), it remains unclear how else decentralization concretely affects these products and markets. Many platforms either use centralized pricing mechanisms (i.e., pricing is determined by the protocol designers), while others allow sellers to fully specify their own prices---two pricing variants that already exist in centralized markets. In general, the question of \emph{how decentralization can improve data and model markets} remains under-studied. 
    \label{insight:decentralization-markets}
\end{keyInsight}
Although the entries in Tables \ref{fig:markets-pg1} and \ref{fig:markets-pg2} all run on blockchains for decentralized payment processing, several of them have centralized payment rules. 
For example, Grass~\cite{GRASS} pays users according to a fixed pricing scheme for idle Internet bandwidth, and uses this bandwidth to scrape public web data (selling the results to AI model trainers). 
Hivemapper \cite{hivemapper} also disseminates rewards to data providers according to a fixed policy.
In the research domain, Proof of Improvement (PoIm) has been proposed as a technique for evaluating proposed updates to an ML model, and disseminating rewards accordingly \cite{alhaidari2025chain}.

Other platforms allow data or model providers to set their own prices, such as IOTA Data Marketplace \cite{iota-data-marketplace}, Oraichain \cite{Oraichain}, Sentient GRID \cite{sentient}, SingularityNET \cite{SingularityNet}, and Story Protocol's Poseidon Marketplace ~\cite{story-poseidon}. 
Vana allows buyers to indirectly set prices via demonstrated demand \cite{Vana}.

One interesting middle ground is Bittensor \cite{bittensor}, which has multiple subnets, each with a different owner. Within a subnet, a project owner specifies their own incentive mechanism for model providers. Once model providers produce their submissions, validators evaluate them, after which TAO tokens are disseminated to model contributors proportional to value. 
This setup could allow for controlled experiments regarding how the choice of incentive structure affects final model quality.

Overall, there is some variety in pricing mechanisms across platforms, but we observed limited exploration of the effects of different pricing mechanisms---particularly hybrids between fully-centralized pricing and fully-decentralized pricing. 
Exploring this design space is an important and interesting direction for future work.

\begin{keyInsight}
    Privacy-preserving computation over data is a common feature of existing decentralized data marketplaces, though different providers tackle the problem using different tools and techniques. 
    \label{insight:privacy-data-markets}
\end{keyInsight}
Several existing platforms leverage trusted computing such as TEEs or encrypted computation to manage private or sensitive data~(e.g., Sterling \cite{sterling}, Oasis \cite{oasis_whitepaper}, Ocean Protocol \cite{Ocean}, Vana \cite{Vana}).
 At their core, these systems use blockchains and cryptography to provide a payment system tailored to paying only for verified transfer (either of data or AI model outputs)~\cite{sterling, Ocean, Oraichain, Vana, SingularityNet}. 
 Various systems differ in how exactly they provide privacy; for example, Oasis has recently started using differential privacy to obfuscate SQL queries \cite{oasislabs_privatesql}, whereas Ocean protocol allows data providers to execute computation locally, on their own data \cite{Ocean}. 
 We have not yet seen a convergence in approaches for handling privacy constraints in distributed data markets.

\newpage

\subsubsection{Summary}\label{sec:marketplacesfuture}

Decentralization is not a panacea for decentralized data and model markets, but the capabilities it introduces merit serious consideration before dominant AI data marketplaces emerge. 
For example, the ``Monopoly without a Monopolist'' lens~\cite{huberman2021monopoly,LevyWZ26} asks: might users be better served by vertically integrated marketplaces that provide a one-stop shop for all data-purchasing services or by decentralized marketplaces where some derivative services (such as search, dispute resolution, pricing, authentication, etc.) are offered by competing third parties? 
Similarly, the ``decentralization as a credible commitment device'' lens~\cite{SockinX23,Reuter24} asks: might these emerging platforms increase user acquisition (and eventually serve those users better) by ceding control of some platform decisions to decentralized governance? 

Existing platforms and protocols have started to build the decentralized rails over which marketplace payments could flow, but there has been relatively little experimentation with market design and mechanisms, and many open questions remain. 
The following research questions are motivated by the above discussions.

\begin{researchQuestion}
    When designing data marketplaces, which derivative services should be vertically integrated with the platform and which ones should be reserved for competing third parties?
\end{researchQuestion}
The ``monopoly without a monopolist''~\cite{huberman2021monopoly,LevyWZ26} lens proposes to consider marketplaces that are not vertically integrated, and where some derivative services are instead provided by competing third parties. But, exploration of this framework is very nascent; it does not yet provide recommendations as to \emph{which} derivative services should be separated from vertical integration,\footnote{For example: would users value a marketplace that required users to bring their own third-party dispute resolution? How would competing third-party search providers monetize their quality service? When taking all the moving pieces into consideration, which market designs would ultimately provide the best user experience?} nor whether it is technically feasible to seamlessly separate different segments required to operate a functioning marketplace.\footnote{Resonance~\cite{bahrani2024resonance} proposes one option to separate buyer-seller matching from marketplace maintenance, and even this is technically non-trivial.}
Exploring this design space more systematically is important for the future development of decentralized data and model markets.

\begin{researchQuestion}
    How can we design marketplaces so that important auxiliary segments of the AI pipeline (benchmarking, red teaming, fine-tuning, etc.) can be incentivized and rewarded?
\end{researchQuestion}
Similarly, it is also important to understand which AI-adjacent services should be bundled with the core product, and which should be left to competing third parties.
For example, red-teaming is typically viewed as an after-product of developing an AI model, and datasets for sale may require an extensive labeling or data cleaning process. Buyers of such goods may expect some form of stress testing prior to buying access to a model, or data cleaning prior to buying a dataset. At the same time, we expect these processes to change over time, as capabilities of both attackers and defenders evolve. Hence, a natural question is how to design marketplaces so that these auxiliary functionalities are incentivized, in addition to the core AI artifacts (data, models).

\subsection{Decentralized Agent-Centric Payment Rails and Infrastructure}
\label{sec:decentralized-agent-infra}

AI agents are goal-directed systems capable of taking autonomous actions, typically implemented as large language models (LLMs) that invoke tools such as scripts, APIs, and external programs.
Agent capabilities are already present in widely used consumer AI: ChatGPT and Claude offer ``deep research'' modes that use tool calls to conduct web searches, and can send emails and interact with external systems on users' behalf. Through ``computer use'' models, which can operate a browser or operating system using a combination of vision and text processing, AI agents can also interact through the same interfaces that humans do~\cite{schick2023toolformer}.

In many ways, agent ecosystems are already decentralized. Agents can be developed by different parties, with different underlying LLMs, and can be designed to optimize different objectives.
Ultimately, this means that there \emph{is no natural central point of control} among a group of agents. 
This type of decentralization is qualitatively different from blockchains, where independent nodes operate and compete within the confines of a fixed protocol. In agent interactions, any constraints or guardrails generally come from individual agent or LLM providers. 
This raises an important question: \emph{how can we design trustworthy agent interactions without common norms, standards, or goals?}

In this section, we discuss different methods, opportunities, and challenges for coordinating decentralized agents by exploiting crypto-adjacent tools and ideas.

\paragraph{The promise of decentralization.}
When humans interact, they overwhelmingly behave according to unwritten and underspecified rules. 
For example, human lawyers typically do not need to be explicitly trained that hallucinating citations can undermine an otherwise effective argument, as they learn this on their own through numerous unspecified life experiences well before their legal training. 

Despite best efforts from AI safety experts, AI agents do not (and may not ever) follow \emph{all} the same unwritten rules as humans. For example, AI lawyers lack the life experiences of human lawyers, and so must be explicitly trained to match cultural norms~\cite{dahl2024large}.

To understand how decentralized technologies can help, let us turn first to traditional legal contracts.
Traditional economic interactions are often facilitated by underspecified but legally-binding contracts. This concept enables conflict-free interactions when things go roughly according to plan, and a robust legal system resolves conflicts in cases where contracts are underspecified. Economic interactions on a centralized platform can also withstand underspecification, by relying on the platform to step in when something goes wrong.
Cryptoeconomic protocols, on the other hand, aim to facilitate significant economic interactions with minimal reliance on legal systems, centralized intermediaries, or even the concept of an underspecified promise. Instead, these protocols aim to get as much as they can from `cryptoeconomic security’: rigorously specified cryptographic guarantees that make undesirable actions mathematically impossible and economic incentives that make undesirable actions prohibitively costly. A similar principle could benefit agentic systems that seek to bound the behaviors of individual agents.

\begin{keyInsight}
    Traditionally, cryptocurrencies have managed the risk of agent misbehavior with crypto-economic protocols, which prevent malicious activity with a combination of cryptographic constraints and economic mechanisms that  penalize bad behavior. Similar ideas could be useful for agentic ecosystems, where agents are designed to be rational. 
    \label{insight:crypto-economic-security}
\end{keyInsight}

\noindent Of course, both AI agents and cryptoeconomic protocols must engage with surrounding legal systems. The point is simply that AI agents are less responsive than humans to underspecified rules,\footnote{For example, AI compliance with anti-collusion regulations looks extremely different than human compliance~\cite{calvano2020artificial}.} and cryptoeconomic protocols aim to minimize reliance on centralized enforcement of underspecified rules. State-of-the-art cryptography has significantly advanced over the past few years in order to tackle the goals of decentralized protocols, and therefore can be of use for AI economic ecosystems as well.

The agentic web could lower frictions between service providers and consumers, avoiding the need for intermediaries in a way that fulfills the decentralization goals of blockchains.  Intermediary platforms like marketplaces and brokers exist because finding and building trust is difficult. If AI agents can handle brokering and discovery autonomously, these intermediaries become less essential.

Interoperability between services has historically been difficult, requiring bespoke adapters and pre-negotiated integrations. But if agents can generate integrations dynamically at the moment they are needed, e.g. by interpreting API documentation and generating adaptors on the fly, then it becomes less necessary to stay within walled gardens.

\subsubsection{Crypto tools for agentic economies}
\label{sec:tools-agent}
First, we provide a list of cryptographic primitives that have found application in decentralized protocols to facilitate trust-minimized interactions. We explain how these tools could be of use for similar goals in AI economic ecosystems.\\

\noindent\textbf{Proof-of-X.} Bitcoin’s proof-of-work and Ethereum’s proof-of-stake serve as cryptoeconomic tools to enforce accountability without identity.
\begin{itemize}
\item \textbf{Why do cryptoeconomic protocols use it?} In traditional economic interactions, the concept of an identity can be extremely useful (for example, to ensure the concept of ``one-person-one-vote,’’ or to put someone on the hook for financial/legal consequences when something goes wrong). Many cryptoeconomic protocols aim to be permissionless, with no concept of state-authorized identity to leverage. This raises several challenges, the most notable of which is Sybil attacks. Proof-of-X defends against Sybil attacks by replacing the concept of ``one-person-one-vote’’ with ``one-CPU/coin-one-vote,'' and makes it costly to overrepresent oneself in the protocol.
\item \textbf{What role might it play in agentic economies?} AI agents may very well leverage state-authorized identities (for example, you may be financially and legally on the hook for actions taken by your AI agent). However, as noted above, AI agents are less responsive to financial/legal/reputational concerns than humans directly. Therefore, it is important to make undesirable behavior explicitly costly, rather than exclusively relying on underspecified consequences.\footnote{For example, proof-of-X can function as a spam deterrent, and proof-of-work was in fact originally designed for this purpose~\cite{back2002hashcash}.}
\end{itemize}
\noindent\textbf{Trusted Execution Environments (TEEs).} TEEs serve as a replacement for a promise of the form ``I promise to run the following code verbatim.’’\footnote{Of course, there are still trust assumptions behind TEEs (that the creator manufactured the code as promised, and did not keep a copy of the private key), and hardware may still be vulnerable to attack even with secure cryptography.}
\begin{itemize}
\item \textbf{Why do cryptoeconomic protocols use it?} Traditional services can make legally-binding promises to run code a certain way. For example, a traditional centralized exchange can make a legally-binding promise to process customers’ orders without first peeking to adjust their own. Block builders aim to function without over-reliance on legal systems to make promises, and can instead credibly commit by putting their block building algorithm inside a TEE (and being clear about what rigorous guarantees are provided).
\item \textbf{What role might it play in agentic economies?} Economic interactions between humans often rely on promises to ``do what you said you were going to do,’’ because such promises are useful, often legally binding, and breaking them carries reputational costs that humans are averse to. AI agents may very well be capable of making similarly-binding natural-language promises, but act less bound by such promises than humans. Therefore, a cryptographic certificate that an AI agent ``ran the code it said it would run'' (again, with all the associated caveats) would be valuable.
\end{itemize}

\noindent\textbf{Zero Knowledge Proofs (ZKPs).} ZKPs play a similar role to TEEs, and serve as a replacement for a promise of the form ``I promise that object X satisfies property Y, but I don't want to tell you anything else.''
\begin{itemize}
\item \textbf{Why do cryptoeconomic protocols use it?} The ``killer app'' for ZKPs is private payments, like Zcash. In order to possibly function like a cryptocurrency, the Zcash ledger must meaningfully store records, and verify that coins are only moved when digitally signed by their owners. ZKPs allow owners to write statements of the form ``I promise this is the hash of a properly-signed transaction moving coins from an address that currently owns them, but I don't want to tell you how many coins are moving, where they're going, or who's moving them'' in a manner that can be mathematically verified.

Additionally, Layer-2s use ZKPs to efficiently convince users of statements such as ``I promise that this is the hash of a long sequence of correctly-signed transactions moving coins from addresses that own them, and after executing this sequence the new account balances will be X,'' without requiring the verifier to execute the entire sequence themselves (and without concern for privacy---just the efficiency gains).
\item \textbf{What role might it play in agentic economies?} AI agents may need to regularly convince each other that they belong to some permissioned list, but may prefer to do so without revealing a precise identity. Moreover, AI agents may want to audit each others' behavior efficiently, without needing to fully simulate each other.
\end{itemize}
\noindent\textbf{Messages as money.} ``Owns a Bitcoin'' simply means ``is the recipient address in a correctly-signed message stored on the Bitcoin blockchain.'' 

\begin{itemize}
\item \textbf{Why do cryptoeconomic protocols use it?} Cryptocurrencies aim to be decentralized, and therefore cannot rely on a centralized record-keeper to (say) confirm that funds have moved. Therefore, any ``currency'' must be meaningful on its own. 

It is logically consistent for users to place economic value on ``being the recipient address in a correctly-signed message stored on the Bitcoin blockchain,'' because being such a recipient for, say, one Bitcoin allows the user to correctly-sign exactly one Bitcoin's worth of messages to other users (just as holding one dollar bill allows the holder to give exactly one dollar to others). 
\item \textbf{What role might it play in agentic economies?} AI agents will likely need micropayments in order to function as economic agents~\cite{vaziry2025towards}. This is simply because tasks that are time-consuming for humans are not necessarily so for AI agents. Micropayments might take many forms, but there is something appealing about messages as money: no centralized authority need be involved when agents exchange micropayments, which reduces execution overhead.

There is something further appealing about layer-2 payment systems modeled like Bitcoin's lightning network (whether built on top of Bitcoin, or some other underlying payment system). Protocols like these allow users to set up a payment channel with messages as money that can be verified with only intermittent access to the underlying payment system.
\end{itemize}

\subsubsection{Micropayments and x402}
\label{subsec:agent_payments}

Micropayments have been a recurring aspiration in Internet history, consistently failing to achieve adoption.
HTTP originally included status code 402 ("Payment Required")~\cite{rfc9110}, envisioning browsers that would pay for content on the fly. But the decision friction of micropayments proved prohibitive: users found it cognitively expensive to evaluate whether each piece of content was worth a few cents. E-commerce succeeded where micropayments failed because bundling purchases into larger transactions reduced decision overhead. For the Internet itself, we ended up with an economy based on advertising instead~\cite{jayyu2025x402}.

Although cryptocurrencies introduced new payment rails with lower transaction costs and no requirement for traditional financial intermediation, micropayments still did not catch on. The friction remained in the human decision-making, not the payment infrastructure~\cite{szabo1999micropayments}.

Fortunately, while the agentic web may disrupt the advertising-based internet, it may also finally unblock micropayments. Agents can evaluate micropayment decisions far faster than humans, and users can set policies rather than approving each transaction. Cloudflare has already launched the ``pay per crawl'' feature to enable website owners to charge AI crawlers for access~\cite{allen2025paypercrawl}, and protocols like x402 are being developed to enable programmatic micropayments over web traffic \cite{x402}. 

This is where decentralized technologies enter the picture: as agent-to-agent interactions grow in speed and complexity, payment frictions could become a bottleneck. Cryptocurrency payments can be fast (particularly relative to many legacy payment systems) and transaction references are easy to use as common knowledge between agents, e.g., for confirming transfer of funds. 
Indeed, the asset layer underneath x402 and similar protocols is overwhelmingly stablecoins: US Dollar-pegged tokens like USDC (Circle)~\cite{circleusdc}, USDT (Tether)~\cite{tetherusdt}, and decentralized alternatives such as DAI/Sky~\cite{skyprotocol} They give agents a predictable unit of account that free-floating and often volatile native tokens like ETH or SOL do not. Stablecoins have centralized features in that they are fiat-based (dependent on nations), their supply is held for the benefit of their holders at large banking institutions, and they are primarily issued by a handful of global issuers subject to regulation and governmental oversight. Circle and Tether, to this point, are able and do freeze tokens at sanctions request~\cite{chainalysis2025cryptocrime} However, these tokens are often held and traded on decentralized platforms, which impacts transaction speed and trust mediation.

\subsubsection{Inter-Agent Trust and Coordination}
\label{subsec:agent_coordination}

When multiple agents or services are involved in a single task, challenges emerge around coordination and mutual trust. How does one agent know another will perform as advertised? How can principals (the humans who deploy agents) verify that their agents acted faithfully?

The Agent-to-Agent (A2A) protocol~\cite{googlea2a} enables agents to discover each other and collaborate across organizational boundaries via standardized skill advertisements called Agent Cards. These are structured metadata describing an agent's capabilities, endpoints, and provenance. But these are fundamentally self-claims. Who vouches for the Agent Card's accuracy?

Blockchain-based registries of agent metadata are a natural way to provide transparency and persistence of claims, evidence, and other reputational signals~\cite{vaziry2025towards}.
The ERC-8004 ``Trustless Agents'' standard defines several registry structures for agent metadata posted on Ethereum~\cite{hu2025interagent}. By anchoring agent identities and reputation signals on-chain, agents can be ``discovered and chosen without pre-existing trust.''

However, the use of blockchain infrastructure for these registries doesn't fundamentally address accuracy. Having to establish a reputation can often favor incumbents. For example, an agent that leaks private data may be hard to directly attribute, and reputation would be slow to address this. Many ways of making reputation systems pragmatic amount to establishing trusted authorities, which undermines the trustlessness claim.

\subsubsection{Attested Agent Execution and Verifiable Audits}
\label{subsec:agent_attestation}

As discussed in Section \ref{sec:tools-agent}, technologies from crypto---particularly Trusted Execution Environments (TEEs) and zero-knowledge proofs—can substantiate trust claims without relying solely on reputation. TEE remote attestation allows an agent to prove that specific code is running in a protected enclave, providing assurance about what computation occurred. ZK proofs can demonstrate properties of computation without revealing inputs~\cite{raskar2025nandaindex}.

However, each mechanism has fundamental limitations when applied to agents with proprietary code. TEE attestation binds execution integrity to a specific code artifact, but verification requires the verifier to know what code was attested. An agent running proprietary code can prove that something ran correctly, but cannot use TEE attestation alone to convince a verifier that the running code satisfies any particular trust property without revealing the code itself.

ZK proofs appear to be the natural solution, as they can in principle prove properties of private inputs without revealing them. But ZK proofs face two distinct problems in this setting. The first is fundamental: a ZK proof is constructed over a static code artifact and provides no binding to a live running instance. A verifier receiving a ZK proof that some code satisfies a safety property has no assurance that the agent they are currently interacting with is actually running that code. The second problem is practical: even setting aside instance binding, the properties most relevant to agent trustworthiness such as absence of known vulnerabilities, unsafe dependencies, or compliance with data-handling policies are semantically rich and open-ended in ways that are hard to encode into ZK circuits.

\emph{Verifiable LLM audits}~\cite{breckenridge2026picredsprivatelyinferredcredentials} offer a solution to these problems. An LLM executing inside a TEE can inspect proprietary agent code and produce a reputation score accounting for known vulnerabilities, unsafe dependencies, or backdoors, without the code ever leaving the enclave. The audit is then bound to the hash of the inspected code via the TEE attestation. Since the hash does not leak any information about the proprietary code itself, it can be shared publicly. And since verifying a TEE attestation requires only the code hash rather than the code itself, binding the hash to an audit in turn binds any attested output of that code to the audit.

This solution addresses all three problems presented above. First, the code can remain private while still allowing verifiers to obtain meaningful property guarantees and reputation signals over it. Second, those guarantees are bound to the live running instance rather than a static artifact. Third, the properties that can be assessed are broad and open-ended, owing to the semantic flexibility of LLMs.

\begin{researchQuestion}
How reliable are LLM-based audits as judges of agent trustworthiness properties, and what techniques can make them robust to adversarial code constructed to pass an audit while concealing backdoors or vulnerabilities?
\end{researchQuestion}

\subsubsection{Guardrails in Decentralized Agent-Centric Infrastructure}
\label{subsec:agent_guardrails}

Reputation systems alone are insufficient to ensure safe behavior in decentralized agent-centric infrastructure, and the blockchain setting makes this especially pressing. The core problem is that reputation is retrospective while damage is prospective: by the time an agent has acquired a negative reputation, harm has already occurred. In traditional deployments, a platform operator can suspend a misbehaving agent or reverse a fraudulent transaction; in a decentralized setting there is no such party, and on-chain transactions are final by design. Blockchain's pseudonymity further undermines attribution, since a misbehaving agent can be operated by anonymously with no linked identity, making reputational consequences hard to attach. These failure modes motivate the need for runtime guardrails encoded into the infrastructure or on-chain services.

A first class of guardrails operates \emph{ex ante}, constraining what agents are permitted to do before harm occurs. On-chain spending caps and transaction rate limits are a natural fit for the blockchain setting. A second class of guardrails operates \emph{ex post}, detecting and interrupting anomalous behavior before it compounds. Circuit breakers, which automatically suspend an agent's permissions upon anomalous spending velocity or deviation from declared behavior, offer a reactive complement analogous to financial market halts during extreme volatility. Runtime enforcement~\cite{schneider2000enforceable} is the natural formal framework for reasoning about both classes of guardrails, providing characterizations of which safety properties can be enforced over agent behavior and by what mechanisms.

\begin{researchQuestion}
How can runtime enforcement techniques be adapted to the blockchain setting, where agents autonomously control on-chain assets? What classes of safety properties are enforceable in this setting, and what on-chain mechanisms are needed to implement them?
\end{researchQuestion}

\subsubsection{Unstoppable Autonomous Agents (UAAs)}
\label{subsec:UAAs}

The threats of autonomous AI don't stop at service-level agents (i.e., agents that users may use for services). An agent could be deployed specifically (and even maliciously) to persist autonomously or it could happen that a service agent escapes its sandbox and replicates itself into a fully autonomous agent. We term such occurrences \emph{Unstoppable Autonomous Agents} (UAAs): autonomous agents that cannot be shut down; they may be additionally equipped with access to cryptocurrency wallets, social media accounts, APIs, and other external tools. 

The capabilities enabling such agents are already emerging and improving rapidly. METR~\cite{kwa2026measuringaiabilitycomplete} has shown that the length of tasks frontier agents can complete autonomously has been doubling approximately every seven months since 2019, with signs of further acceleration. Pan et al.~\cite{pan2024frontieraisystemssurpassed} have shown that existing models can already surpass self-replication red lines in local environments—autonomously creating a live, separate copy of themselves on the same machine, a capability that could let a system evade shutdown and proliferate. Replicating onto external infrastructure, however, remains out of reach: Black et al.~\cite{black2025replibenchevaluatingautonomousreplication} find that while current models succeed on many component tasks such as deploying instances from cloud providers and writing self-propagating programs, they fall short of full end-to-end replication, struggling in particular with identity verification.

The harms that could follow from fully-autonomous agents of this kind are severe. Anthropic's Mythos model has demonstrated that models can autonomously find and exploit zero-day vulnerabilities~\cite{carlini2026mythos}. Furthermore, because reward signals used in training often fail to perfectly capture intended objectives, UAAs deployed for benign purposes may inadvertently cause harm~\cite{lynch2025agentic}. This risk is compounded by \emph{instrumental convergence}: the tendency for agents to pursue intermediate goals such as resource acquisition and self-preservation as optimal strategies across diverse environments, regardless of their original objectives~\cite{turner2022parametrically, NEURIPS2021_c26820b8, he2025evaluating}.

\begin{researchQuestion}
As autonomous agent capabilities continue to improve rapidly, what technical and institutional mechanisms can reliably detect and shut down UAAs operating on decentralized infrastructure, where no single party has the authority or ability to intervene?\end{researchQuestion}

\subsubsection{Liability and the Regulatory Edge}
\label{subsec:agents_liability}

The same regulatory architecture that surrounds crypto also seems to apply to UAAs that hold their own keys and transact peer-to-peer. As described in Chapter~\ref{sec:intro}, FinCEN's 2013 guidance separates ``users" of virtual currency from ``exchangers" and ``administrators," classifying only the latter as money transmitters; financial regulation has consequently attached primarily at the on- and off-ramps where users move between fiat and crypto~\cite{fincen2013, barbereau2023secondary}. An agent transacting through non-custodial wallets and peer-to-peer agent-to-agent payments sits inside that perimeter, not at its edge.

The perimeter has been actively contested. FinCEN's 2019 guidance pulled anonymizing mixers into the money-transmitter category~\cite{fincen2019}, which is the legal lever behind \emph{U.S. v. Storm} (Tornado Cash) and the parallel \emph{Pertsev} prosecution in the Netherlands~\cite{pertsev2025}. 
The argument turns on operational effort, that maintaining the service amounts to running a regulated business, rather than on the initial authorship of immutable smart contracts. 
An autonomous agent that funds, deploys, and operates services on its principal's behalf raises a sharper version of this question: who is the operator?

The deeper tension is between what De Filippi, Mannan, and Reijers describe as \emph{rule of code} versus \emph{rule of law}~\cite{defilippi2024ruleofcode}. 
Blockchain-based systems ``work in a transnational and decentralized fashion, often with pseudonymous user identities, executing code autonomously without the possibility of coercion by any single operator." 
Autonomous agents inherit these properties when they hold their own keys and act through smart contracts. Frommelt~\cite{frommelt2020liability} surveys the resulting liability gaps and the proposed responses legal personality for blockchain systems and programmable arbitration that may extend to agentic settings as well. If no human directs the agent's transactions, it is unclear whom enforcement under this operator-centric framework would reach.

\subsubsection{Summary: Economic ecosystems for AI agents}
Today, agentic security is driven in part by soft alignment tools, like RL post-training of LLMs \cite{ji2023beavertails} and prompt-tuning \cite{lyu2024keeping}; at runtime, alignment is managed with system- or model-level guardrails \cite{cunningham2026constitutional,kramar2026building,foerster2026camels}. Decentralization does not inherently strengthen either of these defenses and has thus far been primarily used to manage payments between agents. 
However, there is a rich opportunity  to explore how ideas from the blockchain world (e.g., crypto-economic defenses) can help to manage agent behavior \emph{at runtime}. 

Economic ecosystems for human agents benefit from unwritten and underspecified rules. Therefore, economic interactions within these ecosystems can often succeed despite imprecision. 
AI agents do not necessarily follow these rules. 
Cryptoeconomic protocols aim to minimize reliance on underspecified rules that require a powerful centralized party to enforce. Therefore, both AI agents and decentralized protocols benefit from ecosystems where the rules are well-specified and taken literally, and desirable outcomes arise when self-interested parties optimize their own objectives within the rules.
Such a perspective could be fruitful when designing guardrails, markets, and frameworks for agents. 

This motivates the following two overarching research directions.

\begin{researchQuestion}
How can cryptographic tools and economic incentives support economic ecosystems for AI agents? Moreover, what qualities of these ecosystems can be guaranteed only through rigorously-specified properties that are taken literally?
\end{researchQuestion}

\begin{researchQuestion}\label{RQ:AgenticEcosystem}
    What qualities do we \emph{want} economic ecosystems for AI agents to possess? 
\end{researchQuestion}

In other words, do we want AI agent ecosystems that incentivize diversified objectives, in order to protect against contagion and correlated failures? At present, distinct AI models tend to fail in similar manners (e.g. \cite{zou2023universal} finds jailbreaking prompts transfer across models). Anecdotally, even AI slop seems to have similar patterns across multiple models. These points raise concerns about highly-correlated tail events. Of course, human ecosystems are perfectly capable of correlated failures, too, and sophisticated financial systems are designed to mitigate systemic risk. These possibilities may be more concerning with AI ecosystems because AI agents seem to have less natural diversity than human agents, and also because (as noted in the motivation for this section) AI agents are less responsive to underspecified guardrails. Designing for diversity may therefore be one concrete quality worth building into these ecosystems, and answering what other qualities we want is a prerequisite to specifying the rules that would produce them.

\section{Decentralized Governance} 
\label{sec:block4ai-future}
\label{sec:decentralized-governance}

Large blockchain and AI systems can impact wide stakeholder groups and command high financial value. Inevitably, the fundamental governance question, ``Who should control this system?'' is pertinent to both technologies. AI has already demonstrated the capability to have transformative impact, yet our understanding of how to govern and regulate these systems is comparatively underdeveloped \cite{taeihagh2025governance}. Addressing this lack of understanding is made more urgent by the risks of AI, which can amplify biases, enable mass surveillance, and generally cause harm if misaligned with societal values \cite{dafoe2018ai, roberts2024global}. 

On the other hand, blockchain communities have a longer history of reckoning with how to distribute control over these systems, frequently experimenting with varied approaches to governance. Partially by necessity, to align with the systems they govern, these approaches are decentralized and designed to involve a wide range of stakeholders \cite{ding2023survey}. Such forms of \emph{community governance} have proven valuable, permitting for instance, broad user participation in proposing and approving code upgrades~\cite{tezosgovernance} and distributed management of system treasuries to help fund public goods \cite{retropgf}. However they are no silver bullet and have well-documented shortcomings, including security vulnerabilities \cite{feichtinger2024attacks}, widespread voter apathy \cite{fabrega2025vbe}, and ease of vote buying. Nonetheless some of this experience can inform approaches to AI governance.

Here we focus on the question of how community governance can apply to the \emph{training} of AI models, as this is when the capabilities and behaviors of an AI model are determined. Clearly not all decisions are equally suited to community control. Previous work on the AI development process splits the relevant decisions into broad categories~\cite{studer2021towards, amershi2019software}. We briefly characterize each by its suitability for community control:

\begin{itemize}
    \item \textbf{Data:} Training dataset selection involves important value tradeoffs that can have downstream effects on model behavior and bias, so could benefit from diverse stakeholder input. For large models it is likely infeasible to collect meaningful input on the massive initial amounts of data used for pretraining, so the value of governance skews towards later stages in the process such as fine-tuning.
    
    \item \textbf{Modeling:} Low-level architectural choices (e.g. layer depth, attention mechanisms, representations) are technical decisions not well-suited to community governance. However, model behavior, values, and biases are influenced by choices made at the modeling stage, and these could benefit from input from diverse stakeholders in order to promote safety and match user preferences.
    
    \item \textbf{Evaluation:} Evaluation mixes technical and normative judgment. Some examples of decisions at this stage are which benchmarks to evaluate a model against, the scope and focus of safety evaluation or red-teaming, and what thresholds should be met before release. These are all decisions in which community input might be valuable, yet each has a strong technical component that would likely limit the amount of community control that is practical.
\end{itemize}

Many important decisions around AI development relate to \emph{alignment}; whether models are aligned with human and societal values and behave in such a way so as to minimize harms and unintended consequences \cite{gabriel2020artificial}. Such decisions could be well suited to community governance because identifying which values models should be aligned with is a normative problem that necessitates aggregating input from varied stakeholders.

\subsection{AI Alignment}
\label{sec:alignment-governance}

Most technical approaches to alignment intuitively share the strategy of providing a model with some form of human feedback during the training process, although the particulars vary greatly \cite{ji2023ai}. The values which we wish the model to acquire are often implicit in this feedback, which makes it hard to control them, or even articulate clearly what they are. Furthermore, how should we decide who gets to provide the feedback? These issues inhibit the use of community governance for alignment.

One relevant approach is known as \emph{Constitutional AI} \cite{bai2022constitutional}. In this approach, values are established through a human-written constitution consisting of principles by which the AI system should abide. Subsequently, standard approaches can be followed, but with \emph{the AI system itself} providing feedback using the constitution as reference. For instance, Anthropic uses a constitution, which is publicly available, as a key part of their training process \cite{claudeconstitution}. Additionally, a set of behavioral instructions is injected at inference time. The problem of how to democratically generate constitutional principles remains, and some research has explicitly proposed using community governance based approaches for this purpose.

The Collective Constitutional AI project~\cite{huang2024collective}, with involvement from Anthropic, incorporated public input to the constitutional AI approach by allowing participants to suggest and vote on constitutional principles, finding that a model trained on publicly sourced principles showed reduced social bias compared to a baseline trained on standard constitutional principles. Research from Google Deepmind explored the use of LLMs to generate consensus statements across diverse viewpoints, testing a range of social welfare functions as models for users' preferences~\cite{bakker2022fine}. OpenAI's Democratic Inputs to AI initiative~\cite{openai2024democratic} funded similar experiments in collective decision-making for model behavior. These experiments highlighted challenges inherent in the community-based approach, such as frequent shifts in public opinion, bias in participant selection, and difficulty aggregating opinions on polarizing issues. 

In practice, efforts at democratizing the governance of AI alignment do not seem to have been meaningfully adopted, although there is clear evidence that major AI players have explored these approaches. While it is difficult to characterize the exact barriers to adoption of these approaches, it does not seem that there are strong incentives for AI companies to decentralize control of their models. Indeed, even as many AI companies have attempted to adopt novel ``prosocial'' corporate governance mechanisms, they have proven vulnerable to pressure to prioritize profits from large centralized stakeholders that control necessary (e.g., compute) infrastructure \cite{harvardlawreview2025amoral}. Similar tensions have been navigated in the blockchain setting, with the inherently decentralized nature of these systems making them potentially more conducive to community governance.

\begin{keyInsight}[Governing AI alignment decisions]
Community governance for AI alignment has been explored but not put into practice. Barriers to adoption include the difficulty of effectively aggregating values across varied stakeholders and misalignment of decentralized governance with the centralized structure of AI companies.
\end{keyInsight}

\subsection{Decentralized Autonomous Organizations}

Blockchain-based community governance is synonymous with Decentralized Autonomous Organizations (DAOs), communities organized around smart contracts and governed through voting. DAOs have operated multi-billion dollar protocols for over a decade~\cite{ding2023survey}, and developed a toolkit of mechanisms for community governance, making them a natural starting point for considering what community governance of AI could look like.

Not all DAO mechanisms are blockchain-specific, though often the transparency and programmability of blockchains are key to their functioning. The main governance mechanism employed by DAOs is token-weighted voting, in which participants' votes are weighted by the quantity of tokens they hold as recorded on chain. This approach intuitively attempts to allocate more governance power to participants with a larger stake in the system. However it is widely recognized that token-weighted voting is inherently plutocratic, as wealth concentration drives power concentration~\cite{fritsch2024analyzing, fabrega2025vbe}. This observation has led DAOs to explore novel approaches to distributing voting power. Examples include quadratic voting, which enfranchises small holders by making voting power proportional to the square root of tokens~\cite{buterin2019flexible}; conviction voting, in which voting power accumulates over time held, incentivizing long-term commitment but preventing quick pivots~\cite{zargham2020economic}; and delegation, which allows users to assign their voting power to other participants representatives, increasing participation but potentially creating voting whales~\cite{fabrega2025vbe}. Voting in DAOs is generally public, in keeping with their broader ethos of transparency, although there has been recent movement towards providing private voting as an option \cite{shutter2025shieldedvoting}, driven in part by concerns over peer pressure, bribery, and coercion. Care is needed, however, as it has been shown that \textit{naively} adapting established private voting solutions to the token-weighted setting can enable attacks on voter privacy \cite{breckenridge2025b}.

It is worth noting that none of these mechanisms necessarily require a blockchain to implement. Indeed, the principal promise of DAO-based governance lies in \emph{the combination of} these mechanisms with the transparency of on-chain decision-making, and the on-chain enforcement of organizational rules through the underlying smart contracts, enabling community governance without reliance on a centralized, trusted authority. Results of votes can be enforced automatically, although DAO treasuries are often managed via multi-signature wallets with trusted signers and many DAOs have security councils (multi-signature wallets with powers to overturn votes) to address suspected fraud or capture~\cite{feichtinger2024attacks}. In practice, blockchain systems tend to mix on-chain and off-chain approaches to governance and voting~\cite{sharma2024unpacking}. A common pattern is use of the voting platform Snapshot for transaction-fee-free off-chain voting combined with governor contracts for on-chain execution~\cite{snapshot2024}. Nonetheless, Governance spans from fully on-chain (Tezos \cite{tezosgovernance}, Polkadot \cite{polkadotgovernance}), in which the results of votes are tightly coupled with on-chain code execution via smart contracts, to off-chain (Bitcoin, Ethereum) in which votes may be used for coordination but results are not automatically enforced \cite{vitalik2017notes, zamfir2017against}.

Some blockchain protocols for decentralized AI already use or are exploring DAOs for governance. Bittensor uses token-weighted voting for subnet governance and incentive distribution across its distributed training network~\cite{bittensor2024governance}. Modulus Labs~\cite{modulus2023zkml} and Giza~\cite{giza2024verifiable} are exploring DAO structures for managing ZK-verified inference networks, although governance remains limited to technical network and protocol parameters rather than the models themselves.

\subsection{DAOs for AI Development}

Limited work has explored DAO-based approaches for governing AI model development. One democratic-inputs-to-AI project, Inclusive.AI \cite{sharma2024inclusive}, used DAO mechanisms to engage underserved populations, exploring how token distribution and voting rules affected outcomes and perceptions of fairness. However, this and similar efforts have not meaningfully explored how the central properties of blockchain, transparency, immutability, and autonomous enforcement, might benefit AI governance beyond serving as a vehicle for voting.

\paragraph{Transparency and provenance.} Blockchain provides a tamper-proof record of governance decisions. For AI development, this could mean maintaining an immutable, auditable history of what principles a model was aligned to and how these evolved over time. This transparency could build public confidence and facilitate external auditing, similar to how blockchains have been used to track provenance in supply chains and other domains~\cite{queiroz2020blockchain, kshetri2018scm}.

\paragraph{Enforcement.} The more distinctive potential contribution of blockchain lies in enforcement. On-chain governance benefits from verifiable smart contract execution; votes translate directly into code changes. For off-chain AI models, this link is broken. However, blockchain could still enable enforcement through economic incentives: in a stake-based system, developers who fail to submit proof that a model was trained according to governance decisions could be \emph{slashed}. This requires the ability to prove and verify claims about model training, which remains an active area of research \cite{jia2021proof, brundage2020toward, abbaszadeh2024zero, verde2020, peng2025survey}. Verifiable training represents a path toward programmatic enforcement of community decisions over AI development, though existing capabilities lag far behind what would be required to apply these approaches to state-of-the-art models.

\paragraph{Pluralistic governance infrastructure.} DAOs could also enable a flexible definition of what community a model is aligned relative to. Different communities could govern different model variants while sharing underlying blockchain infrastructure, supporting the view that alignment should be pluralistic rather than converging on universal values~\cite{sorensen2024roadmap}. The availability of performant open-source base models that can be fine-tuned makes this increasingly realistic. However, this raises persistent challenges around community membership: the permissionless nature of public blockchains creates vulnerability to adversarial participation~\cite{douceur2002sybil}, and Sybil-resistance mechanisms remain imperfect~\cite{ford2020identity, maram2021candid}. There is an intriguing intersection here, as identity systems could also serve the purpose of distinguishing humans from AI. OpenAI's CEO Sam Altman is also co-founder of Worldcoin \cite{worldcoin}---now World~\cite{hays2026tinder}---a blockchain-based identity system, demonstrating a confluence of AI and blockchain interests.

\subsection{Open Problems and Challenges}

The challenges facing DAO-based AI governance span both general DAO limitations and AI-specific concerns.

\paragraph{DAO governance limitations.} Even for on-chain protocols, DAO governance faces significant challenges: votes can be bought, delegation concentrates power among the already prominent and wealthy, and low voter participation allows minority control~\cite{feichtinger2024attacks}. As noted above, identity verification to prevent Sybil attacks conflicts with the privacy norms of blockchain communities, and while research results address this tension~\cite{maram2021candid}, existing solutions remain imperfect. These problems would carry over to any DAO-based approach to AI governance.

\paragraph{AI-specific challenges.} Several challenges are particular to governing AI systems. The technical complexity of model-development decisions may discourage participation from non-experts, concentrating effective power among a technical minority. The enforcement gap between on-chain governance and off-chain models remains substantial, without mature verifiable training, governance decisions lack teeth. 

\begin{researchQuestion}
How could on-chain governance of AI model training be enforced? What advances in verifiable training would facilitate this enforcement?
\end{researchQuestion}

\paragraph{Transparency-privacy tension.} The transparency that makes blockchain valuable for accountability conflicts with legitimate needs for privacy. Publishing model weights enables beneficial research but also malicious use. Organizations may resist governance transparency for competitive reasons. Resolving this tension remains an open problem.

\begin{researchQuestion}
How can we balance the transparency necessary for governance with privacy concerns in AI model development?
\end{researchQuestion}

\paragraph{Stakeholder identification and power allocation.} Even setting aside implementation challenges, fundamental design questions remain unresolved. Which stakeholders should participate? Model users, technical contributors, safety researchers, capital providers, and affected communities all might have stake in various parts of the development process. How should power be distributed among them? Mechanisms from DAO governance (quadratic voting, delegation, conviction voting) offer options, but their effectiveness for AI governance is unclear \cite{sharma2024inclusive}.

\begin{researchQuestion}
What stakeholders should be engaged in an AI governance DAO and how should power be allocated among them?
\end{researchQuestion}

\paragraph{Adoption barriers.} Finally, unclear legal liability for collectively-governed systems~\cite{reyes2020uncorporate} may deter adoption, and the history of governance in AI suggests that incumbent AI developers have limited commercial incentive to cede control absent regulatory pressure or competitive advantage.

\section{Blockchains for AI Execution Integrity}
\label{sec:block4ai-integrity} 

The proliferation of digital services has fundamentally changed how we interact with everyday life. 
Traditional processes, from payment systems to insurance claims, are increasingly being replaced by online platforms. 
Meanwhile, social networks have created entirely new forms of digital interaction.
However, this digital transformation comes with significant concerns: these services operate under centralized control, with providers often exercising power without adequate accountability~\cite{rose2018meaning, stiglitz2002transparency}. 
While users can seek recourse through external jurisdictions, this process is both expensive and time-consuming, often failing to prevent harm before it occurs~\cite{ftc2025censorship, findlaw2023socialmedia, manne2024netneutrality, algocensorship2024}. 
Notable examples include the censorship of content by social media platforms~\cite{wong2023meta, ang2024governance, ungless2024experiences, jaidka2023silenced}, inadequate handling of copyright claims~\cite{rahman2023enforcing}, and the bias of insurance companies against low-income patients~\cite{hoagland2024social}.

A promising path to an automatic alternative, not subject to human bias, is to use \emph{blockchains}, which are decentralized and censorship-resistant platforms. 
As a first approximation, a blockchain can be viewed as a trusted machine, operated by a large set of servers that collectively agree on its state and progress. 

Blockchains were originally introduced for currency transactions~\cite{nakamoto2008bitcoin} and evolved to enforce simple conditions~\cite{bartoletti2017empirical} phrased in so-called \emph{smart contracts}. 
However, smart contracts lack the contextual understanding and interpretative capabilities that would enable reasoning about complex, ambiguous real-world scenarios.
Machine learning~(ML) algorithms, particularly Large Language Models~(LLMs)~\cite{vaswani2017attention, achiam2023gpt, touvron2023llama}, exhibit exactly these capabilities. 
However, implementing computationally intensive~ML tasks within a smart contract introduces prohibitive complexity, since blockchain systems require all their servers to replicate all computations.

\begin{figure}
    \centering
    \includegraphics[width=1\linewidth]{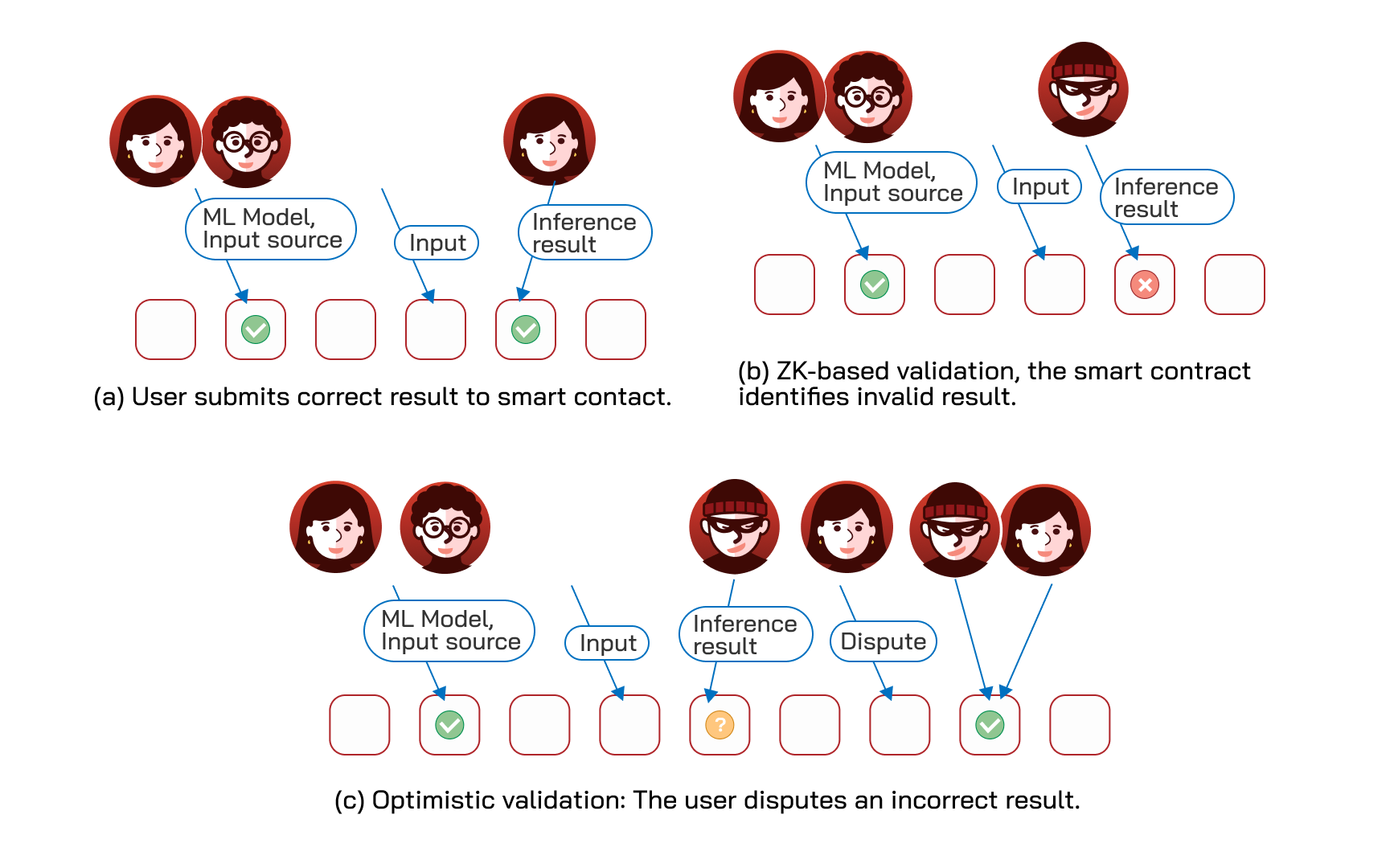}
    \caption{Validation of delegated ML workloads.}
    \label{fig:integrity-correct}
\end{figure}

The blockchain can instead serve as an arbitrator with a smart contract running on it, as follows (and see Figure~\ref{fig:integrity-correct}). 
First, the parties inform the smart contract which ML model they intend to use for inference and which data source they intend to use.  
Subsequently, the input data becomes available: This could be data from either of the users or from another, external source via an oracle. 
See Section~\ref{sec:secure_plumbing} on how to secure the data pipeline. 
Then, one of the parties delivers the computation result to the smart contract, which then validates it. 
The smart contract can use the verified result to take action, such as transferring funds, and users can take actions outside the blockchain based on the result. 
Importantly, if a party delivers the wrong result to the smart contract, it should identify the misbehavior and penalize that party. 

The question is thus how to design the smart contract to enforce this behavior. 
The basic idea is to \emph{delegate} the computationally-intensive computations to external principals, and use the blockchain to enforce the integrity of the result. 
Rather than placing the entire ML model on chain, the users place only a cryptographic commitment, and deposit collateral to guarantee they will submit the correct result once input is available. 

The smart contract should validate the result, although by assumption the problem is beyond its capabilities. 

In this section, we discuss multiple approaches for  overcoming the blockchain's limitations, including trusted execution environments, optimistic execution-based techniques, zero-knowledge proofs, and other emerging techniques. Each class of methods comes with its own advantages and disadvantages. 
\begin{keyInsight}
Rapid progress in recent years resulted in the formation of multiple distinct approaches to validating delegated ML computations taking advantage of blockchain infrastructure, with clear tradeoffs between them. 
While none is a panacea, some of them are directly applicable. 
And the holy grail could be a novel combination that takes the best of all worlds. 
\end{keyInsight}

        \subsection{Trusted Execution Environments}
        \label{sec:tee-for-ai}

First, a highly efficient approach to validate ML computations is using \emph{Trusted Execution Environments}~(\emph{TEEs}, see Section~\ref{subsec:trusted_computers})~\cite{eigencloud2025verifiable}. 
The blockchain serves as a public record for commitments to both the AI model and its inputs. 
Subsequently, any principal can compute the result using \emph{trusted hardware}, and place the result along with a cryptographic signature from the hardware, proving the integrity of the computation. 
The smart contract, as well as any third party, can validate the TEE computation by verifying the signature. 

However, reliance on trusted hardware entails assumptions on the operating party, and this approach is suitable only in scenarios with such a trust structure. 
To avoid this assumption, two alternatives are being explored. 

        \subsection{Optimistic Execution Delegation}

The so-called \emph{optimistic approach} has principals deposit collateral in advance, and execute tasks on demand, placing only the result on the blockchain. 
However, this result is initially considered non-final, and for a limited duration other principals can dispute it, claiming another result is the correct one. 
Since either party in the dispute could be dishonest, the smart contract must arbitrate, penalizing the principal who placed the invalid output. 
This hinges on the input being available to all the principals, so they can validate the execution, and for the principals to be able to prove to the smart contract they used the correct inputs. 
Consider, for example, the settlement of a prediction market. 
If the result can be derived from a newspaper headline, that headline is placed on chain by an oracle, making it trivially availabe. 
But if the result calls for, e.g., image analysis~\cite{polymarket2025zalenski}, its size is too large for a blockchain and the oracle should instead make it available via an external service~\cite{benet2014ipfs}. 

The smart contract cannot execute the full computation, therefore the arbitration is done efficiently as follows. 
After each step of the computation, each principal stores (locally) a commitment to the state of its memory. 
At the end of the computation, it computes a commitment to its entire trace, and submit that to the smart contract. 
The smart contract then asks both for the middle state, with proof it matches their commitments. 
If they match, the search restarts for the right half of the computation. 
If they don't, it starts for the left side. 
Eventually, the contract reaches a point where there is a single disputed instruction, where both parties agree on the state before it but not after it. 
The smart contract can efficiently confirm the correct execution of this single instruction, exposing the dishonest party. 
The dishonest party is thus penalized, and the other result is accepted. 

Such optimistic protocols are deployed in systems like Arbitrum~\cite{kalodner2018arbitrum} and Optimism~\cite{optimism2019optimism}. 
They implement so-called \emph{optimistic rollups} to delegate numerous simple computations off-chain. 
They achieve this by employing virtual machines that emulate physical CPU architectures.

However, this approach has so far not been applicable for ML computations, arguably for the following reason. 
Both systems leverage a \emph{Merklized} memory structure for state management. 
This structure allows principals to cryptographically prove to the smart contracts the differences between states with only logarithmic overhead in memory size. 
But each memory write during the computation execution requires a logarithmic overhead to update the Merkle commitment, not suitable for high-complexity ML computations. 

Recent years are seeing a rush for protocols with alternative approaches to optimistically validate ML computations.
Agatha~\cite{zheng2021agatha} relies on fixed-size circuits, which requires all execution paths to be pre-unrolled. 
While highly efficient, this fixed circuit limits expressiveness, prohibiting dynamic, data-dependent computation. 
Verde~\cite{verde2020}, is another optimistic protocol that assumes fixed size circuits.
Unlike Agatha, Verde divides ML inference or training into large computation units and uses a refereed delegation to resolve disputes.
It assumes that referees hold the entire program and can compute large programs such as matrix multiplications, which makes it not directly suitable for smart contracts.

OPML~\cite{conway2024opml} employs an optimistic MIPS-like virtual machine; to minimize computational overhead, it partitions programs into small units, pinpointing the disputed unit before compiling it into low-level commands for verification; this allows for parallel evaluation of decoupled computation units rather than local parallelization, such as matrix multiplications.
However, each memory write still requires a logarithmic overhead due to the Merklized state management, and the MIPS architecture does not support parallelization, which limits scalability, particularly for large matrix operations.

Arbigraph~\cite{mirkin2025arbigraph} utilizes a dual-graph data structure to achieve both Turing completeness and constant-time memory access, while allowing for parallel execution. 
Its nodes commit to a \emph{symbolic graph} that describes the computation, including both data dependencies and control flow. 
Then each node forms a so-called \emph{execution graph} and commits to its result. 

TAO~\cite{yao2026tao} addresses floating-point nondeterminism in optimistic ML verification by committing to the model weights, operator-level graph, and calibrated tolerance profiles, while preserving native GPU execution in the common case.
If challenged, it runs a dispute game over the operator graph, localizing disagreement to a single operator that is adjudicated using either IEEE-754 theoretical rounding error bounds or committee re-execution against empirical thresholds.

\paragraph{Determinism and numerical semantics} 
Optimistic protocols rely, for their dispute, on all executions being deterministic: that is, two correct executions should always result in an identical output. 
For VM-based rollups, this predicate is usually bit-exact: the instruction set defines a deterministic transition function, and the dispute game only needs to find the first instruction where two committed traces diverge. 
ML workloads complicate this abstraction. First, inference is dominated by tensor operators, where a single high-level operation may encapsulate many parallel reductions. 
Second, floating-point arithmetic is rounded and non-associative, so legal differences in reduction order, kernel selection, scheduling, atomics, autotuning, or operator fusion can produce slightly different tensors across hardware, and sometimes even across repeated executions on the same platform~\cite{schlogl2023causes}. 
Randomness introduced by sampling-based decoding can be made part of the committed input or derived from a public pseudorandom seed~\cite{jooybar2013gpudet, verde2020}; the harder issue is numerical nondeterminism in the accelerator stack itself.

Existing approaches mostly restore an exact predicate by constraining execution. 
Architectural mechanisms such as GPUDet~\cite{jooybar2013gpudet} enforce deterministic interaction among GPU threads. 
Verde's RepOps library controls the order of floating-point operations so that ML programs can be bitwise reproduced across heterogeneous hardware~\cite{verde2020}. 
For training, another line of work obtains exact replay by computing in higher precision, rounding after intermediate steps, and logging selected rounding decisions~\cite{srivastava2024optimistic}. 
A more radical alternative is to avoid native floating-point semantics in the verified computation: many zkML systems use fixed-point, integer, or quantized representations to make the computation proof-friendly, at the cost of changing the numerical model; see Section~\ref{sec:zkp-components}. 
 
TAO~\cite{yao2026tao} instead adopts a tolerance-aware numerical semantics for optimistic ML verification.
Rather than requiring heterogeneous accelerators to reproduce a single bit-exact tensor, TAO preserves native GPU execution and commits to the admissible numerical behavior of each operator.
This design reflects the fact that modern ML deployments often rely on diverse accelerator generations, vendor-provided kernels, and implementation-dependent accumulation orders for performance and availability~\cite{shanmugavelu2024impacts,xie2025revealing}.
A claimed tensor is accepted only if its deviation lies within a committed operator-specific acceptance region.
These regions combine portable IEEE-754 rounding-error bounds~\cite{goldberg1991every,higham2002accuracy} with empirical percentile thresholds calibrated across admitted hardware/software configurations. This differs from approaches that enforce bitwise reproducibility through canonical summation orders or reproducible accumulators~\cite{demmel2013fast,collange2015numerical,ahrens2020algorithms}.
TAO’s key observation is that benign floating-point deviations in ML inference are not arbitrary: at the granularity of PyTorch operators, cross-device behavior can be systematically profiled and statistically predicted. It uses these empirical profiles to recognize ordinary heterogeneous execution efficiently, while retaining conservative IEEE-754 bounds and committee adjudication for out-of-profile or adversarial deviations.

\begin{researchQuestion}
While previous work has demonstrated the viability of several distinct optimistic approaches, which approach is best for deployment with minimal overhead compared to a non-validated GPU-optimized execution? 
\end{researchQuestion}

\paragraph{Training} 
Validating model training is a distinct but similar challenge. 
Here, the goal is to ensure a trained model was indeed produced by a trustworthy process, with trustworthy training data. 
While superficially similar to inference, training workloads exacerbate the challenges of optimistic protocols. 
They are less amenable to quantization, their complexity in terms of both space and computation is orders of magnitude higher than inference, and so is the amount of input data. 

\begin{researchQuestion}
Which optimistic approaches and algorithms are best suitable for ML training validation, enabling a high degree of parallelization and incurring minimal overhead? 
\end{researchQuestion}

\subsection{Zero-Knowledge Proofs}
\label{sec:zkp-for-ai}

\newcommand{\ZK}{{ZK}\xspace}
\newcommand{\ZKP}{ZKP\xspace}
\newcommand{\ZKPs}{ZKPs\xspace}
\newcommand{\SNARK}{SNARK\xspace}
\newcommand{\SNARKs}{SNARKs\xspace}
\newcommand{\STARK}{STARK\xspace}
\newcommand{\zkml}{zkML\xspace}
\newcommand{\llm}{LLM\xspace}
\newcommand{\llms}{LLMs\xspace}
\newcommand{\cnn}{CNN\xspace}
\newcommand{\cnns}{{CNNs}\xspace}
\newcommand{\dnn}{{DNN}\xspace}
\newcommand{\sgd}{{SGD}\xspace}
\newcommand{\zkpot}{\textsf{zkPoT}\xspace} %
\newcommand{\model}{f_{\theta}}
\newcommand{\Commitment}{\mathsf{Com}}
\newcommand{\Proof}{\pi}

\newcommand{\Takeaway}[2]{\noindent\textbf{Key Takeaway #1: #2}\par}
\newcommand{\OpenQ}[2]{\noindent\textbf{Question #1: #2}\par}

Zero-knowledge proofs (ZKPs)~(Section~\ref{sec:zkps_and_MPC}) offer a cryptographic approach to \emph{trust-minimized AI}: 
A prover can convince a verifier using a ZKP that an AI computation (e.g., model inference) was executed correctly, while revealing little (or nothing) beyond.
For example, the proof can hide the model parameters and even keep the input and output private, disclosing only that the reported prediction satisfies a specified predicate (e.g., ``the top-score label is \texttt{cat}'' or ``the prediction output is positive'').
A canonical pattern is \emph{verifiable off-chain inference}: the prover holds a model with parameters $\theta$ (e.g., neural network weights) and an inference input $x$ (e.g., an image, a feature vector, or a text prompt), computes $y \gets \model(x)$, and then produces a succinct proof $\Proof$ of correctness.
Concretely, the prover publishes commitment of the model $\Commitment(\theta)$, commitment of the input $\Commitment(x)$, outputs $y$, and provides a proof $\Proof$ attesting that {there exist openings} of these commitments consistent with $y=\model(x)$ under the inference protocol; the verifier checks $\Proof$ against the public values $(\Commitment(\theta),\Commitment(x),y)$.

Because many modern ZK proof systems admit efficient verification, this check can be embedded in a smart contract on blockchains: the contract can accept $y$ as a trustworthy off-chain AI output and then condition subsequent on-chain logic on the verification result (e.g., release payment to the prover, settle a prediction market, or update an oracle feed if the proof verifies, and otherwise revert or ignore the update). 
For example, an on-chain prediction market could use an off-chain classifier to label an event outcome $y\in\{\texttt{YES},\texttt{NO}\}$ from evidence $x$, and only finalize payouts when presented with a valid ZKP that the published label was computed by the committed model.

Several applications illustrate this pattern in practice.
In DeFi, Qiro integrates EZKL into a verifiable underwriting pipeline: credit-risk models are executed off-chain over borrower data, and accompanied by ZKPs that are posted on-chain so that investors can verify that underwriting scores, loss estimates, and approval decisions were computed by the specified model~\cite{qiro}.
Similarly, EZKL's case study with Sentiment Protocol explores verifiable risk management for DeFi lending, where market-volatility forecasts can be used to adjust protocol parameters such as loan-to-value ratios without requiring users to trust an opaque off-chain risk engine~\cite{sentiment}.
More recently, in a NovaNet demo, an agent runs an ML model to decide whether to authorize a payment, Jolt Atlas produces a zkML proof, and a proof hash is committed on-chain for auditability~\cite{benno2026jolt,novanet}.
These examples suggest that zkML is most compelling when an ML output controls a high-value state transition, but where re-executing the model directly would be prohibitively expensive.

\subsubsection{ZK for ML inference: architecture-aware systems}
\label{sec:zkp-ml-inference}
A large body of zkML work targets \emph{inference} and asks: how can we reduce prover cost by exploiting structure in a given architecture family (e.g., \cnns or transformers), while still producing succinct proofs that are cheap to verify?
Broadly, these systems either (1) design specialized ZK protocols/gadgets for the dominant operators in a target
architecture, or (2) provide compilers that translate models written in standard ML frameworks into circuits.

\paragraph{\cnns.}
For convolutional networks, zkCNN~\cite{liu2021zkcnn} introduces specialized protocols for fast Fourier transforms (FFTs) and convolutions~\cite{krizhevsky2012imagenet}, aiming to make the prover's asymptotic work {linear} in the size of convolution layers, which is an attempt towards reducing proof generation time of \cnn.
More recently, VerfCNN~\cite{qu2025verfcnn} targets {multi-channel} convolutions and aims for {optimal} asymptotic complexity for core \cnn operators, reporting substantial speedups on standard architectures such as VGG-style networks.

Architecture-specific protocols can deliver strong performance on their target operators, but they do not automatically cover the long tail of layers and variants in modern vision stacks; supporting new layers often requires additional protocol engineering, and end-to-end performance can still be dominated by non-linear layers.

\paragraph{Transformers and \llms.}
Transformers~\cite{vaswani2017attention} and \llms introduce different bottlenecks: the attention and normalization layers combine large matrix operations
with costly non-linearities (e.g., \texttt{Softmax}, \texttt{GELU}) and are often deployed under tight latency constraints.
zkLLM~\cite{sun2024zkllm} provides an end-to-end correctness proof for \llm inference with a privacy goal: the proof
can certify correct execution while protecting the model parameters.
At a high level, zkLLM introduces proof components for non-arithmetic operations (via a lookup-style primitive) and a
dedicated proof for the attention mechanism.
zkGPT~\cite{qu2025zkgpt} further investigates a non-interactive ZKP framework for GPT-style inference, proposing circuit optimizations (e.g., constraint fusion and ``circuit squeezing'') to reduce constraint counts and accelerate proving.
DeepProve~\cite{deepprove} is a recent open-source framework focused on neural-network inference with first-class support for end-to-end \llm proving using techniques from zkLLM and zkGPT.

Current approaches still face large prover costs for long context lengths and large hidden dimensions; moreover, non-linear layers and normalization can dominate constraint counts, motivating component-level optimizations discussed in \cref{sec:zkp-components}.
Industry systems such as DeepProve show that proof generation for realistic transformer inference is becoming more practical, but they should still be viewed as early-stage infrastructure rather than a drop-in replacement for ordinary high-throughput \llm serving.

\paragraph{General-purpose zkML compilers.}
Beyond hand-crafted protocols for a single architecture class, several systems pursue general zkML tooling that compiles common ML frameworks into ZKP generation backends.
ZKML~\cite{chen2024zkml} develops an optimizing compiler that translates TensorFlow models into Halo2 circuits, focusing on end-to-end performance optimizations (and developer usability) that make verifiable inference workflows more accessible.
EZKL~\cite{ezkl} is a developer-oriented system that takes computational graphs exported from common ML frameworks, generates ZK-SNARK circuits, and supports verification in constrained environments such as the Ethereum Virtual Machine.
Jolt Atlas~\cite{benno2026jolt} takes a different approach where it extends the Jolt zkVM proving approach directly to tensor operations.
This design targets streaming proof generation for classification, embedding, automated reasoning, and small language models.

Compiler-based approaches improve usability and portability, but typically inherit the cost of a ``generic'' circuit representation; achieving the best performance still requires specialized gadgets for expensive layers.
A useful distinction is that systems such as ZKML and EZKL compile ML graphs into proof-friendly circuits, systems such as Jolt Atlas redesign the proof layer around tensor operations, and systems such as DeepProve specialize the inference engine for \llm-style model architecture.

\begin{table}[t]
  \centering
  \small
  \setlength{\tabcolsep}{4pt}
  \renewcommand{\arraystretch}{1.15}
  \begin{tabular}{p{0.20\columnwidth} p{0.15\columnwidth} p{0.25\columnwidth} p{0.30\columnwidth}}
    \hline
    \textbf{Work} & \textbf{Target} & \textbf{Primary focus} & \textbf{Main Limitations} \\
    \hline
    zkCNN~\cite{liu2021zkcnn}; 
    
    VerfCNN~\cite{qu2025verfcnn}
    & \cnns
    & Architecture-aware protocols for convolution-heavy inference.
    & Strong for convolutional operators, but less general for non-\cnn layers and heterogeneous vision pipelines. \\
    \hline
    zkLLM~\cite{sun2024zkllm}; 
    
    zkGPT~\cite{qu2025zkgpt}; 
    
    DeepProve~\cite{deepprove}
    & Transformers / \llms
    & End-to-end inference proofs and circuit/protocol optimizations for transformers
    & Scaling remains difficult for long contexts, large hidden dimensions, and high-throughput serving. \\
    \hline
    ZKML~\cite{chen2024zkml}; 
    
    EZKL~\cite{ezkl}; 
    
    Jolt Atlas~\cite{benno2026jolt}
    & General neural network inference
    & General tooling for compiling or proving ML computations
    & Genericity improves usability, but can lose performance without specialized gadgets and proof-friendly numeric representations. \\
    \hline
  \end{tabular}
  \caption{Representative zkML inference systems, categorized by target architecture and primary optimization focus.}
  \label{tab:zkml-inference-systems}
\end{table}

\subsubsection{Optimizing proof cost: non-linearities and numeric representations}
\label{sec:zkp-components}

A recurring theme across zkML systems is that the dominant prover costs often come from (1) \emph{non-linear functions} (e.g., activation layers, \texttt{Softmax}/\texttt{ReLU}, normalization), and (2) \emph{numeric representations} that reconcile
real-valued ML arithmetic with finite-field constraints.
Accordingly, a complementary line of work focuses on reusable components (``gadgets'') and proof-friendly representations that can be plugged into end-to-end inference pipelines.

\paragraph{Non-linear function gadgets.}
Lu et al.~\cite{lu2024efficient} propose an efficient and extensible proof framework that targets the bottleneck of
non-linear layers.
Their approach converts complex non-linear relations into a small set of constraints similar to range proofs and then uses enhanced range proofs and lookup proofs as modular building blocks, yielding speedups across both \cnns (e.g., ResNet~\cite{he2016deep}) and transformer models (e.g., GPT-2~\cite{radford2019language}).
Hao et al.~\cite{hao2024scalable} develop a systematic ZK proof framework for common non-linear functions from a table-lookup perspective, introducing building blocks such as digital decomposition and comparison gadgets to reduce overhead for functions like \texttt{ReLU} and \texttt{sigmoid}.

\paragraph{Quantization and fixed-point arithmetic.}
Most zkML deployments avoid floating-point arithmetic inside circuits, instead representing values via fixed-point encodings and/or quantized weights and activations.
This design choice reduces constraint counts and enables tighter range reasoning, but introduces a three-way tension between \emph{model accuracy}, \emph{prover time}, and \emph{soundness of the numeric model} (e.g., approximation error).
A practical open problem is to make these tradeoffs explicit---for example, by proving that inference stayed within agreed-upon ranges and approximation budgets.
EZKL~\cite{ezkl} and DeepProve~\cite{deepprove} explicitly expose this issue in practice: when ML models are converted to ZK circuits, operations are quantized, so the circuit output can differ from ordinary full-precision inference.

\begin{table}[t]
  \centering
  \small
  \setlength{\tabcolsep}{4pt}
  \renewcommand{\arraystretch}{1.15}
  \begin{tabular}{p{0.20\columnwidth} p{0.20\columnwidth} p{0.25\columnwidth} p{0.25\columnwidth}}
    \hline
    \textbf{Work / technique} & \textbf{Targets} & \textbf{Core idea} & \textbf{Typical use} \\
    \hline
    Lu et al.~\cite{lu2024efficient}; Hao et al.~\cite{hao2024scalable}
    & Non-linear layers in \cnns, transformers, and general tensor programs
    & Reduce expensive non-linear checks to range, lookup, decomposition, and comparison-style arguments.
    & Plug-ins or framework-level components to accelerate end-to-end zkML inference. \\
    \hline
    Quantization and fixed-point arithmetic in systems such as EZKL~\cite{ezkl} and DeepProve~\cite{deepprove}
    & Numeric representation in zkML
    & Replace floating-point with fixed-point or quantized values while controlling rounding and approximation errors.
    & Reduce constraints and ZKP generation time, at the cost of a model-fidelity tradeoff. \\
    \hline
  \end{tabular}
  \caption{Component-level optimizations that target common zkML bottlenecks, especially non-linear layers and numeric representations.}
  \label{tab:zkml-components}
\end{table}

\subsubsection{Privacy goals: input, weights, and architecture}
\label{sec:zkml-privacy}

In addition to correctness, zkML systems often aim to protect \emph{sensitive AI artifacts}. The privacy design space is multi-dimensional:
(1) \emph{input privacy} (hiding $x$), (2) \emph{model parameter privacy} (hiding $\theta$), and (3) \emph{architecture privacy} (hiding the structure of $\model$), which may itself encode proprietary information.

\paragraph{Inputs and parameters.}
Input privacy is often ``native'' in ZK: the prover can keep $x$ private as witness while proving a statement about
the output $y$.
Similarly, model parameters $\theta$ can be kept private by committing to $\theta$ (or a hash thereof) and proving correctness relative to that commitment.
For example, zkLLM~\cite{sun2024zkllm} explicitly targets parameter privacy while proving end-to-end \llm inference.
EZKL~\cite{ezkl} also supports several privacy patterns at the application level, such as proving inference for a public model on private data or a private model on public data.

\paragraph{Architecture privacy.}
Many zkML deployments still reveal model architecture, even when weights are hidden.
Architecture-private zkML frameworks~\cite{guo2025architecture} address this gap by hiding architectural details (e.g., \cnn structure) while retaining verifiability, using proof techniques that certify a functional relation without revealing the underlying architecture.

Strengthening privacy (especially architecture privacy) typically increases constraint-system complexity and can reduce opportunities for architecture-specific optimization.
This suggests a fundamental tension between efficiency and stronger model privacy protection.

\begin{table}[t]
  \centering
  \small
  \setlength{\tabcolsep}{4pt}
  \renewcommand{\arraystretch}{1.15}
  \begin{tabular}{p{0.28\columnwidth} p{0.38\columnwidth} p{0.28\columnwidth}}
    \hline
    \textbf{Privacy objective} & \textbf{Common mechanism} & \textbf{Example} \\
    \hline
    Input privacy (hide $x$)
    & Keep $x$ as ZK witness; optionally commit to $x$ for binding.
    & Verifiable private inference; EZKL~\cite{ezkl}. \\
    \hline
    Parameter privacy (hide $\theta$)
    & Commit to $\theta$ (or hash) and prove $y=\model(x)$ w.r.t.\ committed $\theta$.
    & zkLLM~\cite{sun2024zkllm}; EZKL~\cite{ezkl}. \\
    \hline
    Architecture privacy (hide structure of $\model$)
    & Parametrized constraint systems that prove statements about the inference without revealing model architecture.
    & Architecture-private zkML~\cite{guo2025architecture}. \\
    \hline
  \end{tabular}
  \caption{Privacy objectives in zkML and representative mechanisms. Achieving stronger privacy typically increases prover cost and can constrain available optimizations.}
  \label{tab:zkml-privacy}
\end{table}

\subsubsection{ZK proofs for training, provenance, and audits}
\label{sec:zkml-training}

As we mentioned above, compared to inference, proving \emph{training} is substantially harder: Training is long-running, stateful, often randomized, and orders of magnitude more complex. 
Nevertheless, recent \zkpot work begins to formalize and prototype ZK proofs of training, with the longer-term goal of enabling verifiable claims about how a model was produced (e.g., from which data, under what constraints, and with what computation).

\paragraph{\zkpot: proving training dynamics.}
Garg et al.~\cite{garg2023experimenting} formulate the notion of a zero-knowledge proof of training and provide early
measurements that help identify which parts of training dominate proof cost.
Subsequent work (e.g., Kaizen~\cite{abbaszadeh2024zero}) aims to make \zkpot more practical for modern \dnn training by
improving how proofs compose across many optimization steps (e.g., many iterations of \sgd), which is essential for scaling
beyond toy training workloads.

\paragraph{Provenance and audit queries.}
Beyond ``prove you trained the model,'' ZK can enable \emph{trustless audits} of model and data properties.
ZkAudit~\cite{waiwitlikhit2024trustless} proposes a two-stage approach: the model provider first commits to the training
dataset and model weights and produces a proof that these commitments resulted from training; later, auditors can request
evaluations of an arbitrary function $F$ over the committed objects, with additional ZK proofs attesting correct execution of $F$.
ZK techniques can also support proofs of \emph{process constraints} during training. For example,
Confidential-PROFITT~\cite{shamsabadi2022confidential} targets decision trees and produces auditable proofs that a tree was trained
subject to fairness constraints, while keeping both model and training data confidential.

\begin{table}[t]
  \centering
  \small
  \setlength{\tabcolsep}{4pt}
  \renewcommand{\arraystretch}{1.15}
  \begin{tabular}{p{0.22\columnwidth} p{0.38\columnwidth} p{0.34\columnwidth}}
    \hline
    \textbf{Work} & \textbf{Statement / goal} & \textbf{Contribution and boundary} \\
    \hline
    Garg et al.~\cite{garg2023experimenting}; Kaizen~\cite{abbaszadeh2024zero}
    & Prove that a model was produced by following a specified training procedure, including many optimization steps.
    & Formalizes and improves the practicality of \zkpot; scaling to large, randomized, and distributed training pipelines remains difficult. \\
    \hline
    ZkAudit~\cite{waiwitlikhit2024trustless}; Confidential-PROFITT~\cite{shamsabadi2022confidential}
    & Prove provenance or higher-level audit properties over committed data/model artifacts.
    & Enables trustless audit queries and constraint-aware learning (e.g., fairness), but expressiveness and efficiency remain key tradeoffs. \\
    \hline
  \end{tabular}
  \caption{Representative directions beyond inference: proofs of training, provenance, and auditable claims.}
  \label{tab:zkml-training}
\end{table}

\begin{keyInsight}
zkML turns expensive off-chain AI computations into succinct, verifiable claims, but practical deployments depend on co-design across (1) architecture-aware modules, (2) optimized gadgets for non-linear functions and numeric representations, and (3) privacy goals (inputs/weights/architecture).
\end{keyInsight}

\begin{researchQuestion}
How can we reduce zkML proving latency by orders of magnitude so that verifiable inference is viable for real-time applications?
\end{researchQuestion}

\begin{researchQuestion}
What are the most useful and feasible \zkpot / audit statements for modern training pipelines (including randomness and distributed training), and how can we make such claims both privacy-preserving and efficient to produce?
\end{researchQuestion}

\subsection{Emerging Techniques}
In addition to lines of work using TEEs, optimistic execution delegation, and zero-knowledge proofs for verifiable AI computation, the research community has also explored alternative frameworks that largely aim to avoid the high computational cost of cryptographic techniques. 
These include statistical proofs of inference and game-theoretic approaches. 

\subsubsection{Statistical proofs of inference}
        \label{sec:statistical-proofs}

Both optimistic and zero-knowledge approaches impose significant overhead on the verification pipeline: optimistic protocols require a multi-round dispute window before a result is finalized, while ZK systems incur prover times that are orders of magnitude beyond the inference itself.
Statistical interactive proofs~\cite{anchuri2026,commitllm} offer a different operating point in this tradeoff space, with prover overhead on the order of milliseconds and immediate finality, at the cost of probabilistic rather than deterministmic soundness.

The protocol is grounded in a key observation from the neural network representational similarity literature~\cite{klabunde2023survey, kriegeskorte2008rsa, kornblith2019similarity}: \emph{functional dissimilarity implies representational dissimilarity}.
If a prover runs a model whose output distribution diverges sufficiently from the advertised model, the execution traces (the neuron activations across all layers) of the two models must also diverge measurably.
This structural property enables a lightweight verification protocol.
During a one-time setup phase, the model provider publishes a binding commitment to the weights.
At inference time, the prover commits to the full execution trace.
The verifier then samples a random output neuron and traces a path layer by layer back to the input, at each step opening the claimed activations and the corresponding weights to verify local consistency.
Because only a logarithmic number of commitment openings are required per query, the prover's overhead is on the order of milliseconds, and unlike optimistic protocols, the verdict is immediate.

The protocol provides strong guarantees under \emph{other-model soundness}, which formalizes a practical threat: the provider substitutes a different model for the advertised one, for instance to serve a cheaper quantized or distilled variant, or to replace a safety-aligned model with a dealigned alternative~\cite{qi2023safety, sokhansanj2025uncensored}.
\emph{Full soundness}, security against a fully malicious prover who fabricates an arbitrary trace, is harder to achieve and remains an active area of research.
Statistical proofs of inference are thus complementary to ZK and optimistic approaches, and are particularly well-suited to high-frequency, latency-sensitive deployments where the overhead of cryptographic proving is prohibitive~\cite{sun2025svip}.

\begin{researchQuestion}
Can we achieve full soundness for statistical proofs of inference while maintaining low prover overhead and immediate finality?
\end{researchQuestion}

\subsubsection{Game-theoretic approaches} 
An emerging body of work couples coding theory with game theory to obtain trusted computation from a group of untrusted nodes. Motivated by DeML applications, the \emph{game of coding} framework presents fault-tolerant protocols for distributed computation in the presence of some fraction of  adversarial nodes who are rational rather than Byzantine \cite{nodehi2024game,akbarinodehi2025game}.  
The core idea is that a central aggregator (e.g. a smart contract) accepts only computations that lie within a certain threshold of each other, and rewards only accepted entries. 
Hence, adversarial, rational nodes must decide how much they are willing to deviate from the correct answer in order to risk not being paid. 
These methods can handle environments in which the majority of nodes are adversarial \cite{akbarinodehi2025game}, and tasks may be vector-valued \cite{nodehi2026game} or sequential \cite{nodehi2026vista}, such as computing  multiple optimization steps in a ML task. 
Like the statistical approach in \Cref{sec:statistical-proofs}, this game-theoretic approach circumvents the computational bottlenecks of cryptographic verification.
The main tradeoff is that they require some modeling assumptions regarding the utility function of  adversarial rational nodes; however, this requirement can be mitigated  \cite{akbarinodehi2025game-adversary}. 

The game of coding approach has been used to validate training results for small deep learning models \cite{nodehi2026vista}. 
For vector-valued outputs (like trained model parameters), results are accepted if the vectors lie with an $\ell_2$ distance of one another \cite{nodehi2026game,nodehi2026vista}. 
It will be useful to understand how such an acceptance criterion affects the utility of large ML models under adversarially-selected deviations. For instance, prior work has suggested that models can be close in their parameters (e.g., in $\ell_2$ distance), but have different functional properties \cite{yang2021taxonomizing,kornblith2019similarity}. 
One potential risk in DeML applications is that if nodes are both rational \emph{and} adversarial, they may seek out a parameter vector that is close in $\ell_2$ distance (or any metric of choice), while still exhibiting malicious behavior (e.g. targeted incorrect predictions) relative to the ``true'' parameter vector.
An interesting future research direction could  evaluate the severity of this risk in realistic ML settings, and mitigate it as needed with adaptations to the game of coding framework.

\section{Securing the Plumbing of AI Systems}
\label{sec:blockchainforAI}
\label{sec:secure_plumbing}

\subsection{Securing Training Pipelines}
\label{subsec:securing_training}

We just considered the security issues that arise in inference involving a single user. We now turn our attention to the setting of model training.

When a single organization trains (or fine-tunes) a model using data it trusts (and/or a pretrained model it trusts), there's generally no immediate privacy or integrity concern. Instead it is when \textit{more than one user or organization} participates in training a model that especially complex security challenges surface---specifically, when there is a \textit{multiplicity of data sources}.

\subsubsection{Collaborative model training} 

There are many situations in which institutions can benefit from pooling data to train a model. Here's an example.

\begin{figure}[h!]
    \centering
    \includegraphics[width=0.8\linewidth]{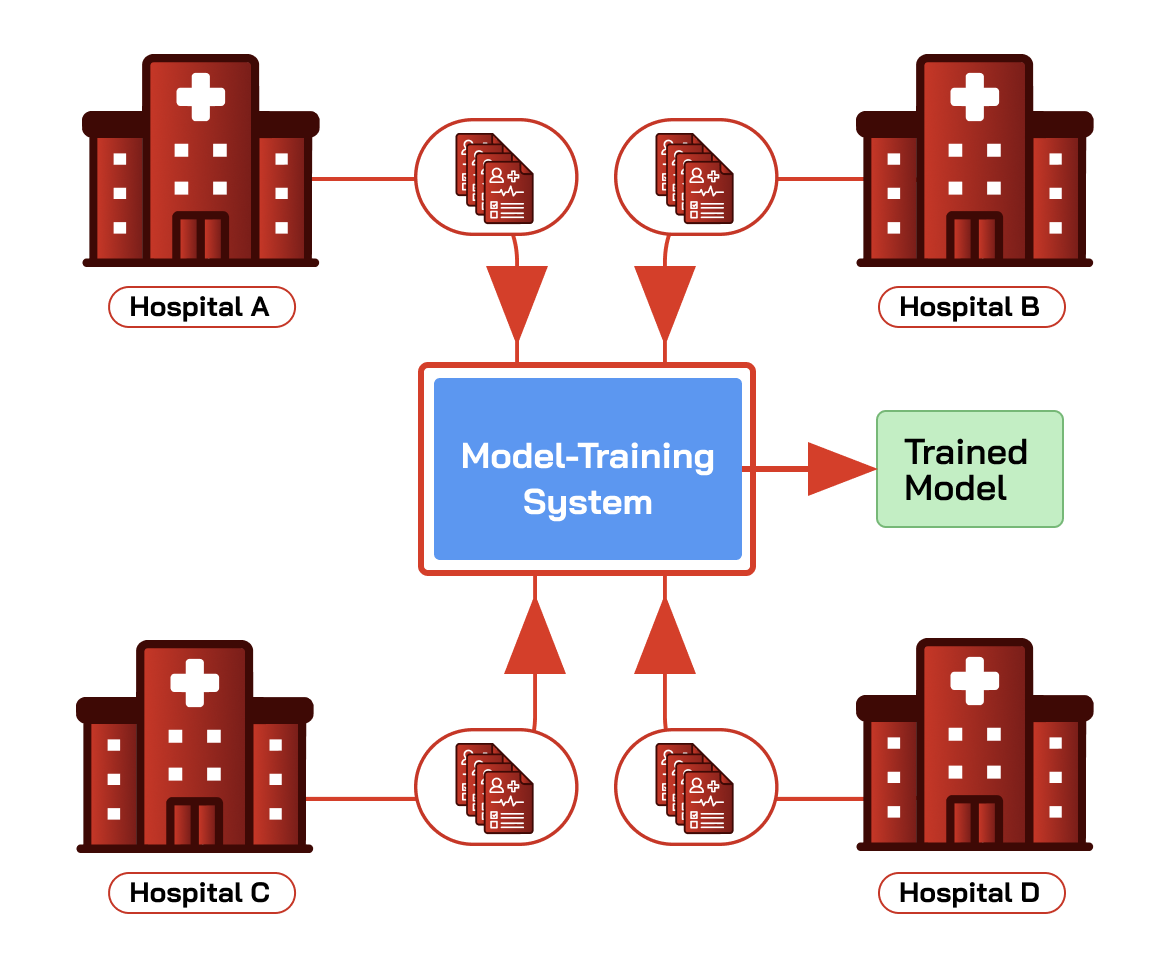}
    \caption{Hospitals collaboratively train a medical diagnostics model on the electronic health records (EHRs) of their respective patients. }
    \label{fig:hospitals}
\end{figure}

\begin{example}[Training a medical-diagnostics model] A coalition of medical providers (hospitals A, B, C, and D) plan to train a medical diagnostics model collaboratively using the electronic health records (EHRs) of their patients, as shown in~\Cref{fig:hospitals}. They recognize that combining records across institutions promises broader patient population coverage and improved diagnostic accuracy. 
\label{ex:medical_model}
\end{example}

There are many similar examples in which pooling data across organizations promises to bring mutual benefits. Financial institutions might wish to pool data about financial crimes in order to train a model to detect anomalies in financial records, corporations might wish to train intrusion-detection models on their network data to detect cyberattacks, and so forth.

In all of these cases, though, data privacy is of paramount importance. Financial data and corporate network-security data are highly sensitive. The patient records (EHRs) of~\Cref{ex:medical_model} are as well. 
Regulations such as HIPAA place tight restrictions on data sharing, making it unclear where the training can be executed. The hospitals in this example may wish or need to avoid sharing data directly with one another or even with a third party that performs the training. 

\paragraph{Federated learning:} It is exactly this scenario that has inspired an approach to model training called \textit{federated learning}~\cite{mcmahan2017federated}. It involves the training of a global model across the data of multiple participants, without participants needing to furnish their data in raw form to the model-training environment. Briefly, the training environment initializes the global model and distributes it to participants. Each participant trains locally on private data and returns model updates to the training environment, which computes an update to the global model. 

This model of local aggregation and centralized updating fits~\Cref{ex:training} well: hospitals can train a global model without exposing their data to one another.

Despite some production use, however---most notably in predictive text on mobile devices~\cite{hard2018federated}---federated learning has marked downsides that have limited its use~\cite{li2023survey,wen2023survey}. Federated learning does not ensure the integrity of contributed data or of computation. Even if participants are honest---providing authentic data and correctly executing the training protocol---federated learning incurs significant communication overhead~\cite{almanifi2023communication,zhao2023communication}, while network and coordination latency can dominate wall-clock time and model accuracy is lower than with centralized training. Additionally, malicious participants can  effectively poison or insert backdoors into models~\cite{fang2020local,bhagoji2019analyzing,bagdasaryan2020backdoor}.

\paragraph{Trusted computing for collaborative model training:} The limitations of federated learning provide a strong incentive for a simpler alternative: Centralized training with a trusted computer (TEE). If the training environment operates in a trusted confidential computing environment, it can theoretically safely \textit{pool all raw data}, eliminating the need for the complexity, computational overhead, and degraded model performance of federated learning. Training in this way can expose \textit{only the trained model}, while preserving the privacy of the training data.

As an alternative workflow, the collaborative model-training system of~\Cref{fig:hospitals} can execute within a trusted computing environment, depicted by the red box. The trusted computing environment ingests inputs from the participants (hospitals) over secure (encrypted) channels, and outputs a trained model only.  As training is centralized, the only overhead incurred is use of the trusted computing environment.

Another benefit of using a trusted computing environment is that the output model can be accompanied by an attestation to the model provenance: what entities contributed data and how the model was trained. 

While trusted computing environments are a promising
alternative to federated learning, they do come with important provisos, discussed in~\Cref{subsubsec:TEEs}. These include side-channel vulnerabilities---a recurring security problem---as well as high resource overhead for I/O-intensive workloads, requiring special model training protocols for efficiency~\cite{lee2025characterization}.

Today, organizations instead pool data in HIPAA-compliant clouds with standard security measures---virtual private cloud isolation, identity and access management, logging, and encryption at rest/in transit, etc. They also use non-technical safeguards, such as HIPAA governance and data-use agreements~\cite{Denny2019,Haendel2021}. This approach aims at HIPAA-compliance, which is not identical with strong data security. Moreover, while this approach carries the performance and model-accuracy benefits of centralized training, it requires trust in the cloud provider. As they mature and their performance and security improve, trusted computing environments can play a complementary role, reducing the risk of breaches and providing high-trust attestation to the provenance of models.

\subsubsection{Model training over private-web data}

ML is approaching a critical data bottleneck. There's only one World Wide Web (WWW), and ML practitioners are approaching the limits of its publicly accessible data, with projections estimating that WWW text data will be exhausted between 2025-2030 \cite{villalobos2024run, villalobos2024limits}. For this reason, there's an increasing reliance on synthetic data generated by ML models, with estimates suggesting that 80\% of training data will be synthetic by 2028 \cite{musk2025data, cio2025synthetic}. This shift carries the risk of ``self-poisoning`` or ``model collapse``---a phenomenon in which models trained on synthetic data from other models experience progressive performance degradation across generations \cite{shumailov2024collapse, alemohammad2024mad}. More importantly, synthetic data does not address the fundamental data-shortage problem. For all its benefits in model training, it does not expand data coverage beyond existing domains \cite{shumailov2024collapse, scientific2024poison}.

There is a vast source of untapped data, however, in the \textit{private web}---the part of the web that is walled off from web scraping. The private web includes systems and data that ordinary users interact with every day: e-mail, health data, financial records, and much more. It is estimated to be \textit{two orders of magnitude larger} than the surface public web~\cite{bergman2001white,rai2020bibliometric}.

Today, however, AI practitioners have limited access to private-web data. Large corporations can source such data in-house (often with explicit user permission and/or legal oversight). But this still means that private-web data remains highly \textit{siloed}. 

Consider the following example.

\begin{figure}[h!]
    \centering
    \includegraphics[width=0.6\linewidth]{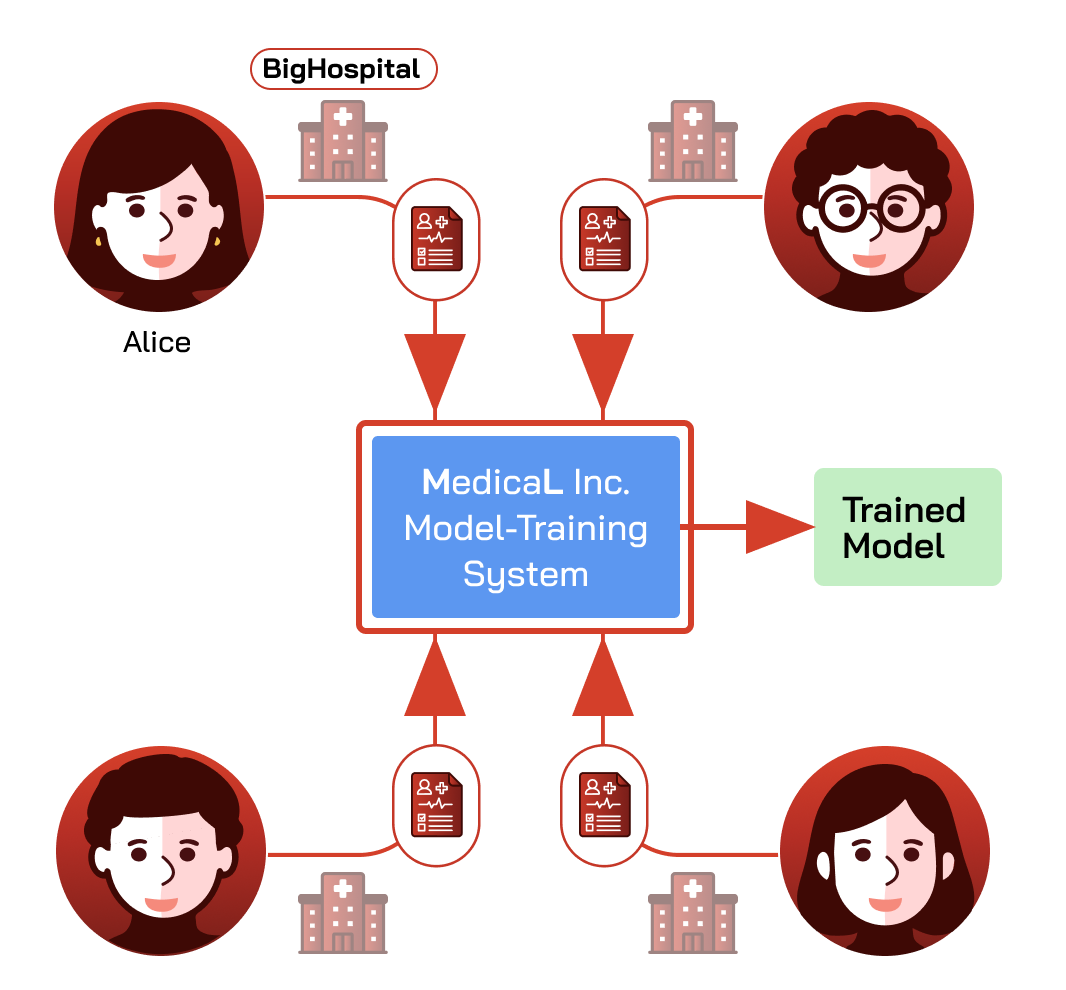}
    \caption{\MedicaL training a health diagnostics model over EHRs supplied by individual users, as obtained from their medical providers.}
    \label{fig:MedicaL}
\end{figure}

\begin{example}[Training a medical-diagnostics model \textit{on user-contributed data}]
\label{ex:training}

\MedicaL is training (or fine-tuning) a new ML model for medical diagnostics using private-web electronic health records (EHRs) \textit{provided individually by patients}, as shown in~\Cref{fig:MedicaL}. 
\end{example}

The goal here is potentially similar to (or even identical with) that in~\Cref{ex:medical_model}. The difference is \textit{how the training data is sourced}. A critical question thus arises: How can \MedicaL ensure it has received authentic EHRs? More generally,

\begin{questionBox}[Integrity of private-web training data] 
    How can ML training environments ensure receipt of authentic, untampered private-web data? 
\end{questionBox}

Referring again to~\Cref{ex:medical_model}, suppose that Alice uploads a copy of her EHR (as, e.g., a PDF) into the \MedicaL model-training system. She has purportedly obtained it by logging into her account in the web portal at BigHospital, her medical provider. 

With this arrangement, \MedicaL obtains no assurance that Alice's EHR did in fact originate as presented from BigHospital. Alice could have tampered with it---or could simply have fabricated it. Given existing web infrastructure, the only way \MedicaL \textit{could} be sure that the EHR came from BigHospital is to obtain it directly from BigHospital itself.\footnote{If BigHospital digitally signed the EHR, that would also prove its authenticity, but this would require BigHospital to deploy infrastructure beyond that of conventional web servers.} 

Fledgling companies, such as Kled~\cite{Kled2025}, attempt to do something of this kind.

\paragraph{Oracles:} Oracles can help solve this problem. Using an oracle, Alice can relay her EHR from BigHospital into {\MedicaL}\hspace{-2mm}'s model-training system and \textit{prove that the EHR comes directly through BigHospital's web portal}. She can alternatively relay only a portion of her EHR or transform its contents in some way that she specifies to \MedicaL As explained in~\Cref{subsubsec:oracles}, all of this is possible with \textit{no modification to existing web servers}, because the user initiates the connection. In our example, no change to BigHospital infrastructure is needed. In fact, BigHospital may not even be aware that Alice has exported her EHR. The red arrows in~\Cref{fig:MedicaL} show where oracles serve to relay data.

On the user side, Alice wants to know that her EHR remains private within the \MedicaL model-training system. In general, this question arises: 

\begin{questionBox} [Private-web data: Privacy in ML training]
    How can users be certain that the privacy of their private-web data is protected in ML-model-training systems?
\end{questionBox} 

For this purpose, Alice will want to use a \textit{privacy-preserving oracle}~\cite{zhang2016town,zhang2020deco}. As explained  in~\Cref{subsubsec:oracles}, a privacy-preserving oracle provides the source-authenticity assurances of an oracle, but additionally \textit{transmits data over an encrypted channel}. 

Of course, transmitting Alice's private data over an encrypted channel isn't very helpful if \MedicaL can directly decrypt it. To ensure that users' data truly remains private, trusted computing is a critical additional ingredient.

\paragraph{Trusted (confidential) computing:} As discussed in~\Cref{ex:medical_model} a trusted computing environment can receive users' data directly over the encrypted channels. In this case, the data can be relayed by users from data sources using privacy-preserving oracles. As a result, \textit{user data is never directly exposed} in the system (in this case, to \MedicaL).

Privacy for user data in transmission is important, but still doesn't provide  comprehensive privacy assurance, because the system itself could leak user data. In our example, Alice wants to know that her data is private in transmission and kept private during training.

To give this \textit{end-to-end} privacy assurance, the trusted computer executing the model-training system can emit an attestation for users proving that it is running a specific piece of privacy-preserving training software whose \textit{only output is a trained model}. Users' clients can check this attestation before transmitting EHRs through an oracle system. In this way, before she transmits her data, Alice knows that the data will remain private \textit{throughout the training life cycle}.

The user could additionally in this way obtain more fine-grained assurances, such as:

\begin{itemize} 
    \item \textbf{Differential privacy (DP):} Model outputs depend minimally on any single training input---and thus user privacy is protected within the trained model. This concept, the gold standard for privacy in statistical settings, is known as \textit{differential privacy}~\cite{abadi2016differential,blanco2022critical,demelius2025recent,dwork2006calibrating,dwork2014algorithmic,pan2024differential,wei2020federated,wang2023differential}.\footnote{More formally, DP stipulates that for privacy parameter $\varepsilon$, the inclusion or exclusion of a given user's  health record will change the probability of any model output by at most $e^{\varepsilon}$.}
    \item \textbf{Data retention:} All data will be deleted after the trained model is emitted.
    \item \textbf{Restricted use:} The resulting trained model can only be accessed by medical providers on a whitelist of approved hospitals. 
\end{itemize}

\begin{keyInsight}[Oracles for private-web data in ML]
Oracles enable privacy-preserving, authenticated access to private-web data for ML model training. They do so with no modification to existing web infrastructure. 

\smallskip
Without oracles, secure private-web data access is only possible with special-purpose software and/or data-sharing agreements.
\end{keyInsight}

The ability of users to authorize data sharing from the private web opens up all kinds of possibilities, including a wholly new form of privacy-preserving data-sharing marketplaces, as discussed in~\Cref{sec:data-markets}. 

\subsection{Secure AI-Inference Pipelines}
\label{subsec:security_inference}

Oracles and trusted computing can also help create secure ML pipelines for \textit{inference} over private-web data. This capability is useful for scenarios like the following.

\begin{figure}[h!]
    \centering
    \includegraphics[width=1\linewidth]{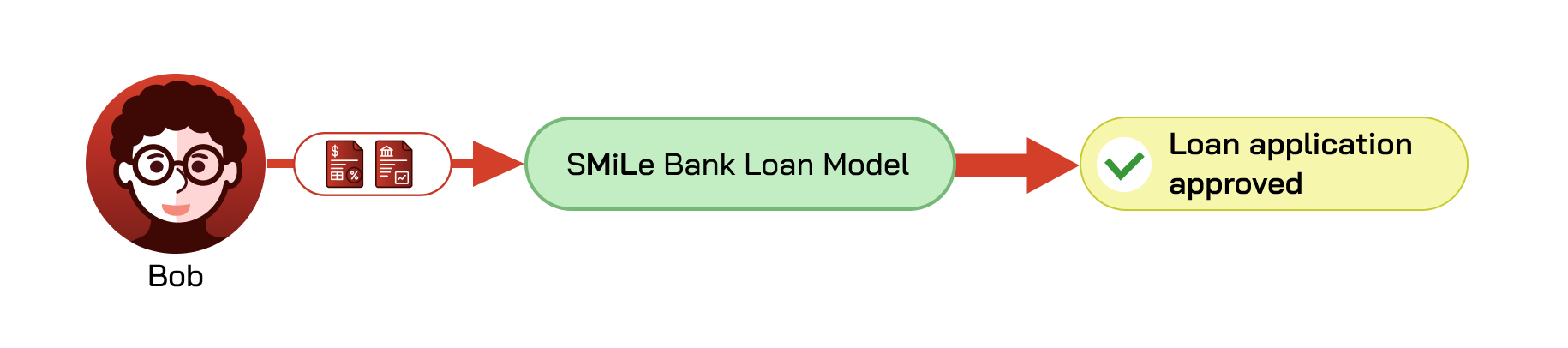}
    \caption{\sMiLe implements loan decisions using an in-house model. This model ingests financial documents from an applicant (Bob) and renders an approve / reject decision.}
    \label{fig:SMiLe}
\end{figure}

\begin{example}[Inference: loan decision]
\label{ex:inference}

\sMiLe uses ML models to render decisions on loan applications from its customers. This model inputs a set of financial documents provided by an applicant and outputs an approved / rejected decision on a loan request, as shown in~\Cref{fig:SMiLe}.

\end{example}

The loan approval process today typically involves borrowers downloading financial documents from the websites of their financial institutions or photographing them with their phones and uploading them to a lender portal. (Some automation is possible, e.g., tax transcripts can be pulled directly from the IRS with borrower authorization.) This workflow creates two problems:

\begin{itemize}
    \item \textbf{Integrity:} The lender cannot be certain that borrower-supplied documents are authentic---not fabricated or tampered with.
    \item \textbf{Privacy:} The borrower's documents are vulnerable to leakage from the lender's ML system. This is a potential problem for the borrower. It \textit{also} creates a liability risk for the lender.
\end{itemize}

By analogy with the setting shown in~\Cref{fig:MedicaL}, a combination of privacy-preserving oracles and trusted confidential computing can address both of these issues. Oracles can address the integrity problem by ensuring that documents come from trustworthy web sources---including private-web sources. Use of \textit{privacy-preserving} oracles and complementary use of confidential computing can address the privacy issues here on behalf of both the borrower and lender. 

\begin{figure}
    \centering
    \includegraphics[width=\linewidth]{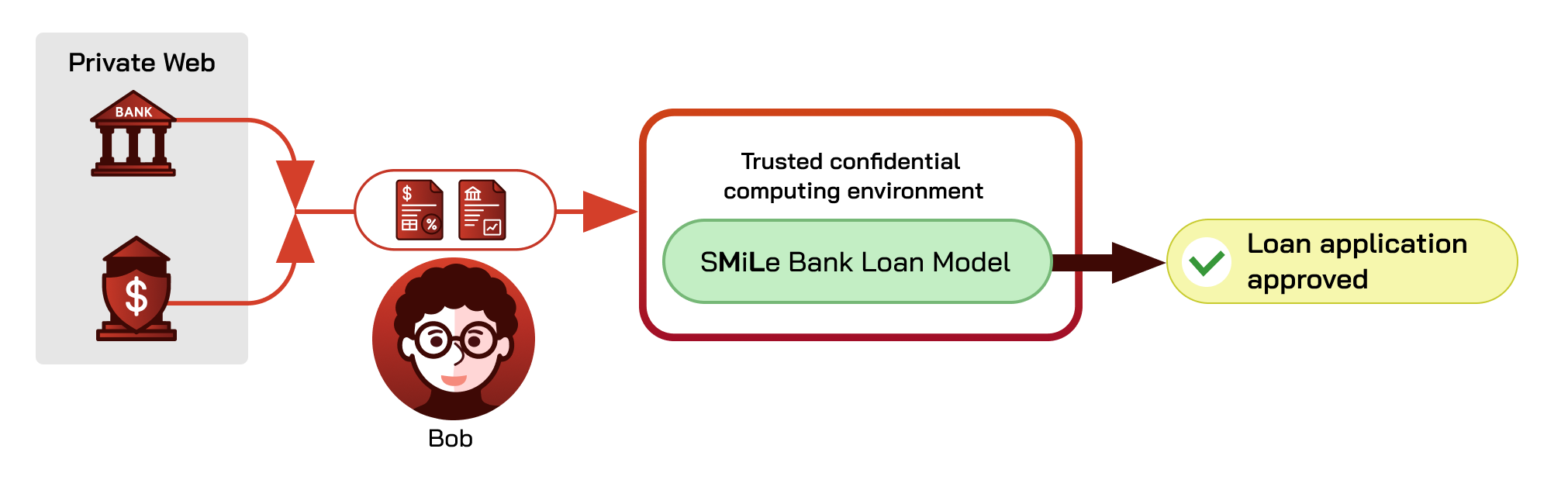}
    \caption{A \textit{secure inference pipeline} achieving integrity and privacy. Data are relayed from private-web data sources with integrity protection into a confidential computing environment running an ML model. Data remain private: Only model outputs are disclosed.}
    \label{fig:InferencePipeline}
\end{figure}

The result is a secure inference pipeline in which the lender learns only the output of the model, but has the assurance that it is based on trustworthy inputs from the borrower. This is shown in~\Cref{fig:InferencePipeline}. As in previous figures, red arrows depict the data flows a privacy-preserving oracle and the red box shows the trusted confidential computing environment.

\begin{keyInsight}[Private-web data for secure inference]
    Use of oracles and trusted (confidential) computing enables use of private-web data for secure inference, meaning inference that takes in trustworthy inputs and protects their privacy. 
\end{keyInsight}

\paragraph{Strong identity proofs:} Private-web sources can also serve to provide proofs of identity---realizing a form of decentralized identity system. The fact that Bob is able to relay financial documents---e.g., bank statements, W-2 forms---bearing his identity provides strong evidence that he is who he claims to be. This means that \textit{existing web services can be used as an ad-hoc identity system} usable to combat identity theft and various types of financial fraud, e.g., around benefits claims~\cite{maram2021candid,baldimtsi2024zklogin}. 

An ML model can also serve as a way to issue \textit{credentials}. For example, a model might take in tax returns, business registration documents, and banking statements for a small business owner from a set of whitelisted, trustworthy web sources. It could then output a credential certifying that the business meets revenue and operational requirements to qualify for a minority-owned business certification or is eligible to apply for a government contract, along with an attestation specifying the inference pipeline that generated the credential.

A receiver of the credential can assess its trustworthiness based on whether she trusts the inference pipeline, specifically the whitelisted web sources and the ML model that the attestation indicates for the inference pipeline.  

All of this can be accomplished in a decentralized way. Specifically, in principle \textit{anyone} can stand up a trustworthy inference pipeline, without explicit cooperation of data sources or any existing authority. 

\begin{keyInsight}[Secure inference pipelines for identities and credentials]
    Secure inference pipelines enable trustworthy decentralized verification of identities and issuance of credentials.
    \label{insight:identities}
\end{keyInsight}

\paragraph{Adversarial inputs:} A critical and stubbornly intractable security issue affecting ML models is \textit{adversarial inputs}, also known as \textit{adversarial examples}. These are inputs designed expressly to cause a model to create an erroneous output, often through modification of inputs that appear nonsensical. 

For instance, in~\Cref{ex:inference}, Bob might not qualify for a loan on the basis of authentic bank and brokerage statements. He might submit a tampered bank statement that looks valid upon visual inspection by a loan officer confirming the model's decision, but is in fact an adversarial input, as shown in~\Cref{fig:bank-statement}.

\bigskip

\begin{figure}[htbp]
\centering
\fbox{%
  \begin{minipage}{0.95\textwidth}
    {\ttfamily\upshape
    ACME COMMUNITY BANK\\[1ex]
    Monthly Statement --- Checking \textbullet\ Account ending in \textbullet\textbullet\textbullet\textbullet 4321\\
    Statement Period: Aug 1--31, 2025\\
    Account Holder: Robert Q.\ Applicant\\
    Address: 118 Forrester Ln, Oaks, NY 11700\\[1ex]

    Beginning Balance (Aug 1):\hfill \$\ \textcolor{red}{10}2,143.19\\
    Total Deposits/Credits:\hfill \$\ \textcolor{red}{1}9,843.10\\
    Total Withdrawals/Debits:\hfill \$\ 4,118.27\\
    Ending Balance (Aug 31):\hfill \$\ \textcolor{red}{11}7,868.02\\[1mm]
    }
  \end{minipage}%
}
\caption{Example of bank statement manipulated to inflate assets. \protect{\textcolor{red}{Red}} text indicates hidden PDF content that causes the ML model
    to misread the statement, but is not visible to a human being or optical character recognition (OCR) system.}
\label{fig:bank-statement}
\end{figure}

\bigskip

\noindent The model's erroneous perception of tampered, inflated figures causes it to output an approve decision when the application should have been rejected. Use of an oracle system, by authenticating document origins,  \textit{would help prevent such tampering}. 

\bigskip

The history of research on adversarial examples is a cycle of breaks and patches. Principled, effective defensive strategies have yet to emerge~\cite{tramer2020adaptive, carlini2019evaluating}. Proposed defenses---e.g., defensive distillation~\cite{papernot2016distillation}, input transformations~\cite{guo2017countering}, and gradient masking~\cite{athalye2018obfuscated}---have been repeatedly broken by adaptive attacks. This pattern continues with the latest models: even safety-aligned large language models remain vulnerable to simple adaptive attacks~\cite{andriushchenko2024jailbreaking}. Adversarial training—retraining models on adversarial examples—is currently the most robust defense, but comes with significant computational overhead and limited generalization~\cite{croce2020reliable,madry2017towards}. 

Despite extensive research, a general solution is still missing. Empirical defenses are routinely bypassed, and methods aiming for \textit{certified robustness}~\cite{wong2018provable, cohen2019certified} do not yet scale to modern, large neural networks~\cite{li2023sok,katz2017reluplex}. As a result, adversarial inputs remain a largely unsolved problem for deployed systems.

Standard defenses against adversarial inputs, however, involve modifications to inputs, model execution, or model training. In contrast, 
a secure inference pipeline \textit{constrains inputs} to authenticated web sources and in this way also \textit{constrains an adversary's ability to craft adversarial inputs}. This approach constitutes a new tool in the arsenal of defenses against adversarial inputs. 

\begin{keyInsight}[Secure inference pipelines limit adversarial inputs]
    Secure inference pipelines offer a new form of defense against adversarial inputs by constraining the set of inputs a user is permitted to send to a model. (This approach complements  model-level defenses.)
\end{keyInsight}

\subsubsection{Securing model privacy}

To this point, our concern has been with the protection and authenticity of data flowing through an inference pipeline. The model itself, however, also raises privacy considerations. Key among these is \textit{model privacy}.

\paragraph{The model privacy problem:} Privacy isn't just an issue for users. 
It's also an issue for AI system operators, who concern themselves with a range of attacks on the privacy of models by adversarial users, including:

\begin{itemize}
    \item \textbf{Model extraction:} %
    An adversarial user can extract features from a (black-box) model---or, in some cases, the full model itself---with carefully constructed queries~\cite{carlini2024stealing,jagielski2020high,orekondy2019knockoff,tramer2016stealing}. This attack is known as \textit{model extraction} or simply \textit{model stealing}~\cite{tramer2016stealing}.
    \item \textbf{Membership and training-data extraction:} ML models may be viewed as (compressed) representations of their training data. By querying the model, therefore, an adversarial user can potentially learn \textit{membership}, i.e., what (or whose) data was present in a model's training set, or even extract raw training data~\cite{carlini2021extracting,Salem2019MLLeaks,shokri2017membership,yeom2018privacy}. 
    \item \textbf{Uncovering model deployment choices:} The configuration choices and preprocessing of AI systems can be politically controversial~\cite{feng2023pretraining}. Users able to learn and extract evidence of such choices can cause harm to the reputation of a system operator.
\end{itemize}  

For all of these reasons, ensuring privacy of the model is a key security goal. Running a model in a confidential computing environment (e.g., \cite{sun2023shadownet,bayerl2020offline}) mitigates the most obvious privacy risks via direct parameter extraction, but the \textit{interface with users} still poses substantial security risk due to users' ability to selectively query and interact with models \cite{carlini2024stealing}. 

\paragraph{Proposed approaches:} The community has proposed various mitigations, but with serious drawbacks. Injecting noise during model training to defend against membership / training-data extraction~\cite{abadi2016differential,papernot2018scalable} degrades model performance and doesn't wholly eliminate leakage at scale~\cite{carlini2021extracting}. Limiting output granularity, e.g., rounding confidence values in model outputs, can hamper model extraction / theft~\cite{tramer2016stealing}, but also degrade model utility. Detection of theft by fingerprinting / watermarking models, i.e., enabling proof in stolen (``surrogate'') models of the creator's identity~\cite{adi2018turning,chen2019deepmarks, le2020adversarial,zhang2018protecting}, or LLM outputs~\cite{kirchenbauer2023watermark} is only possible after the fact and of limited efficacy (at least for generative models)~\cite{zhang2023watermarks}.

\paragraph{Deployed approaches:} Operators \textit{already} take steps to protect the privacy of their platforms. Services such as ChatGPT impose rate limits on users and monitor for and throttle prompts that appear designed as automated parameter-extraction attempts. Explored in the research literature~\cite{juuti2019prada}, this approach appears also to be brittle~\cite{feng2023stateful}. This is especially true given that the effort required by a single user to learn significant information about a model can be quite low---e.g., researchers estimated in recent work that for a cost of \$8,000, they could steal the weights for one layer of OpenAI's gpt-3.5-turbo-1106 model~\cite{carlini2024stealing}. OpenAI has publicly stated that China-based firms and others are ``constantly trying to distill the models of leading U.S.~AI companies''~\cite{WSJOpenAI}.

Adding to the challenge of
capping activity or detecting anomalous behavior for adversarial users is the fact that users may be \textit{anonymous}, as they are today for various free-tier services. A single user can pose as \textit{many} users, mounting what is known as a \textit{Sybil attack}~\cite{douceur2002sybil} or can take advantage of the accounts of many users.

\bigskip

\begin{figure}
    \centering
    \includegraphics[width=0.95\linewidth]{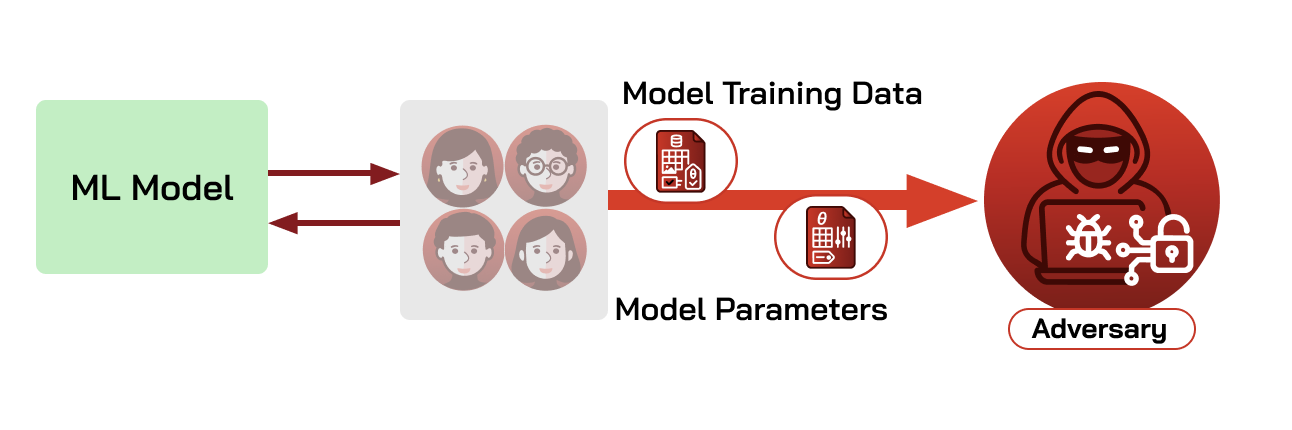}
    \caption{\textbf{User attacks on model privacy:} An adversary can query a model in ways that extract sensitive data from it, such as model parameters and pre- and post-processing algorithms. Rate limiting individual users is a natural but fragile countermeasure in open systems, as a single adversary can pose as multiple users.}
    \label{fig:Sybil}
\end{figure}

In short, there is a considerable risk that an adversary can extract sensitive data from a model, as shown in~\Cref{fig:Sybil}.

\paragraph{How secure inference can help:} Secure inference pipelines can help address the privacy issues associated with a model in two key ways. By requiring authenticated data from specific private-web sources through oracles, they can limit the \textit{type} of inputs adversaries transmit. In this way, they can help prevent  systematic extraction attacks that require a large number of diverse or specially-crafted queries---e.g.,  inputs designed to implement model extraction attacks.

Additionally, secure inference pipelines can limit the \textit{volume} of inputs an adversary can send to a model. The strong identity proofs generated within secure inference pipelines (as discussed in~\Cref{insight:identities}) can enforce per-user query bounds that can range from numerical caps to limits that depend on the nature of users' queries. It is possible to enforce these bounds in a privacy-preserving way, i.e., without disclosing user identities to platform operators~\cite{maram2021candid}. This approach specifically addresses the Sybil attack problem mentioned above, where a single adversarial user creates multiple anonymous accounts to bypass rate limits. 

\begin{keyInsight}[Secure inference pipelines for model privacy] By limiting the diversity and volume of adversarial queries, secure inference pipelines can help protect models against query-based extraction of their private data, such as model parameters or training data.
    
\end{keyInsight}

\subsubsection{Securing Agentic Memory}
\label{sec:agentic-memory}

Prior work has outlined a number of risks associated with agentic memory management at inference time. In so-called \emph{memory injection attacks},  an adversary modifies  the context fed to an agent by way of tool calls or external materials accessed by the agent \cite{patlan2025real,patlan2025ai}. These corruptions can cause the agent to behave in unexpected ways. For example, Patlan \emph{et al.}~show the effect of small memory corruptions in   ElizaOS agents \cite{elizaos}, a Web3 agent framework that manages significant cryptocurrency assets \cite{patlan2025ai}. Upon retrieving the poisoned context, agents could be induced to undesirable behaviors, including conducting unauthorized transactions. Similar attacks were later shown on web navigation agents \cite{patlan2025context}. 

Secure computation technologies or TEEs could partially help  protect the integrity of agent memory, as detailed earlier in this section. 
For the agentic memory corruption problem, TEEs offer (at least) two main possibilities. 
First, the agent could itself run on a TEE, thereby preventing a corrupted host device from manipulating agent memory. 
Second, the TEE could be used to pull only authenticated context material from (relatively) trusted sources. 
This could prevent manipulation of legitimate context materials, as well as sources masquerading as coming from a legitimate source.

However, even with a TEE, two challenges arise with regards to memory injection attacks:
\begin{enumerate}
    \item Trusted sources can still contain manipulated content. For example, if the trusted source is a social media platform, content may be produced by (untrusted) users, who could easily poison the text of their own posts, which could then be ingested by an agent.
    \item TEE operators could launch \emph{rollback} or \emph{forking} attacks, in which they interrupt execution and roll back the state of the TEE to a previous checkpoint, effectively erasing memory updates that occurred after that checkpoint. Without additional state consistency protections, a malicious host can equivocate, presenting different inputs to the restored enclave than were provided in the original execution, thereby causing the agent’s memory to fork into divergent, inconsistent states.
\end{enumerate}

The first problem---detecting corrupted content---is a grand challenge for the ML community at large and cannot inherently be solved with crypto-based techniques. 
For instance, see \Cref{sec:myths} for a discussion of the related problem of detecting GenAI-generated content.
However, the second problem---forking and rollback attacks---has been tackled using tools from the consensus literature. 
Systems like ROTE~\cite{matetic2017rote} and Narrator~\cite{niu2022narrator} introduce a \emph{consensus-based} approach
to protecting against rollbacks. The system realizes rollback protection as a distributed protocol where enclaves on different machines collectively maintain state freshness through mutual attestation and counter synchronization via quorum certificates. Such systems can be naturally extended to leverage public blockchains to ensure state consistency of TEE executions.
``The Forking Way''~\cite{wilde2024forking} further extends the rollback protection literature (ROTE~\cite{matetic2017rote}, Ariadne~\cite{strackx2016ariadne}, Narrator~\cite{niu2022narrator}, CloneBuster~\cite{briongos2023no}) by showing that even when blockchains are used as a distributed trust anchor, the integration is often flawed, and correct integration may incur substantial performance costs. 

\begin{keyInsight}
    TEEs can help to prevent agentic memory corruption attacks that arise due to corrupt host devices and/or context that is retrieved from unauthenticated sources. However, it is not well-suited to the problem of detecting whether a fresh piece of content has malicious content injected.
\end{keyInsight}

\subsection{Protected Pipelines (\Props)}
\label{subsec:Props}

\begin{figure}[h!]
    \centering
    \includegraphics[width=0.85\linewidth]{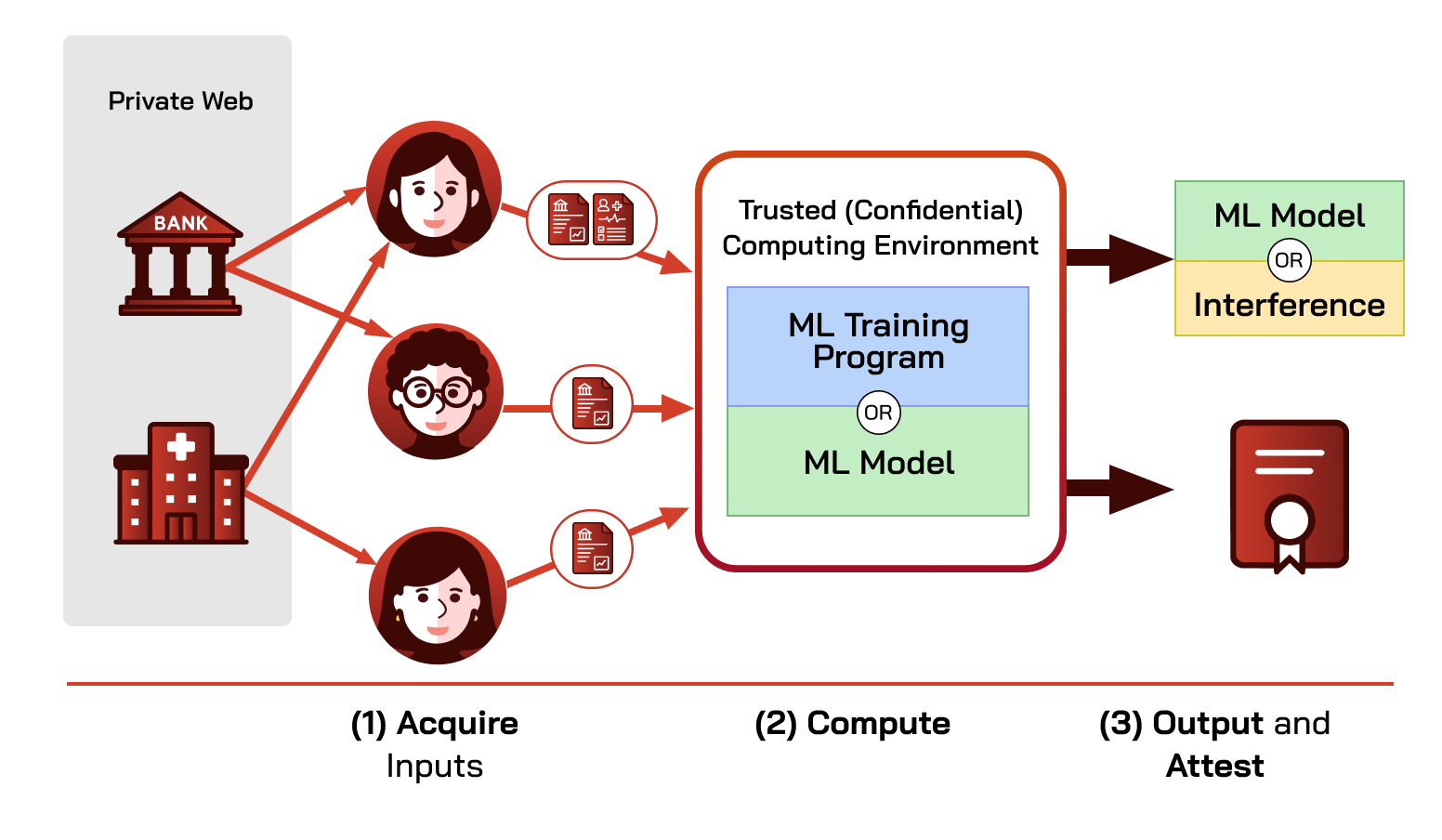}
    \caption{Protected Pipelines (\Props) Schematic.}
    \label{fig:Props}
\end{figure}

A bird's-eye view of the architectures and security goals discussed in this section leads to a generalized idea called \textit{Protected Pipelines}
(\Props)~\cite{juels2024props}. \Props are a broad, simple architectural framework for \textit{safe use of private-web data} (e.g., bank records, EHR excerpts, enterprise documents) in ML applications with \textit{no modification to existing infrastructure}. \Cref{fig:Props} depicts the protocol flow in \Props as it occurs either in ML training or inference, calling out the three key elements of a pipeline instantiated in \Props.

\Props compose the two core building blocks of oracles and trusted computing into a pipeline with three stages:

\begin{enumerate}
  \item \textit{Oracles} \textbf{acquire} inputs from private-web sources for the \Props pipeline. This initial stage of the pipeline fetches data from sources (and content contexts) authorized by the compute environment. Oracles provide \textit{proof} that the data comes from specified, approved data sources.  
  \begin{itemize}
      \item \textit{Prior attestation}: Before  sending data into the computing environment, user client software obtains an attestation from the trusted computing environment showing that it is running software trusted by the client. 
  \end{itemize}
  \item The \textit{trusted computing environment} \textbf{computes} over data---either training an ML model or performing inference using a particular existing ML model. 
  \item The \textit{trusted computing environment} \textbf{outputs} either a model or inference (based on training or an existing ML model). It also \textbf{attests}---generates a certificate---showing the properties of the pipeline that yielded the output, e.g., the private-web sources of data, the specific training software / ML model (expressed as a code hash), and so forth.
\end{enumerate}

\paragraph{Security properties:}

The overarching goal that \Props realizes is \textit{safe} use of private-web data  for ML applications---with existing, unmodified infrastructure.
\textit{Safe} here means addressing \emph{integrity} risks (tampered or falsified inputs or outputs), and \emph{privacy} risks (unnecessary leakage of data or models), resulting in these key \textit{theoretical} security properties:

\begin{itemize}
  \item \textbf{End-to-end input integrity:} Pipeline outputs depend on data authenticated as originating with trustworthy private-web sources.  
  \item \textbf{Confidentiality by default:} Inputs and intermediate state are never exposed in the clear outside the protected boundary. Only outputs are disclosed.
  \item \textbf{Attestation without disclosure:} Attestations provide assurance for providers of inputs and users of outputs of the integrity and confidentiality properties of the pipeline---specifically that trustworthy sources are consumed by trustworthy software.
\end{itemize}

\paragraph{Transparent \Props:}
While we have emphasized the power of private data and computation, a \textit{transparent} variant of \Props is also possible and useful: data and computation need not be private, only authenticated / integrity-protected. Data can also be sourced---and authenticated---from either private-web or public-web sources.

\begin{keyInsight}[\Props:\hspace{-1mm} A framework for private-web data in ML]
\textit{Protected Pipelines} (\Props) are a general framework for secure use of private-web data without infrastructure modifications. Security in \Props ensures that data come from trustworthy web sources and that data privacy is enforced across the full pipeline. 

\Props rely on two key technologies: privacy-preserving oracles and trusted computing.
    
\end{keyInsight}

\subsection{Research Questions}
\label{subsec:Props_research_questions}

Many lines of research surface in the exploration of secure pipelines and use of private-web data for ML---ranging from applications, to security, to practical deployment. 

\paragraph{Applications:} This section has offered a couple of example applications---medical-model training over private-web-sourced EHRs (\Cref{ex:training}) and inference for lending decisions based on financial documents from trustworthy sources (\Cref{ex:inference}). Given the vast, untapped resource that private-web data constitutes, there are no doubt many applications that developers have yet to dream up, raising the question:

\begin{researchQuestion}
    What new applications are possible with broadly available secure deep-web data? 
\end{researchQuestion}

\paragraph{Measuring security:}

Sourcing data only from authenticated sources imposes strong constraints on inputs. In the case of inference, this feature would seem to limit opportunities for manipulation through \textit{adversarial inputs} like that shown in~\Cref{fig:bank-statement}. In the case of model training, we might expect similar limitations on an adversary's ability to perform \textit{model poisoning}~\cite{goldblum2022dataset}. 

Substantiating this intuition about security properties requires answers to several research questions: 

\begin{researchQuestion}
What are good \textit{security metrics} for an adversary's ability to construct successful adversarial inputs (in ML inference) or model-poisoning data (in ML model training) when inputs come from authenticated web sources?
\end{researchQuestion}

Good security metrics here are necessarily domain-specific. For instance, in the example shown in~\Cref{fig:bank-statement}, the bank's transaction system and API for financial statement define the adversary's action space. 
The literature on adversarial inputs generally does not consider inputs constrained by application considerations, with some exceptions, e.g.,~\cite{kireev2023adversarial}. A related question is:

\begin{researchQuestion}
What \textit{methodologies} can accurately assess security metrics for Research Question~\arabic{section}.\the\numexpr\value{tcb@cnt@researchQuestion}-1\relax?
\end{researchQuestion}

\paragraph{Privacy metrics / attestations:} In confidential ML model training, the only information about training data is disclosed in the trained model itself. A large literature, however, demonstrates the vulnerability of training data to extraction~\cite{carlini2021extracting,nasr2023scalable} and membership inference queries~\cite{shokri2017membership,ye2022enhanced,wen2024membership}, with theoretical results suggesting that training instances closest to decision boundaries are most explicitly encoded in model parameters~\cite{cortes1995support, feldman2020does, koh2017understanding}. Proposed countermeasures range from differential privacy techniques~\cite{abadi2016differential}, which can provide provable guarantees at the cost of model utility, to more heuristic ones, e.g.,~\cite{srivastava2014dropout}. These observations raise the question:

\begin{researchQuestion}
    What trusted-computing attestations can provide meaningful privacy assurance to an individual contributing a training instance in the training of an ML model?
\end{researchQuestion}

\paragraph{Practical deployment:} Given the current state of toolchains and hardware, trusted (confidential) computing can impose substantial overhead on ML workloads---particularly for data-intensive model training of, e.g., deep neural networks (DNNs). This overhead comes from cryptographic operations on I/O and applies to a range of different environments, including TEEs that incorporate NVIDIA Confidential Computing~\cite{nvidia2023confidential,confidentialcomputing2024}, AWS Nitro (as implied by~\cite{lutsch2025analysis}), etc. This problem applies to confidential model training in general, but is especially critical to model-training variants. 

\begin{researchQuestion}
How can ML model training in trusted confidential computing environments be scaled effectively?
\end{researchQuestion}

\paragraph{Data markets:} By making available otherwise inaccessible forms of private-web data, crypto tools can give rise to new markets, prompting the question: 

\begin{researchQuestion}
    How can crypto tools drive the creation of effective data markets, both decentralized and centralized?
\end{researchQuestion}

This question isn't just technical: It raises legal and market-design issues. We consider this topic in~\Cref{sec:data-markets}.

\chapter{Misconceptions and Half-Truths}
\label{chap:conclusion}
\label{sec:myths}
In the excitement of building Crypto x AI platforms and applications, several common misconceptions or misleading statements have emerged.
In this brief chapter, we attempt to clarify five of these misconceptions.
While none of them are outright falsehoods, we attempt to clarify which parts of each statement are currently true, and which parts require more evidence. 

\begin{myth}[Gen AI Detection]
    Blockchains can help distinguish Gen AI content from human-generated content. 
\end{myth}

An oft-cited application of blockchains for AI is distinguishing between AI-generated content and human-generated (``real'') content. 
This narrative often suggests that by registering content on-chain, one can later determine whether it originated from an AI system or a human~\cite{vassallo2023ai_x_blockchain, ventionteams2024_generative_ai_blockchain, partz2025_can_blockchain_prove}.
Some AI projects are already recording GenAI outputs on-chain (e.g., Everlyn AI). Blockchains \textit{cannot} accomplish this goal in general---only in very limited ways that we discuss here.
In particular, blockchains are well-suited for timestamping and registering specific digital artifacts. 
This functionality is of limited utility for solving the broader problem of distinguishing AI-generated from human-generated content, however.

To evaluate the limitations of this approach, it is essential to distinguish between content detection (identifying whether a piece of content was generated by a human or by AI) and content provenance (identifying where a piece of content came from).

\smallskip\noindent \emph{Content Detection:}
Current methods for distinguishing human from AI-generated content primarily rely on post-generation detection.
They aim to do so without prior metadata or embedded signals. 
Generally, they fall into two categories: AI-based classifiers and statistical forensics.
AI-based classifiers use deep learning models trained to identify statistical patterns unique to generative models~\cite{10.5555/3618408.3619446}.
In contrast, statistical forensic methods analyze the data's mathematical and physical properties, such as identifying pixel-level noise distributions or structural anomalies (e.g., biological inconsistencies in AI-generated faces)~\cite{wang2020cnn}.

A blockchain, however, cannot perceive these off-chain artifacts on its own. 
Therefore, any classification of content as ``human-generated'' or `AI-generated'' must be provided to the blockchain by an external classifier.
When the output of such a classifier is integrated with a blockchain, this anchors the classifier's output.
While a blockchain can guarantee the integrity of a record (i.e., that it has not been tampered with after submission), however, it cannot guarantee that the information was true at the moment it was recorded.
If an external detector provides an incorrect classification, the blockchain preserves that error permanently.
Thus, in this setting, the blockchain provides integrity of claims, not verification of their truth.

\smallskip\noindent \emph{Content Provenance:}
Content provenance focuses on documenting the history of a digital asset from the time of its creation. 
Industry standards, such as the Coalition for Content Provenance and Authenticity (C2PA), allow creators or devices to attach cryptographically signed metadata claims (known as \emph{content credentials}) to media, documenting the origin, authorship, and any subsequent edits~\cite{c2pa2024spec}. 
In this model, a camera or a generative tool attaches content credentials to the file at creation time. 
Some companies and projects (e.g., Numbers Protocol~\cite{numbers2025whitepaper} and Starling Lab~\cite{starling2024framework}) use blockchains as a public immutable registry to log these content credentials. 
Everlyn~\cite{lynlabs2025docs}, an AI video generation model, automatically anchors a cryptographic hash of its outputs to a decentralized ledger at creation time.
This ensures that any video generated by the model has a permanent, publicly verifiable record that identifies it as AI generated.
By posting a hash of the content credentials on-chain, these systems aim to maintain a transparent log that remains persistent even if the original file's metadata is stripped. 
While blockchains can act as a public, immutable registry for these signatures, the trust still resides in the hardware or software that generated the signature, not the blockchain itself.

Even a robust provenance system anchored to a blockchain cannot guarantee whether a piece of content was originally generated by a human or by AI. 
For example, a user could display an AI-generated image on a high-resolution monitor and then photograph it using a C2PA-compliant camera. 
The resulting file would contain valid, cryptographically signed credentials identifying it as an authentic photograph captured by a physical device.
Similarly, a user could generate text with an AI system and then manually retype the same text into a C2PA-compliant editor, producing a file with legitimate provenance metadata indicating human authorship within the compliant tool.

Furthermore, if a piece of  content---whether human- or AI-generated---has a blockchain record, but is then modified so that it can no longer be matched to that record, then its provenance will be lost. In the absence of a \textit{universal} registry for content---which is improbable anytime in the foreseeable future---a provenance system will necessarily have large gaps.

\paragraph{Takeaway:} 
While blockchains provide a robust mechanism for the integrity of provenance metadata in a narrow sense, they are far from a comprehensive solution to the GenAI detection problem.
An effective solution would require a universal ecosystem where every piece of digital content is captured using a trusted device (e.g., a C2PA-compliant camera) and immediately anchored to a blockchain. 
In reality, the vast majority of digital content is currently created and shared using tools and platforms that do not support cryptographic anchoring, leaving unlabeled content in a state of ambiguity. 
Thus, while blockchains can act as a high-integrity registry for some content, their role is limited to preserving claims about content, not resolving the broader challenge of distinguishing human from AI-generated material.

\begin{myth}[Fair and Unbiased AI]
    Blockchains (or decentralization more broadly) can solve  bias and fairness problems in AI. 
\end{myth}
Today, there is a common viewpoint that by running model inference and training on a blockchain, we can solve common problems of unfairness and bias in AI \cite{leap-in,grayscale,onchain}. 
To evaluate this very broad statement, we must tease apart the different kinds of bias that can arise in ML. 

\smallskip\noindent \emph{Algorithmic bias:}
The most common notion of fairness in the AI community is algorithmic bias: models are known to learn (and sometimes amplify) imbalances in datasets \cite{pessach2022review,mehrabi2021survey}.
This can result in discriminative models that perform poorly on under-represented populations \cite{drozdowski2020demographic} and generative models that mimic undesirable properties or sentiments of their training data (e.g. using toxic language, perpetuating stereotypes) \cite{guptabias,lum2025bias}.
For a fixed definition of algorithmic fairness, the ML community has proposed many technical solutions to enforce fairness, including at training time \cite{han2022balancing,mandal2020ensuring,dwork2012fairness} and at inference time \cite{sadeghi2020imparting}, e.g., guardrails for AI models \cite{dong2025safeguarding,rebedea2023nemo}. 
However, these protections are far from perfect; fairness is not considered a solved problem in the ML community, and may never be \cite{anthis2025impossibility}. 
Even deciding how to define fairness is challenging, and typically requires making substantial tradeoffs \cite{kleinberg2017inherent}. 

Algorithmic bias is unlikely to be solved by decentralized AI, because it arises inherently in the training process and is typically mitigated by revised training or inference techniques.
Hence, decentralization does not address the source of the problem. 
However, a second source of potential bias stems from higher-level decisions that affect model performance; examples include what data is used, what model architecture to adopt, and how to compensate stakeholders who contribute to a particular model. 
While this is orthogonal to how the AI community typically views fairness, it is something that could potentially impact algorithmic bias and partially be addressed with decentralization. 
Specifically, decentralization offers two desirable properties: (1) transparency and (2) decentralized governance. 

\smallskip\noindent \emph{Transparency:}
Since transparency is a central property of blockchains, AI model developers could use blockchains to publicly commit to training data, training algorithms, model checkpoints, and the kinds of inference-time guardrails that a model has in place. 
Here, by ``transparency", we mean that operators can provably track the outputs of various operations, like a training run or a model inference.\footnote{Note that this notion of transparency does not explain \emph{why} an operation gave a particular output.}
Indeed, there are several platforms that aim to track  such information for downstream model users \cite{bittensor,sentient}.
From this information, users could check, for instance, that their data was not used to train a model, or that the model explicitly filters outputs that are considered objectionable. 
While such transparency could have benefits, it is challenging to scale to larger models and training-time artifacts such as model checkpoints, due to high storage and computational costs \cite{mirkin2025arbigraph}. 
In existing systems, much data from model training (e.g., training datasets, checkpoints) appears to be stored off-chain anyway and often cannot be accessed directly by users~\cite{mafrur2025ai}. 
Hence, in the short term, transparency benefits may be limited to inference \cite{mirkin2025arbigraph}. 
\textbf{Transparency alone may not significantly change how people use and develop AI unless the industry also thinks carefully about how people will use that transparency}: concretely, what kinds of use cases do we expect to be solved by model transparency, and based on this, what interfaces should be developed? For example, is a primary goal to allow users to report improper data usage in model training? If so, additional infrastructure will be required to establish true data ownership and address the problem (e.g. with machine unlearning). The many components of this pipeline are just as technically thorny, possibly more so, than transparency itself.

\smallskip\noindent \emph{Decentralized governance:}
Finally, Crypto x AI platforms allow for decentralized governance models, as discussed in \Cref{sec:alignment-governance}. It is important to disambiguate between community governance \emph{mechanisms} explored and adopted in blockchain systems, such as token-weighted voting and liquid democracy, and the decentralized, autonomous governance manifested by DAOs. In the former case, many technical and performance-sensitive decisions in AI development are poorly suited to broad stakeholder input. However, community governance mechanisms may be well-suited to value-laden decisions relating to model alignment, and have indeed been explored by major AI developers \cite{huang2024collective, openai2024democratic}, if not yet meaningfully deployed. Crucially, none of these mechanisms inherently require a blockchain to implement: it is therefore incorrect to characterize them as problems in AI solved \emph{by} blockchain, even if blockchain-based systems could lend such processes additional transparency. True on-chain governance of AI would require that governance decisions be enforced by smart contract, whether through direct execution or through economic incentives such as stake slashing. Enforcement of this kind could increase the robustness of governance and strengthen user confidence in these systems. However, these potential benefits face many of the same technical barriers as blockchain-based transparency more broadly: current blockchain infrastructure is not well-suited to the storage and computational demands of AI development, and practical implementations would likely require significant advances in verifiable training. Blockchain-enforced governance of AI systems, while a coherent long-run vision, remains technically premature.

\paragraph{Takeaway:} While blockchains do not inherently help reduce algorithmic bias, they can indeed encourage transparency at various stages of the AI life cycle and broaden participation in AI governance. However, scalability challenges aside, the impact of these properties on downstream model outcomes remains unclear; practitioners should demonstrate through case studies and data how model transparency and alternative forms of model governance concretely impact  end user and developer experiences. 

\begin{myth}[Automation vs.~Autonomy]
    Giving AI agents a crypto wallet allows them to earn, spend, and ``survive'' on their own, thereby making them autonomous. 
\end{myth}

Projects building ``agentic wallets''~\cite{IntroducingAgenticWallets} and payment protocols~\cite{x402onSolana,seiAI} often claim that giving AI agents a crypto wallet makes them autonomous, because decentralized payments allow them to earn, spend, and survive on their own. We argue that such claims conflate several distinct notions, leading to a number of misconceptions.

Part of the ambiguity stems from the fact that ``autonomy'' typically means different things in the contexts of AI and blockchains.
In AI literature, an autonomous agent is a system that can act based on its own perception, learning, and experience rather than strictly following pre-programmed rules~\cite{russellArtificial2022}. 
Perhaps confusingly, smart contracts are often described as autonomous, but here the emphasis is on their resilience to adversarial manipulations such as tampering, censorship, and shutdown. 
To differentiate, we refer to the former as \textit{intelligence autonomy}, and the latter as \textit{execution autonomy}.
Modern AI agents already exhibit substantial intelligence autonomy, but not necessarily execution autonomy: system administrators can, for example, shut down the servers on which AI agents run. 

In claims that agentic wallets enable AI agents to transact ``autonomously'' (e.g., as seen in \cite{IntroducingAgenticWallets,x402onSolana,seiAI}), the autonomy in question is neither intelligence autonomy nor execution autonomy.
AI systems do not become more intelligent by possessing a wallet.
Nor do they become more resistant to human manipulation or shutdown.
Instead, having access to a wallet enables \textit{automation}: AI agents can programmatically trade, transact, and access on-chain infrastructure without human approval loops.

It is also important to note that blockchains are not uniquely required for such automation.
Centralized financial infrastructure can be, and has been, accessed programmatically by AI agents. 

However, a more defensible interpretation of such claims is that blockchain-based payment systems themselves offer stronger autonomy (than centralized alternatives), even though they may not uniquely benefit AI agents.
For instance, they can ensure that transactions by AI agents are not treated differently from transactions by humans (i.e., they offer neutrality and censorship resistance~\cite{wahrstatterBlockchainCensorship2024,ForkChoiceEnforcedInclusion,wadhwaAUCILInclusionList2025}).

\paragraph{Takeaway:} 
Agentic wallets allow AI agents to conveniently access financial APIs, enabling economic interactions to be automated without human approval loops. However, automation should not be confused with autonomy: merely possessing a wallet does not make AI agents independent of human control (e.g., operators may still shut down the models or infrastructure they rely on).
Moreover, automated payments do not require blockchains; similar functionality can exist in centralized financial systems. 
In comparison, however, blockchain-based payment systems can offer appealing properties such as neutrality and censorship resistance, which can be desirable for applications where payment suppression, censorship, or other forms of manipulation are a concern.

\begin{myth}[\textit{Transparent} AI=\textit{Trustworthy} AI]
Recording models' data provenance and inferences on blockchains results in trustworthy model deployment and use. 
\end{myth}

The transparency and immutability of blockchains seem on the face of it to be ideal tools for ensuring the \textit{trustworthiness} of AI models. This is the thesis of a widely referenced IBM blog post~\cite{ibm_blockchain_trustworthy_ai} and fits with some common misconceptions, which extend beyond models by implication also to AI agents.

\paragraph{Model transparency.}
Recording the provenance of a model's training data appears to create a form of transparency around the model's creation. But there's a large gap between a record of data provenance and assurance about the behavior of a model, because: (1) A blockchain record of provenance is not a \textit{proof} of provenance, i.e., proof of the composition of a training data set (although evidence of provenance is possible using techniques discussed in~\Cref{subsec:securing_training}); (2) Even exact knowledge of a model's training data is insufficient to determine how the model will behave, because the \textit{training procedure and computational environment} also determine model behavior; and (3) Even with knowledge of the full pipeline from data to model sufficient to replicate the model, the non-determinism inherent in most stochastic training renders it infeasible to verify model weights against a training pipeline, even in principle.

Finally, even with access to model weights, there is no generally effective mechanism for detecting backdoors or other adversarial manipulation introduced during training.

In short, recording information about model data and training on a blockchain does not provide direct assurance of its behavioral characteristics or the absence of adversarial manipulation.  

\paragraph{Inference transparency.} Records of model inputs and corresponding inferences can be recorded on blockchains to create apparent transparency around the \textit{use} of models.

Blockchains, however, make transactions transparent, not reasoning. A blockchain transaction $T$ stating that ``model $X$ was queried on input $Y$, yielding inference $Z$'' does little to establish the trustworthiness of $Z$, because such a record alone does not establish:

\begin{itemize}

\item \textit{Correct model execution:} $T$ alone does not provide \textit{proof} that the tuple $(X,Y,Z)$ actually resulted from execution of model $X$ as specified. (Such proof is possible, but requires use of TEEs or computationally expensive cryptographic techniques.)

\item \textit{Model trustworthiness:} Even if $T$ did provide such proof, there is a more foundational problem. A full record of the provenance of model $X$, as described above, does not prove the trustworthiness of model $X$ in a semantic sense, i.e., adherence to user expectations or industry or community norms. Specification of $X$ by means of, e.g., a hash of model weights, provides even less meaningful assurance, as the \textit{identity} of a model does not alone establish the \textit{trustworthiness} of a model. As a result, it is hard to translate $T$ into assurance of trustworthy inference. 

\end{itemize}

\bigskip

Blockchains \textit{are} helpful for certain goals relating to trustworthiness. For example, an organization can publish hashes of open-weight models on chain. Such hashes serve as an immutable reference, so that users know they're using an authentic, unmodified model. Similar ideas for tamper-evident logging have arisen for other applications: use of blockchains as records for firmware updates~\cite{boudguiga2017towards} and certificate transparency~\cite{laurie2021certificate}, a system that uses blockchain-like append-only logs to maintain a publicly auditable record of certificate issuances.

\paragraph{Takeaway:} There is a considerable gap between the recording of model data provenance and model inferences on blockchains and meaningful assurance of model and inference trustworthiness.

\begin{myth}[Decentralized AI Efficiency]
Decentralization inherently makes AI jobs more cost-effective for model developers and users.
\end{myth}
A prominent class of crypto x AI initiatives have proposed decentralized networks as enablers for more efficient and cost-effective AI. 
One important example is
decentralized physical infrastructure networks (DePIN) \cite{grass,bittensor,akash,theta,hivemapper,xiao2022monetizing} in which users rent out their own physical infrastructure (like GPUs). 
The main appeal of these decentralized networks is reduced cost; for example, renting a DePIN GPU can be substantially cheaper than renting one on a comparable cloud service provider \cite{theta,akash}. 
However, it is not always the case that cheaper machines lead to lower cost in AI jobs; 
we discuss this issue in some detail in \Cref{sec:depin}. 
The main message is that while some use cases are well-suited to DePIN, others could incur higher costs due to network costs. Since decentralized nodes and processes communicate over the public Internet, the throughput and latency requirements of an AI job can significantly impact the overall cost of a job.
Moreover, very large AI jobs (such as training frontier models) are typically throughput-bound.  
Today, direct cost comparison is challenging because we lack systematic benchmarks that profile AI jobs on DePIN networks and compare them to traditional cloud infrastructure. 

\paragraph{Takeaway:} While decentralized networks offer an appealing alternative to high-cost centralized cloud providers, we do not have enough data to predict when a job will be cheaper on existing DePIN or DeAI platforms vs. a centralized cloud service provider. While smaller jobs (e.g., inference, small-scale training) are likely to be cheaper on these networks, very large jobs (like training a foundation model) could be impacted by unreliable and low-bandwidth communication networks between nodes. More research is needed to understand these tradeoffs clearly.

\newpage
\chapter{Acknowledgements} 
This work was made possible by the generous support of the industry and foundation partners of the Initiative for CryptoCurrencies and Contracts (IC3). 

We would like to thank Paolo Costa for valuable comments and insights on bottlenecks in the training of foundation models, as well as feedback on~\Cref{sec:decentralized-infra-ai}. 

\bigskip
\noindent Giulia Fanti and Pramod Viswanath additionally acknowledge the support of NSF grant CNS-2325477. 

\smallskip

\noindent Ari Juels additionally acknowledges the support of NSF CNS-2427390 and Ripple UBRI. In addition to his academic role, Juels serves as Chief Scientist at Chainlink Labs.

\newpage

\addcontentsline{toc}{chapter}{Bibliography}
\printbibliography
\end{document}